\documentclass{article}

\usepackage[square,numbers]{natbib}
\usepackage{amssymb}

 \usepackage[main, final]{neurips_2026}

\usepackage[utf8]{inputenc} 
\usepackage[T1]{fontenc}    
\usepackage{hyperref}       
\usepackage{url}            
\usepackage{booktabs}       
\usepackage{amsfonts}       
\usepackage{nicefrac}       
\usepackage{microtype}      
\usepackage{xcolor}         
\usepackage{amsthm}
\usepackage{amsmath}
\usepackage{graphicx}
\usepackage{enumitem}

\newtheorem{theorem}{Theorem}
\newtheorem{proposition}{Proposition}

\newtheorem{corollary}{Corollary}
\newtheorem{definition}{Definition}
\usepackage{pifont}

\newcommand{\cmark}{\ding{51}} 

\definecolor{lightgray}{RGB}{180,180,180}
\newcommand{\xmark}{\textcolor{lightgray}{\ding{55}}}

\title{Selective Disk Bispectrum: A Complete and Rotation Invariant Image Descriptor}

\author{Adele Myers Lantow \\
Department of Physics\\
University of California, Santa Barbara\\
\And
Nina Miolane \\
Department of Electrical and Computer Engineering \\
University of California, Santa Barbara \\
}

\begin{document}

\maketitle
\begin{abstract}
    The classical bispectrum is a signal representation with two fundamental properties desirable for machine learning. First, it is invariant to translation of the signal, and second the signal can be reconstructed (inverted) from its bispectrum coefficients, up to a translation. These properties, along with its ability to capture higher-order frequency correlations, have made the bispectrum a fixture for signal processing across science and engineering. For image processing, the relevant symmetry is rotation, and the disk bispectrum has been proposed for this domain. However, its size and computation time grow almost cubically with the number of basis functions, making its tractability difficult on downstream tasks. Furthermore, no inversion or completeness proof had previously been proposed. We introduce the \emph{Selective Disk Bispectrum} (SDB), a structured subset of disk bispectrum coefficients whose size grows linearly with the number of basis functions. We derive an explicit inversion algorithm proving that the SDB is complete, since an image can be reconstructed, up to a rotation, from its SDB coefficients alone. We verify the inversion numerically and demonstrate the SDB's practical benefits for (i) classification and (ii) multi-reference alignment on translated and noisy MNIST, FashionMNIST, and cell microscopy images. (i)~Classification: Across datasets, the SDB fed to an MLP classifier matches learned baselines. (ii)~Rotational MRA: Across datasets, the SDB inversion recovers an image from $N$ noisy rotated copies. Our results bring to images what the classical bispectrum has long provided for 1D signals: an invariant that is complete, efficient, and invertible, and has proven advantages across downstream tasks.
    \end{abstract}
    
    \section{Introduction}
    \label{sec:intro}
    
    Introduced in the mid-20th century in signal processing~\citep{tukey1953spectral, brillinger1991some, nikias1993signal}, the bispectrum is a third-order statistic constructed from triple products of Fourier coefficients of an input signal. Its ability to encode higher-order frequency correlations has led to its widespread use across the sciences, from ocean waves~\citep{elgar1985observations} and seismic data~\citep{geophysics_2004}, to cardiac~\citep{martin2021bispectral} and brain signals~\citep{zhang2000bispectrum} (see~\citet{koukiou2024identifying} for a review). 
    
    The bispectrum has two fundamental properties desirable in machine learning tasks: translation-invariance and invertibility. Translation-invariance guarantees that bispectrum statistics will not be affected by the global phase of the signal, and will instead only be affected by shape changes. Bispectrum invertibility means a generic signal can be reconstructed from its bispectrum coefficients, up to translation~\citep{giannakis1989signal}, guaranteeing the bispectrum encodes all information about the original signal, other than its global phase. This makes it a so-called \textit{complete} invariant. 
    
    In \emph{classification}, completeness guarantees that two signals that differ by more than a translation receive different descriptors, hence no discriminative information is lost before learning starts. Integrating bispectra in neural networks leads to up to $+30\%$ accuracy over baselines under fixed parameter budgets, equal accuracy with 10× fewer parameters~\citep{mataigne2024selective}, and provable resistance to the invariance-based adversarial attacks that fool standard models~\citep{sanborn2023general}.  In \emph{multi-reference alignment} (MRA), where a signal is recovered from many translated, noisy observations of itself by taking their mean in bispectrum space, invertibility guarantees that the resulting mean can be used to reconstruct the average, de-noised signal~\citep{bendory2018}. 
    
    For images, the relevant transformation is often rotation rather than translation. Many images carry no meaningful orientation, like a cell under a microscope, a particle in cryo-electron microscopy, or a galaxy in a telescope. The \textit{disk bispectrum}~\citep{zhao2014rotationally} extends the bispectrum construction for rotation-invariance on images. Its coefficients achieve rotation-invariance by replacing the Fourier basis with the disk harmonic basis. However, it has seen limited use for three reasons. First, it is too big. The number of coefficients grows almost cubically with the number of basis functions. Second, it is too slow, for the same reason. Third, no inversion has been derived, and its completeness is unproven. Therefore, its tractability and discriminative power for classification is not guaranteed, and it cannot be used for MRA. 
    
    \begin{figure*}[t]
      \centering
       \includegraphics[width=\textwidth]{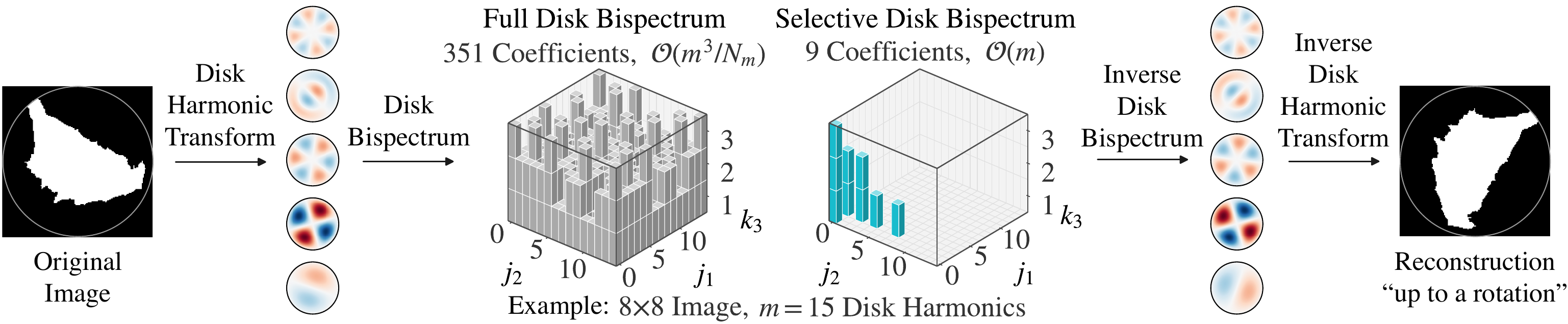}
        \caption{\textbf{Our selective disk bispectrum grows linearly with the number $m$ of disk harmonic basis functions.} For a small $8\times8$ image at the library default bandlimit ($m=15$ DH coefficients), the full disk bispectrum has 351 coefficients, whereas the SDB has 9. For a $28\times28$ image, these numbers rise to 152,025 and 97, respectively, for $m=185$ DH coefficients.}
       \label{fig:spatial-complexity}
    \end{figure*}
        
    \textbf{Contributions.} We propose a computationally tractable selective disk bispectrum, along with a disk bispectrum inversion. We provide an open sourced implementation for the forward and inverse SDB maps at \texttt{https://github.com/geometric-intelligence/bispectrum/}, and demonstrate their utility in downstream tasks.
    \begin{itemize}[leftmargin=*]
        \item \textbf{Disk bispectrum inversion (main theoretical contribution).} We derive the first analytic inversion of the disk bispectrum. This inversion proves that the disk bispectrum contains all information in an image, up to a rotation. We verify this inversion numerically by mapping MNIST, FashionMNIST, and cell microscopy images to and from their SDB coefficients.
    
        \item \textbf{The Selective Disk Bispectrum (SDB).} Our inversion reveals that most disk bispectrum coefficients are redundant: only a minimal subset is needed for complete image reconstruction. We call this subset the SDB. Because an image can be reconstructed from its SDB coefficients, the SDB is a \emph{complete invariant} --- two images have the same SDB if and only if they differ by a planar rotation alone. Using the SDB instead of the full disk bispectrum reduces spatial complexity from $\mathcal{O}(m^3/N_m)$ to $\mathcal{O}(m)$ and time complexity from $\mathcal{O}(L^3+m^3/N_m)$ to $\mathcal{O}(L^3+m)$, for an $L\times L$ image represented by $m$ disk harmonic basis functions with maximum angular frequency $N_m$. Time complexity can be further reduced to $\mathcal{O}(L^2\log L+m)$ with a recent approximation of disk harmonic coefficients \cite{fast_dhcs_2023}. Spatial complexity improvements are shown visually in Fig.~\ref{fig:spatial-complexity}.
    
        \item \textbf{Accuracy guarantees and noise analysis.}
        We provide SDB accuracy guarantees when using \cite{fast_dhcs_2023}'s disk harmonic coefficient approximation. We also derive its expectation and variance under i.i.d.\ noise and prove that most coefficients are asymptotically unbiased, the guarantee required to apply the descriptor in lower SNR regimes.
    
        \item \textbf{Classification of noisy and rotated MNIST, FashionMNIST, and cell microscopy images.} Using the SDB as input to an MLP classifier achieves comparable accuracy as pooled G-CNNs, harmonic networks, $SE(2)$ Networks or frozen DINOv3, EVA-02, and OpenCLIP features. Relative success between representations seems to be dataset-dependent.
    
        \item \textbf{MRA on noisy and rotated MNIST, FashionMNIST, and cell microscopy images.} Bispectrum-based MRA requires an analytic inversion, which did not exist for any rotation-invariant bispectrum until this work. We demonstrate that our inversion makes rotation-MRA possible, and propose an alteration to make the inversion more robust against noise propagation during averaging.
    \end{itemize}
    Together, these contributions characterize a new disk bispectrum whose newfound tractability opens doors for utilization in signal processing, science, and engineering.

    \section{Background}
    \label{sec:background} 
    We first review existing bispectra that motivated the definition of our new selective disk bispectrum.
    \paragraph{Classical Bispectrum.} The \emph{classical 1D translation-invariant bispectrum} ~\citep{tukey1953spectral} is defined as
    \begin{equation}
    B_{\nu_1,\nu_2} = F_{\nu_1} F_{\nu_2} F^*_{\nu_1+\nu_2},
    \end{equation}
    where $F_{\nu} = |F_\nu| e^{-i \nu \phi_\nu}$ are Fourier coefficients of $f: \mathbb{Z}/n\mathbb{Z} 
    \to \mathbb{C}$, with frequencies $\nu$. Under cyclic translation $g \in \mathbb{Z}/n\mathbb{Z}$, $f$ transforms as $\tilde{f}(x) = f(x - g)$, with Fourier coefficients transforming equivariantly as 
    $\tilde{F}_\nu = e^{-i\nu g} F_\nu$, so phases cancel in the bispectrum product:
    \begin{equation}
    \tilde{B}_{\nu_1,\nu_2} = e^{-i\nu_1 g}F_{\nu_1} \cdot e^{-i\nu_2 g} F_{\nu_2} 
    \cdot e^{i(\nu_1+\nu_2)g} F^*_{\nu_1+\nu_2} = B_{\nu_1,\nu_2}.
    \end{equation}
    Unlike the power spectrum $|F_\nu|^2$, which discards phase entirely, the bispectrum preserves relative phase $\phi_\nu$ between coefficients, making it a \emph{complete} invariant. This was proven by \cite{giannakis1989signal}, who demonstrated $f$ can be reconstructed from $B$ up to a global translation, as shown in Appendix~\ref{app:classical-inversion}.
    
    \paragraph{$G$-Bispectrum}  Mathematically, a group is a pair $(G, \circ)$, where $G$ is a set and $\circ$ a binary operation on that set. Groups must satisfy axioms of closure, associativity, inverse, and identity. The classical bispectrum is invariant to the group of 1D translations $(\mathbb{Z}/n\mathbb{Z}, +)$, where $+$ is addition. This concept is extended by the \emph{group-invariant $G$-Bispectrum} \citep{Kakarala2009CompletenessOB}, which is constructed from the Group-Fourier Transform (G-FT) \citep{rudin1962fourier}. For discrete groups, the G-FT is defined $F_{\rho} = \sum_{g \in G} \rho(g) f(g)$, where $\rho(g)$ is an irreducible representation of a group element $g \in G$, defined in Appendix~\ref{app:definitions}. Note that for the G-FT, $f : G \to \mathbb{C}$, meaning the domain of $f$ is an abstract group. Consequently, utilization of this bispectrum is not always straightforward, as most signals are defined over domains such as the line (1D signals), plane (2D images), sphere (earth models), or volume (3D voxels). 
    
    \paragraph{Disk Bispectrum} Because of the domain restrictions of the $G$-Bispectrum, the rotation-invariant disk bispectrum was developed for signals whose domain is planar, enabling utilization directly on images. In particular, it is convenient to consider an image as being a square integrable real-valued function $f: \mathbb{D} \to \mathbb{R}$ defined on the disk $\mathbb{D}$. This can be implemented for any $L \times L$ image by masking any pixels outside radius $L/2$. The disk bispectrum \cite{zhao2014rotationally}, defined as
    \begin{equation}
        b_{j_1, j_2, k_3} = a_{n_{j_1},k_{j_1}} \cdot a_{n_{j_2},k_{j_2}}\cdot a_{n_{j_1}+n_{j_2},k_3}^* \in \mathbb{C},
    \end{equation}
    \label{eq:disk_bispectrum}
    is a product of disk harmonic coefficients  \cite{Verrall_Kakarala_1998},
    \begin{equation}
        a_{nk} = \int_0^{2\pi} \int_0^1 f(r, \theta) \, \psi^*_{nk}(r, \theta) \, r \, dr \, d\theta,
    \end{equation}
    where $\psi_{nk}$ are \emph{disk harmonics} (DH), defined as $\psi_{nk}(r, \theta) = c_{nk} J_n(\lambda_{nk} r) e^{i n \theta}$. Here, \( (n, k) \in \mathbb{Z} \times \mathbb{Z}_{>0} \), \( J_n \) is the $n$-th order Bessel function of the first kind, \( \lambda_{nk} \) is its \( k \)-th root of the $n$-th Bessel function, and \( c_{nk} \) is a normalization constant. 
    
    Disk harmonics are a natural basis for a rotation-invariant bispectrum. Just as Fourier modes transform equivariantly under 1D translation, DH coefficients transform equivariantly under rotation, i.e. $a_{nk} \;\longrightarrow\; e^{i n \phi} a_{nk}$ for $\phi \in SO(2)$, the group of 2D rotations.  Moreover, just as Fourier modes $e^{-ivx}$ (see classical bispectrum) and frequencies $v$ are eigenfunctions and eigenvalues of the Dirichlet Laplacian over $S^1$, the domain of continuous cyclic signals, disk harmonics and their ``frequency'' \( \lambda_{nk} \) are eigenfunctions and eigenvalues of the Dirichlet Laplacian over $\mathbb{D}$. 
    
    Because disk harmonics have a natural notion of frequency, the Nyquist criterion dictates a frequency cutoff or \textit{bandlimit} of $\frac{\lambda}{2\pi} \leq \frac{L}{4}$ for an $L \times L$ image, where higher-frequency components correspond to features beyond the image resolution \citep{zhao2013fourier}. When \cite{fast_dhcs_2023} implemented the first tractable DH coefficients, they truncate according to this frequency cutoff. They order $\lambda_{nk}$ from lowest frequency to highest $\lambda_0 \leq \cdots \lambda_j\cdots \leq \lambda_{m-1}$, listing $-n$ before $n$ when $\lambda_{-n,k} = \lambda_{n,k}$. The index $j=0, ..., m-1$ represents this order. Here, notation shifts to accommodate indexing by $j$. Concretely, since DH frequency is unique to a $(n, k)$ pair, $\lambda_{n,k}$ can be equivalently written as $\lambda_j$ or $\lambda_{n_j k_j}$,  DHs $\psi_{n,k}$ can be written as $\psi_j$ or $\psi_{n_j k_j}$, and DH coefficients $a_{n,k}$ as $a_j$ or $a_{n_j k_j}$, where $n_j, k_j$ are the $(n,k)$ corresponding to the $j^{th}$ frequency $\lambda_j$.
    
    Then, the number of DH coefficients $m$ in an expansion $f(r, \theta) = \sum_{n \in \mathbb{Z}} \sum_{k \in \mathbb{Z}_{>0}} a_{nk} \, \psi_{nk}(r, \theta)$ is the number of DHs within the bandlimit, $m = \#\{(n,k) : \lambda_{nk} \leq \lambda\}$. Similarly,  \( n \in \{-N_m, \ldots, N_m\} \) where \( N_m = \max\{n_j \in \mathbb{Z} : j \in \{0, \ldots, m-1\}\} \) and $k \in \{1, ..., K_n  \}$ where \( K_n = \max\{k \in \mathbb{Z}_{>0} : \lambda_{nk} \leq \lambda\} \).  Table~\ref{tab:indices} in Appendix~\ref{app:dh-truncation} shows an example of this correspondence between $j$ and $(n,k)$ for an $8 \times 8$ image expanded in the truncated DH basis. In the disk bispectrum definition, $-N_m \leq n_{j_1}+n_{j_2} \leq N_m \text{, and } k_3 \le K_{n_{j_1} + n_{j_2}}$.
    
    
    \section{Related Works}
    We review the existing bispectra in Tab.~\ref{tab:bispectra-summary} and compare common rotation equivariant and invariant descriptors in Tab.~\ref{tab:rotation-summary}.
    
    \begin{table}[t]
    \centering
    \footnotesize
    \setlength{\tabcolsep}{4.5pt}
    \renewcommand{\arraystretch}{1.15}
    \begin{tabular}{lcccccc}
    \toprule
    \shortstack{\textbf{Rotation-Equivariant (Equiv.)}\\\textbf{\& Invariant (Invar.) Descriptors}}&
    \textbf{Equiv.} &
    \shortstack{\textbf{Local}\\\textbf{Invar.}} &
    \shortstack{\textbf{Global}\\\textbf{Invar.}} &
    \shortstack{\textbf{Magnitude}\\\textbf{is Invar.}} &
    \shortstack{\textbf{Invar. if}\\\textbf{ Pooled}} &
    \shortstack{\textbf{Invertible}} \\
    \midrule
    Selective Disk Bispectrum (SDB) (Ours) & \xmark & \xmark & \cmark & \cmark & \xmark & \cmark \\
    Full Disk Bispectrum \cite{zhao2014rotationally}  \\
    \midrule
    G-CNNs \cite{cohenc16} & \cmark & \xmark & \xmark & \xmark & \cmark & \xmark \\
    Spherical Convolutions \cite{article} \\
    $SE(2)$ Steerable Networks \cite{weiler2021general} \\
    Capsule Networks \cite{sabour2017dynamic} &  \\
    \midrule
    Disk Harmonic Transform \cite{fast_dhcs_2023} & \cmark & \xmark & \xmark & \cmark & \xmark & \xmark \\
    Fourier Mellin Transform \cite{reddy1996fft} \\
    Tensor Field Networks \cite{thomas2018tfns}  \\
    Harmonic Networks \cite{worrall2017harmonic} \\
    \midrule
    SIFT: Scale–Invariant Feature Transform \cite{lowe2004sift} & \xmark & \cmark & \xmark & \xmark & \xmark & \xmark \\
    SURF: Speeded–Up Robust Features \cite{bay2006surf} \\
    ORB: oFAST \& Rotated BRIEF \cite{rublee2011orb} \\ 
    AKAZE: Accelerated KAZE \cite{alcantarilla2012kaze}  \\
    \midrule
    Hu Moments \cite{hu1962visual} & \xmark & \xmark & \cmark & \xmark & \xmark & \xmark \\
    \midrule
    Zernike Moments \cite{teague1980zernike} & \xmark & \xmark & \xmark & \cmark & \xmark & \xmark \\
    \bottomrule
    \end{tabular}
    \caption{\textbf{The SDB (ours) is the only rotation-invariant descriptor that is simultaneously globally invariant, magnitude-preserving, \emph{and} invertible --- meaning it retains all image information up to a global rotation.} Equivariant methods achieve invariance only after pooling or taking a norm, costing of structural information. Local invariants discard global structure. Other global invariants are not invertible and therefore not complete. }
    \label{tab:rotation-summary}
    \end{table}
    
    \textbf{G-Bispectra and Bispectra on G-Homogeneous Spaces.} As discussed in Sec. \ref{sec:background}, the primary challenge of the $G$-Bispectrum is its assumption that the signal domain is the space of the group itself $G$. Thus, the $SO(2)$ invariant G-bispectrum is only well-defined for signals $f: SO(2) \to \mathbb{C}$, whose domain is the group $SO(2)$. This makes the $G$-Bispectrum ill-defined as a rotation-invariant image descriptor, because images $f: \mathbb{R}^2 \to \mathbb{C}$ lie in the plane $\mathbb{R}^2$. \cite{mataigne2024selective} circumvented this problem by taking $SO(2)$ group convolutions over an image, thus taking the $G$-Bispectrum of the convoluted signal whose domain is the group, rather than the image itself. \cite{kakarala2012} worked towards addressing the domain problem with a ``Homogeneous $G$-Bispectrum'', relevant for signals defined over spaces homogeneous for the group, $\mathcal{H}$. A space that is homogeneous for a group is one where for any $x, y$ there is a group element $g \in G$ s.t. $g \circ x = y$. This is helpful for $SO(3)$ invariant bispectra whose signals $f:S^2 \to \mathbb{C}$ are defined on the sphere $S^2$. 
    
    \textbf{Bispectra on Non-Homogeneous Spaces} 2D domains like planes and disks are not homogeneous for the group $SO(2)$, because $SO(2)$ group action can only ``move a point'' along the circle $S^1$. This necessitated the creation of the disk bispectrum \cite{zhao2014rotationally}, defined in Sec.~\ref{sec:background}, which is the first bispectrum defined for a signal not homogeneous to the invariant group. Unfortunately, the ``full'' disk bispectrum is both spatially and computationally expensive. To reduce spatial complexity and enable utility on classification tasks, \cite{zhao2014rotationally} resorted to PCA on bispectrum coefficients. Authors reported a feature-vector length of $\mathcal{O}(m^3 / N_{m})$, where $m$ is the number of disk harmonic coefficients, and $N_m$ the maximum angular frequency. Furthermore, they offered no disk bispectrum inversion, meaning that there was no guarantee that the disk bispectrum was a \emph{complete invariant}.
    
    \begin{table}[t]
    \centering
    \footnotesize
    \setlength{\tabcolsep}{4.5pt}
    \renewcommand{\arraystretch}{1.15}
    \begin{tabular}{lccccc}
    \toprule
    \shortstack{\textbf{Bispectrum Name}}&
    Invariant to &
    Signal Domain &
    Constructed From &
    Inversion \\
    \midrule
    Classical Bispectrum &$SO(2)$ &$\mathbb{R}$ & Fourier Coefficients &\cite{giannakis1989signal} \\
    Classical 2D Bispectrum &$SO(2)\times SO(2)$ &$\mathbb{R}^2$ & 2D Fourier Coefficients & \cite{giannakis1989signal} \\
    $G$-Bispectrum &$G$ &$G$ &$G$-Fourier Coeffs. \cite{rudin1962fourier} &\cite{mataigne2024selective} \\
    Selective $G$-bispectrum \cite{mataigne2024selective} &$G$ &$G$ &$G$-Fourier Coeffs.\cite{rudin1962fourier} &\cite{mataigne2024selective}\\
    $G$-Bisp. on Homogeneous Spaces $\mathcal{H}$ \cite{kakarala2012} &$G$ & $\mathcal{H}$ &$G$-Fourier Coeffs.\cite{rudin1962fourier}  &None \\
    Full Disk Bispectrum \cite{zhao2014rotationally} &$SO(2)$ & Disk $\mathbb{D}$ & Disk Harmonics \cite{Verrall_Kakarala_1998, fast_dhcs_2023} &\textbf{(Ours)}  \\
    Selective Disk Bispectrum (SDB) \textbf{(Ours)} &$SO(2)$ & Disk $\mathbb{D}$ & Disk Harmonics \cite{Verrall_Kakarala_1998, fast_dhcs_2023} &\textbf{(Ours)}  \\
    \bottomrule
    \end{tabular}
    \caption{\textbf{The SDB (ours) is the first bispectrum on a non-homogeneous space to admit an analytic inversion, establishing it as a complete invariant.}}
    \label{tab:bispectra-summary}
    \end{table}
    
    \textbf{Rotation-invariant descriptors} Existing rotation-invariant descriptors (Tab.~\ref{tab:rotation-summary}) do not possess the disk bispectrum's mathematical properties of completeness and inversion. Hand-crafted descriptors such as SIFT~\cite{lowe2004sift}, ORB~\cite{rublee2011orb} are locally invariant to rotation but  Zernike moment magnitudes~\cite{teague1980zernike}, and Fourier--Mellin magnitudes~\cite{reddy1996fft} are fast and require no training, but they are not complete representations; the original signal cannot be recovered from their coefficients. This means that two images with very different visual content could map to identical descriptors, making them potentially less reliable for fine-grained recognition tasks. $G$-CNNs~\cite{cohenc16} and related equivariant architectures~\cite{cohen2018spherical, thomas2018tfns} have achieved great success in image processing. However, they obtain rotation-invariance through pooling, which is a lossy operation, or through data augmentation, which spends data and compute to approximate a symmetry that is known a priori. Our experiments will compare our selective disk bispectrum to these descriptors and show that its properties of completeness and inversion provide a learning-free way to achieve rotation-invariance and discriminative power in classification, and they are indispensable in other tasks such as MRA, which requires an invertible rotation-invariant descriptor.

    \section{Theory: The Selective Disk Bispectrum and Its Inversion}
    \label{sec:theory}
    We propose the selective disk bispectrum (SDB), defined as a specific subset of disk bispectrum coefficients. Proofs for all theorems are in Appendices~\ref{app:sdb-proofs}, \ref{app:approx-proofs}, and \ref{app:noise-proofs}.
    \begin{definition}[Selective Disk Bispectrum (SDB)]\label{def:selective-bsp}
    Let $f: \mathbb{D} \to \mathbb{R}$ be a real-valued square-integrable function and let $a_{n, k}$ be its disk harmonic coefficients, bandlimited by $\lambda$. The selective disk bispectrum is defined as the following subset:
    \begin{equation} \label{eq:selective_disk_bispectrum}
    b^f = \left[
    \begin{array}{cccccccc}
    b_{0,0,1}  \cdots b_{0,0,K_0} \\
    b_{2,0,1}  \cdots b_{2,0,K_{1}} \\
    \ldots\\
    b_{2,N_m-1,1} \cdots b_{2,N_m-1,K_{N_m }}
    \end{array}
    \right]
    =\left\{
    \begin{aligned}
        b_{0, 0, k} 
        &= a_{0, 1}^2 \cdot a_{0, k}^*,  \text{ for } k \in \{1, \dots, K_{0}\},\\
        b_{2, n, k}  
        &= a_{1, 1} \cdot a_{n,1} \cdot a_{n+1,k}^*, \\
        & \text{ for } k \in \{1, \dots, K_{n+1}\}, n \in \{0,\dots, N_m-1\}.
    \end{aligned}
    \right.
    \end{equation}
    \end{definition}
    The SDB coefficients fix $k_{j_1} = k_{j_2} = 1$ and limit $j_1\in\{0,2\}$, allowing simplified indexing by $j_1,n,k_3$, as opposed to $j_1, j_2, k_3$. Choosing only a subset of disk bispectrum coefficients leads to dramatic improvements in space and time complexities. Both descriptors first require the same disk harmonic transform, costing $\mathcal{O}(L^3)$. The full disk bispectrum then forms $\Theta(m^3/N_m)$ products, for a space complexity of $\mathcal{O}(m^3/N_m)$ and a time complexity of $\mathcal{O}(L^3+m^3/N_m)$. In contrast, the SDB forms only $\Theta(m)$ products.
    
    \begin{proposition}[Complexity of the SDB]
     Consider an image of size $L\times L$, represented with $m$ DH coefficients $a_{n,k}$. With $N_m$ denoting the maximum angular frequency $n$, the space complexity of the SDB is $O(m)$, and the time complexity is $O(L^3 + m)$.
    \end{proposition}
    
    Despite this reduction in complexity, the selected coefficients retain enough information to completely represent the image, up to a rotation and bandlimit, as proven by Th. \ref{th:inversion}. This means that most of the coefficients in the original disk bispectrum are, in fact, redundant.
    
    \begin{figure}[t]
      \centering
      \includegraphics[width=0.9\linewidth]{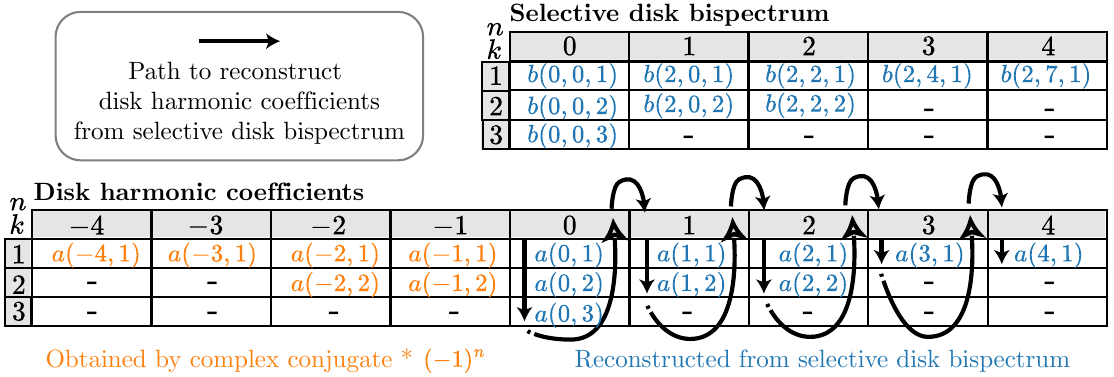}
      \caption{\textbf{The SDB inversion recovers each disk harmonic coefficient $a_{n,k}$ from exactly one SDB coefficient and previously recovered ones, column by column.} Arrows give the recovery order.}
      \label{fig:bispecrum_inversion}
    \end{figure}
    
    \begin{theorem}[Inversion] \label{th:inversion}
        Let $f: \mathbb{D} \to \mathbb{R}$ be a real-valued square-integrable function and let $a_{n, k}$ be its disk harmonic coefficients, bandlimited by $\lambda$. Assume that $a_{n, 1} \neq 0$ for $n \in \{0, ..., N_m-1\}$. Then, its selective disk bispectrum $b^f$ can be inverted, that is $f$ can be reconstructed from $b^f$ up to a planar rotation and frequency $\lambda$. With $f$ discretized on an $L \times L$ image, the time complexity of the inversion is $O(m+L^3)$.
    \end{theorem}
    
    An analysis of round trip error, from image to SDB and back, is in Appendix~\ref{sdb-round-trip-error}, with examples shown in Appendix Fig.~\ref{fig:reconstruction-examples}. The full inversion proof is in Appendix~\ref{app:sdb-proofs}. We summarize here and provide a graphical illustration of the reconstruction path in Fig.~\ref{fig:bispecrum_inversion}. First, reconstruct $a_{n,k}$ with $n=0$ with
    \begin{equation}
       \|a_{0,1}\| = \|b_{0,0,1}\|^{1/3}, \;
    \arg(a_{0,1}) = \arg(b_{0,0,1}); \;
    \space a_{0,k} = \left(\frac{b_{0,0,k}}{a_{0,1}^2}\right)^* \text{ for } k \in \{2,\ldots,K_0\}.
    \end{equation}
    Next, reconstruct $a_{n,k}$ with $n=1$ with
    \begin{equation}
     \|a_{1,1}\| = \left\|\frac{b_{2,0,1}}{a_{0,1}}\right\|^{1/2}, \;
    \arg(a_{1,1}) = 0; \;
     \space a_{1,k} = \left(\frac{b_{2,0,k}}{a_{0,1} \cdot a_{1,1}}\right)^* \text{ for } k \in \{2,\ldots,K_1\}. 
    \end{equation}
    
    Then, reconstruct $a_{n,k}$ with $n>1$ with
    \begin{equation}
        a_{n,k} = \left(\frac{b_{2,n-1,k}}{a_{1,1} \cdot a_{n-1,1}}\right)^* \text{ for } k \in \{1,\ldots,K_n\}.
    \end{equation}
    Notationally, $b_{2,n-1,k} \to b_{2,i(n-1,1),k}$ for the full disk bispectrum inversion. Finally, reconstruct negative coefficients and invert the DH coefficients to a reconstructed image \cite{Verrall_Kakarala_1998}
    \begin{equation}
        a_{-n,k} = (-1)^n a_{n,k}^* \text{ for } n \in \{1,\ldots,N_m\}; \quad f(r,\theta) = \sum_{n=-N_m}^{N_m} \sum_{k=1}^{K_n} a_{n,k}\,\psi_{n,k}(r,\theta).
    \end{equation}
    As the SDB coefficients are the exact coefficients used in our inversion Th. \ref{th:inversion}, our main result shows that despite discarding many coefficients from the full disk bispectrum, the SDB coefficients are sufficient to completely represent the image, up to a global rotation and bandlimit. With this, we proceed to establish the SDB as a \emph{complete invariant}, meaning it retains all information about the image, while being invariant to global rotation.
    
    \begin{theorem}[Complete Invariance]
    Let \( f \) and \( f' \) be a pair of real-valued, square-integrable functions on the disk. Let the bispectrum be defined as in Definition~\ref{def:selective-bsp}. Assume that the DH coefficients \( a_{n,1} \) of \( f \) are non-zero for all \( n \in \{0, \dots, N_m - 1\} \). Then $b^f = b^{f'}$ if and only if \( f' = f \circ R_\phi \) for some 2D rotation \( \phi \in SO(2) \). Thus, the selective bispectrum is a complete invariant.
    \end{theorem}
    The SDB's completeness comes from the inversion proof. Its invariance comes from the equivariance of the DH coefficients, as shown in Appendix~\ref{app:sdb-proofs}.
    
    \subsection{Fast Selective Disk Bispectrum Approximation}
    
    To further reduce the time complexity of the SDB, we propose using a recent numerical approximation of the DH coefficients \citep{fast_dhcs_2023} that computes the disk harmonic transform with time complexity $O(L^2 \log L)$ instead of $\mathcal{O}(L^3)$.
    
    \begin{proposition}[Complexity of the approximate SDB]
        Denote $L \times L$ the size of the image, $m$ the number of DH coefficients. The time complexity of the approximate SDB, computed using the approximate DH coefficients of \citet{fast_dhcs_2023}, is $O(L^2 \log L + m)$.
    \end{proposition}
    
    We provide guarantees on the accuracy of the approximate SDB coefficients and their invariance.
    
    \begin{theorem}[Guarantee for Approximate Accuracy]\label{th:accuracy}
    Let \( 0 < \varepsilon \leq 1 \) be a target accuracy, and assume that \( \lambda \leq \sqrt{\pi L^2} \) and \( |\log \varepsilon| \leq L \), where $\lambda$ is the DH coefficient bandlimit for an image $f$ of size $L \times L$. Let \( b^f \) and \( \tilde{b}^f \) denote the exact and approximate SDB of \( f \). Then,
    \[
    \| \tilde{b}^f - b^f \|_{\infty} \leq C\varepsilon \| f \|_{1},
    \]
    for a constant \( C \) depending only on a bound of \( f \).
    \end{theorem}
    
    As a corollary, we find that the bispectrum coefficients are approximately invariant with controlled guarantees.
    
    \begin{corollary}[Guarantee for Approximate Invariance]
        Let $f, f'$ be a pair of real-valued square integrable functions on the disk such that there exists a $\phi \in SO(2)$ such that $f' = f \circ R_\phi$. Under the notations and assumptions of Th.~\ref{th:accuracy}, we have:
        \begin{equation}
            \| \tilde b^{f} - \tilde b^{f'}\|_\infty \leq 2C \varepsilon \|f\|_1,
        \end{equation}
        for a constant \( C \) depending only on a bound of \( f \).
    \end{corollary}
    
    Linking the accuracy to $f$ is useful, as it allows us to control accuracy through number of pixels $p$. If $|f|$ and $C$ are large, but we still want to reach a given accuracy, then we can choose $p$ accordingly such that $\log(\epsilon) < \sqrt{p}$ to match the desired accuracy. 
    
    \subsection{Noisy Selective Disk Bispectrum Statistics}
    
    Real-world images are often corrupted by noise, raising the question of whether the SDB remains informative in this setting. We derive closed-form expressions for the expectation and variance of each SDB coefficient when additive i.i.d. Gaussian noise is added to the image, and show that most SDB coefficients are asymptotically unbiased. This makes the mean SDB a reliable and noise-robust representation. We also propose an adaptation to our recursive analytic inversion to make it more robust to noise. In the theorems that follow, consider a noiseless signal $f$ with bispectrum $b$ and DH coefficients $a_{n,k}$. Upon corruption with Gaussian noise, $\tilde{f}(x) = f(x) + \eta_x$ for $\eta_x \sim \mathcal{N}(0, \sigma^2)$ added i.i.d. at each pixel.
    
    \begin{theorem}[Expectation]\label{th:expectation-value-main-text}
     The expectation of the noise-corrupted SDB $\tilde{b}$ is given by the following.
    \begin{equation}
    \mathbb{E}[\tilde{b}_{00k}] = b_{00k} + \sigma^2 a^*_{0,k} + \delta_{k,1}2 \sigma^2 a_{0,1}; \quad \mathbb{E}[\tilde{b}_{201}] = b_{201} + \sigma^2 a_{0,1}; \quad \text{Else, } \mathbb{E}[\tilde{b}_{2nk}] = b_{2nk},
    \end{equation}
    where $\delta$ denotes the Kronecker delta.
    \end{theorem}
    
    The variance of $\tilde{b}$ is given by the following.
    \begin{theorem}[Variance]\label{th:variance}
    The variance of $\tilde{b}$ is given by
    \begin{equation}
            Var(b_{j_1 n_2 k_3}) = A_2 \sigma^2 + A_4 \sigma^4 + A_6 \sigma^6,
        \end{equation} 
    where $A_6$ is a scalar, $A_4$ is a linear combination of $|a_{n,k}|^2$, and $A_2$ is a linear combination $|a_{n,k}|^4$. Exact $A_2, A_4, A_6$ are provided in Appendix~\ref{app:noise-proofs} and depend on $j_1, n_2, k_3$ of the SDB coefficient.
    \end{theorem}
    Together, the expectation and variance tell us precisely how noise distorts the SDB. The expectation shows where the noisy SDB distribution is centered relative to the ground truth, and the variance shows how much noisy SDB observations fluctuate around that center. In practical terms, given an estimate of the noise level $\sigma^2$ and the signal's DH power spectrum, a practitioner can predict the reliability of any SDB coefficient before using it downstream.
    
    Analytic inversion of the SDB proceeds by the recursion of Th.~\ref{th:inversion}, in which each DH coefficient $a_{n,k}$ is recovered by dividing the SDB coefficient $b_{2,n-1,k}$ by the previously recovered product $d_n = a_{1,1}\,a_{n-1,1}$. In the noiseless setting this division is exact even when $d_n$ is small, because the numerator $b_{2,n-1,k} = d_n\,a_{n,k}^*$ contains the divisor as a factor and the two cancel. In multi-reference alignment, numerator and divisor are instead estimated from finitely many noisy observations and carry separate errors that do not cancel, so a divisor that falls near zero makes the quotient arbitrarily large. The same instability is known for the analogous recursion in one-dimensional multi-reference alignment, leading \citet{bendory2018} to abandon the recursion for a least-squares fit of all bispectrum coefficients. We keep the recursion and instead use Tikhonov regularization (ridge regression)~\cite{tikhonov1977solutions} at each recursion step.
        \begin{theorem}[Damped inversion]\label{th:damped-inversion}
        Let $\hat{b}$ be the SDB averaged over $N$ noisy copies $\tilde{f}_i = R_{\theta_i} f + \eta_i$, $\eta_i \sim \mathcal{N}(0,\sigma^{2})$, of a template $f$ with unknown rotations $\theta_i$, and let $\hat{d}_n = \hat{a}_{1,1}\hat{a}_{n-1,1}$ be the estimated divisor in the recursion of Th.~\ref{th:inversion}. For $\varepsilon > 0$, the damped inversion replaces each division $\hat{b}_{2,n-1,k}/\hat{d}_n$ by
        \begin{equation}
          \hat{a}^{\varepsilon}_{n,k} = \left( \frac{\hat{b}_{2,n-1,k}\,\hat{d}_n^{*}}{\lvert \hat{d}_n\rvert^{2} + \varepsilon^{2}} \right)^* .
        \end{equation}
        The damped quotient has the same phase as $\hat{b}_{2,n-1,k}/\hat{d}_n$, and its magnitude is $\lvert \hat{b}_{2,n-1,k}\rvert$ times $\lvert \hat{d}_n\rvert/(\lvert \hat{d}_n\rvert^{2} + \varepsilon^{2})$ instead of $1/\lvert \hat{d}_n\rvert$. The two agree when $\lvert \hat{d}_n\rvert \gg \varepsilon$, but the damped factor never exceeds $1/(2\varepsilon)$, whereas the exact one is unbounded as $\hat{d}_n \to 0$.
        \end{theorem}
    The derivation is in Appendix~\ref{sec:damping}, where we also set the frequency-dependent scale $\varepsilon_n$ using the variance (Th. \ref{th:variance}) of SDB coefficients in this setting, and  concepts from Wiener filtering~\cite{wiener1949extrapolation}.

    \section{Experiments: Image Classification}
    
    We classify noisy and rotated MNIST \citep{lecun2010mnist}, FashionMNIST \citep{fashionmnist} and real mouse osteosarcoma cell images ($N=650$) \citep{alizadeh2019tismorph, li2023using} to see where theoretical guarantees of Th.~\ref{th:inversion}--\ref{th:variance} (completeness, rotation invariance, and noise stability) translate into practical advantage. We use the SDB as a fixed, parameter-free descriptor and benchmark it against the DH power spectrum, Harmonic Networks \cite{worrall2017harmonic}, and convolutional and transformer baselines. In each setting, features are classified by an MLP. We vary MLP parameter counts $p \in \{10k, 50k, 500k\}$ and image resolutions $L\times L \in \{28\times 28, 56\times 56, 112\times 112\}$. We train each classifier pipeline on one NVIDIA \texttt{A100} GPU. 
    
    Convolutional baselines are $SE(2)$ steerable networks \citep{weiler2021general}, max and average pooled G-CNNs~\citep{cohenc16}, the Selective $G$-Bispectrum~\citep{mataigne2024selective}, and the G-Triple Correlation~\citep{sanborn2023general, mataigne2024selective}, all with 24 filters. For G-CNNs, we pool the final convolutional layer before passing features to the MLP classifier. Convolutional filters are trained end-to-end jointly with the MLP classifier. Vision encoder baselines are pre-trained transformers \texttt{DINOv3}~\citep{dinov3} (\texttt{ViT-B/16}), \texttt{EVA-02}~\citep{eva} (\texttt{ViT-B/16}), and \texttt{OpenCLIP}~\citep{clip} (\texttt{ViT-L/14}) with backbones \texttt{vit\_base\_patch16\_dinov3.lvd1689m}, \texttt{eva02\_base\_patch14\_224.mim\_in22k}, and \texttt{laion2b\_s32b\_b82k} respectively, used as frozen feature extractors. For \texttt{DINOv3}, \texttt{EVA-02}, and \texttt{OpenCLIP} we must up-sample each image to $224 \times 224$.  
    
    In the cell dataset, each image has a label indicating control, or one of two cancer drug treatments (\texttt{jasp} or \texttt{cytd}), for three classes in total. The train/test sets explicitly test rotation invariance. Train images are randomly rotated between $0^\circ$ and $90^\circ$ and test images between $90^\circ$ and $180^\circ$, ensuring that performance reflects true rotational invariance rather than interpolation within the training distribution. We add Gaussian noise $\mathcal{N}(0, \sigma^2)$ to each image pixel in both train and test sets with $\sigma^2 \in \{0, 0.2, 0.4\}$. We use canonical train/test sets for MNIST and FashionMNIST and 70/30 split for the osteosarcoma dataset. We repeat across 5 random seeds.
    
    On the cell dataset, we also compare against the full disk bispectrum descriptor reduced by PCA, as was used in \cite{zhao2014rotationally}. We don't include this descriptor in the MNIST or FashionMNIST experiments because computing it for every image across seed, image size, and noise levels would take about 220 GPU-days per dataset.
    
\begin{figure}[t]
      \centering
      \includegraphics[width=\linewidth]{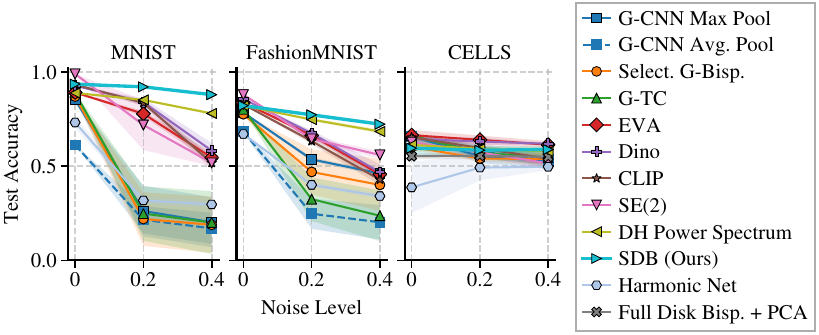}
      \caption{\textbf{Using the SDB as a rotation-invariant descriptor, MLPs can accurately classify noisy rotated MNIST digits, FashionMNIST classes, and cancer cell treatments.} The SDB descriptor achieves accuracy comparable to transformer descriptors and SE(2) network representations. It is more robust to noise than convolutional descriptors. Here, results are shown for $L=28$ with a $500{,}000$-parameter MLP, and markers show the mean over five seeds, with standard deviation shaded. Appendix Fig.~\ref{fig:classification-appendix} shows results for every combination of MLP and image size.}
      \label{fig:classification}
    \end{figure}

    \textbf{Results.} The SDB outperforms convolution-based descriptors when any amount of noise is added to the image. This is true across dataset, image size, and MLP parameter budget, as shown in Fig.~\ref{fig:classification}. Their relatively low accuracy reflects the difficulty of generalizing rotation invariance to a set of unknown angles, on noisy data. It is known that max and average pooled convolutional layers are not robust to noise \cite{convnoise2024}. On MNIST and FashionMNIST, the SDB outperforms transformer representations when significant up-sampling is required on very noisy images, mimicking realistic conditions for practitioners wishing to use pre-trained transformers on small, noisy datasets. The advantage narrows when transformer representations are classified using sufficiently large MLPs, and disappears at image size $112 \times 112$ and for all image resolutions of the cell dataset.
    
    The DH power spectrum is a strong baseline because it retains the magnitude of every disk harmonic coefficient $|a_{n,k}|^2$. This is a fine-grained map of how the image's ``energy'' is distributed over angular frequency and radius, and discards only the phases that encode how these components are oriented relative to one another. The power spectrum seems to match the SDB on FashionMNIST and CELLS, where classes differ mainly in coarse shape and radial energy profile. Its classification accuracy is slightly worse on MNIST, where digits share similar strokes and differ in how those strokes are arranged. The SDB, which is complete up to rotation, thus matches the power spectrum where magnitudes suffice and outperforms it where classes differ in arrangement.
    
    The SDB requires zero learned parameters, while convolutional and transformer encoders require thousands and hundreds of millions of learned parameters respectively. Despite this asymmetry in total system complexity, the SDB achieves competitive accuracy, validating that its theoretical guarantees translate into practical usage.

    \section{Experiments: Multi-Reference Alignment} \label{sec:mra}
    
    Multi-reference alignment (MRA) is an active field of research for the 1D translation bispectrum \cite{bendory2018, chen2018spectral, herring2019gauss}. Given multiple noisy, randomly translated observations of a ground truth signal, MRA seeks to recover the ground truth without explicit alignment. All noisy translated images are mapped to their bispectra. Then, the mean bispectrum is computed and inverted to signal space. Bispectrum-based MRA has previously been impossible on rotated images, as no inversion existed for a rotation-invariant bispectrum.
    
    Here, we show that SDB MRA is possible via our inversion algorithm (Th.~\ref{th:inversion}) and suggest an MRA-specific damping technique to stabilize the inversion in this case (Th. \ref{th:damped-inversion}). Starting from an initial noiseless $56 \times 56$, $112 \times 112$, or $256 \times 256$ image, we generate $n_x \in \{ 20, 50, 100\}$ randomly rotated and noisy $\sim \mathcal{N}(0, \sigma^2)$ observations under varying noise levels $\sigma^2 \in \{0.0, 0.1, 0.2, 0.4\}$ and evaluate reconstruction quality by comparing the relative error between the original image and the reconstructed image. The bispectrum inversion yields the original signal in an arbitrary orientation, so we register the reconstructed image to the original before computing the relative error. 
    
    \textbf{Results.} Our damped inversion (Th. \ref{th:damped-inversion}) successfully recovers templates from averaged noisy and rotated observations, as shown in Fig. \ref{fig:mra-results} across all three datasets. Simple reconstruction results in Appendix~\ref{sdb-round-trip-error} directly verify our SDB inversion and thus our choice of SDB coefficients. MRA is a more complex setting, and successful reconstruction relies on our expectation (Th. \ref{th:expectation-value-main-text}) and variance (Th. \ref{th:variance}) derivations, in order to properly account for bias and coefficient spread induced by noise. These results constitute the first demonstrated instance of bispectrum-based multi-reference alignment on rotationally-varied images, a task that was impossible prior to the inversion derived in this work.
    
    \begin{figure*}[t]
      \centering
       \includegraphics[width=\textwidth]{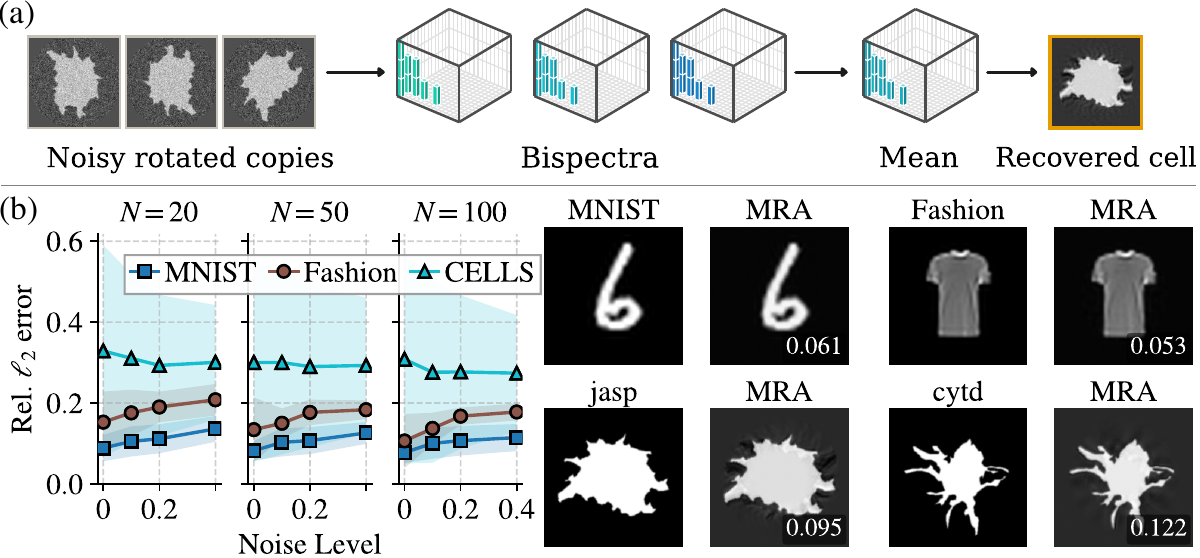}
        \caption{\textbf{The damped SDB inversion recovers templates from averaged noisy, rotated observations on all three datasets.} (a) The MRA task considers noisy images in unknown orientations and seeks to recover their common underlying signal by mapping each image to its bispectrum coefficients and inverting the images' average bispectrum. (b) Reconstructions for $L=256$. In Appendix Fig.~\ref{fig:mra-appendix}, we report these results for image sizes $L\in\{56, 112\}$.}
       \label{fig:mra-results}
    \end{figure*}
     
    \section{Conclusion}
    We introduce a compact rotation-invariant image descriptor, the SDB, by deriving a minimal number of disk bispectrum coefficients necessary to reconstruct the image. We also provide accuracy and invariance guarantees for its approximation, as well as summary statistics and inversion stabilization techniques under Gaussian noise. Classification tasks and SDB MRA on rotated images demonstrate rotation-invariance and noise robustness of the SDB descriptor and viability of our SDB inversion (Th. \ref{th:inversion}), which in turn validates our choice of SDB coefficients.
    
    \textbf{Limitations.} The SDB, like all bispectra, is invariant to one (global) group action: planar rotation. The SDB map is not invariant to rotation of local features, global translation, or reflection. Therefore, SDB is best used on centered single-object images, however we do discuss the effect of global translation in Appendix~\ref{sdb-translation}. In addition, SDB seems to correctly group signal deformations \textit{to an extent}, as it can classify MNIST and FashionMNIST, but a more thorough analysis of object deformations in bispectrum space would be interesting future work.

    \section*{Acknowledgements}
    A. M. L. and N. M. acknowledge partial support from NSF grant 2313150, NSF Career grant 2240158 and from the Noyce foundation.
    
    A. M. L. also acknowledges support from an NSF GRFP Fellowship and NSF Grant No. DGE-2125644.
    
    N. M. also acknowledges a Goodfire Research Grant for credits using Silico, Goodfire’s interpretability research agent.

\medskip

{
    \small
    \bibliographystyle{ieeenat_fullname}
    \bibliography{main}

@book{rudin1962fourier,
  title={Fourier Analysis on Groups},
  author={Rudin, Walter},
  year={1962},
  publisher={Courier Dover Publications},
}

@misc{dinov3,
      title={DINOv3}, 
      author={Oriane Siméoni and Huy V. Vo and Maximilian Seitzer and Federico Baldassarre and Maxime Oquab and Cijo Jose and Vasil Khalidov and Marc Szafraniec and Seungeun Yi and Michaël Ramamonjisoa and Francisco Massa and Daniel Haziza and Luca Wehrstedt and Jianyuan Wang and Timothée Darcet and Théo Moutakanni and Leonel Sentana and Claire Roberts and Andrea Vedaldi and Jamie Tolan and John Brandt and Camille Couprie and Julien Mairal and Hervé Jégou and Patrick Labatut and Piotr Bojanowski},
      year={2025},
      eprint={2508.10104},
      archivePrefix={arXiv},
      primaryClass={cs.CV},
      url={https://arxiv.org/abs/2508.10104}, 
}

@misc{clip,
  author       = {Ilharco, Gabriel and
                  Wortsman, Mitchell and
                  Wightman, Ross and
                  Gordon, Cade and
                  Carlini, Nicholas and
                  Taori, Rohan and
                  Dave, Achal and
                  Shankar, Vaishaal and
                  Namkoong, Hongseok and
                  Miller, John and
                  Hajishirzi, Hannaneh and
                  Farhadi, Ali and
                  Schmidt, Ludwig},
  title        = {OpenCLIP},
  month        = jul,
  year         = 2021,
  note         = {If you use this software, please cite it as below.},
  publisher    = {Zenodo},
  version      = {0.1},
  doi          = {10.5281/zenodo.5143773},
  url          = {https://doi.org/10.5281/zenodo.5143773}
}

@article{eva,
   title={EVA-02: A visual representation for neon genesis},
   volume={149},
   ISSN={0262-8856},
   url={http://dx.doi.org/10.1016/j.imavis.2024.105171},
   DOI={10.1016/j.imavis.2024.105171},
   journal={Image and Vision Computing},
   publisher={Elsevier BV},
   author={Fang, Yuxin and Sun, Quan and Wang, Xinggang and Huang, Tiejun and Wang, Xinlong and Cao, Yue},
   year={2024},
   month=sep, pages={105171} }

@Article{convnoise2024,
AUTHOR = {Zafar, Afia and Saba, Noushin and Arshad, Ali and Alabrah, Amerah and Riaz, Saman and Suleman, Mohsin and Zafar, Shahneer and Nadeem, Muhammad},
TITLE = {Convolutional Neural Networks: A Comprehensive Evaluation and Benchmarking of Pooling Layer Variants},
JOURNAL = {Symmetry},
VOLUME = {16},
YEAR = {2024},
NUMBER = {11},
ARTICLE-NUMBER = {1516},
URL = {https://www.mdpi.com/2073-8994/16/11/1516},
ISSN = {2073-8994},
DOI = {10.3390/sym16111516}
}

@article{zhao2013fourier,
  title={Fourier--Bessel rotational invariant eigenimages},
  author={Zhao, Zhizhen and Singer, Amit},
  journal={Journal of the Optical Society of America A},
  volume={30},
  number={5},
  pages={871--877},
  year={2013},
  publisher={Optical Society of America}
}

@inproceedings{mataigne2024selective,
 author = {Mataigne, Simon and Mathe, Johan and Sanborn, Sophia and Hillar, Christopher and Miolane, Nina},
 booktitle = {Advances in Neural Information Processing Systems},
 editor = {A. Globerson and L. Mackey and D. Belgrave and A. Fan and U. Paquet and J. Tomczak and C. Zhang},
 pages = {115682--115711},
 publisher = {Curran Associates, Inc.},
 title = {The Selective G-Bispectrum and its Inversion: Applications to G-Invariant Networks},
 url = {https://proceedings.neurips.cc/paper_files/paper/2024/file/d1a1e8713fcd5626656553c82f7c3b26-Paper-Conference.pdf},
 volume = {37},
 year = {2024}
}

@article{zhao2014rotationally,
  title={Rotationally invariant image representation for viewing direction classification in cryo-EM},
  author={Zhao, Zhizhen and Singer, Amit},
  journal={Journal of structural biology},
  volume={186},
  number={1},
  pages={153--166},
  year={2014},
  publisher={Elsevier}
}

@article{koukiou2024identifying,
  title={Identifying System Non-Linearities by Fusing Signal Bispectral Signatures},
  author={Koukiou, Georgia},
  journal={Electronics},
  volume={13},
  number={7},
  pages={1287},
  year={2024},
  publisher={MDPI}
}

@article{zhang2000bispectrum,
  title={Bispectrum analysis of focal ischemic cerebral EEG signal using third-order recursion method},
  author={Zhang, Ji-Wu and Zheng, Chong-Xun and Xie, An},
  journal={IEEE transactions on biomedical engineering},
  volume={47},
  number={3},
  pages={352--359},
  year={2000},
  publisher={IEEE}
}

@article{martin2021bispectral,
  title={Bispectral analysis of heart rate variability to characterize and help diagnose pediatric sleep apnea},
  author={Mart{\'\i}n-Montero, Adri{\'a}n and Guti{\'e}rrez-Tobal, Gonzalo C and Gozal, David and Barroso-Garc{\'\i}a, Ver{\'o}nica and {\'A}lvarez, Daniel and Del Campo, F{\'e}lix and Kheirandish-Gozal, Leila and Hornero, Roberto},
  journal={Entropy},
  volume={23},
  number={8},
  pages={1016},
  year={2021},
  publisher={MDPI}
}

@article{fast_dhcs_2023, title={Fast Expansion into Harmonics on the Disk: A Steerable Basis with Fast Radial Convolutions}, volume={45}, ISSN={1064-8275}, DOI={10.1137/22M1542775}, number={5}, journal={SIAM Journal on Scientific Computing}, publisher={Society for Industrial and Applied Mathematics}, author={Marshall, Nicholas F. and Mickelin, Oscar and Singer, Amit}, year={2023}, month=oct, pages={A2431–A2457} }

@article{nikias1993signal,
  title={Signal processing with higher-order spectra},
  author={Nikias, Chrysostomos L and Mendel, Jerry M},
  journal={IEEE Signal processing magazine},
  volume={10},
  number={3},
  pages={10--37},
  year={1993},
  publisher={IEEE}
}

@article{elgar1985observations,
  title={Observations of bispectra of shoaling surface gravity waves},
  author={Elgar, Steve and Guza, RT},
  journal={Journal of Fluid Mechanics},
  volume={161},
  pages={425--448},
  year={1985},
  publisher={Cambridge University Press}
}

@article{geophysics_2004, title={Earthquake Relocation Using Cross-Correlation Time Delay Estimates Verified with the Bispectrum Method}, volume={94}, ISSN={0037-1106}, DOI={10.1785/0120030084}, abstractNote={Cross-correlation (CC) determined relative time delays, or related differential travel times, between pairs of seismic events at the same station are often used as input data to improve earthquake relocation results. Researchers generally select those time delays with associated CC coefficients larger than a chosen threshold. When two similar time series are contaminated by correlated noise sources, the relative time delay between them calculated with the CC technique is sometimes not reliable. Noise sources at a station for different events can be partially correlated or just randomly correlated. In this work, we use the bispectrum (BS) method, which works in the third-order spectral domain, to check the reliability of the CC determined time delay. We calculate two time delays with the BS method, one using the bandpass-filtered waveforms and the other with the raw data, and use them to verify (select or reject) the CC estimate computed with the filtered waveforms. We apply this technique to obtain bispectrum-verified CC differential times for 822 New Zealand earthquakes in the Wellington area. Our work demonstrates that the CC time delays verified with the BS method provide improved (smaller root mean square residual and more clustered) earthquake relocation results compared to those selected with the standard threshold criterion.}, number={3}, journal={Bulletin of the Seismological Society of America}, author={Du, Wen-Xuan and Thurber, Clifford H. and Eberhart-Phillips, Donna}, year={2004}, month=jun, pages={856–866} }

@article{tukey1953spectral,
  title={The spectral representation and transformation properties of the higher moments of stationary time series},
  author={Tukey, J.},
  journal={Reprinted in The Collected Works of John W. Tukey},
  volume={1},
  pages={165--184},
  year={1953}
}

@article{brillinger1991some,
 ISSN = {10170405, 19968507},
 URL = {http://www.jstor.org/stable/24304021},
 author = {David R. Brillinger},
 journal = {Statistica Sinica},
 number = {2},
 pages = {465--476},
 publisher = {Institute of Statistical Science, Academia Sinica},
 title = {Some history of higher-order statistics and spectra},
 urldate = {2024-10-18},
 volume = {1},
 year = {1991}
}

@inproceedings{sanborn2023general,
 author = {Sanborn, Sophia and Miolane, Nina},
 booktitle = {Advances in Neural Information Processing Systems},
 pages = {67103--67124},
 publisher = {Curran Associates, Inc.},
 title = {A General Framework for Robust G-Invariance in G-Equivariant Networks},
 url = {https://proceedings.neurips.cc/paper_files/paper/2023/file/d42523d621194ba54dda098669645f91-Paper-Conference.pdf},
 volume = {36},
 year = {2023}
}

@inproceedings{sanborn2023bispectral,
  title={Bispectral Neural Networks},
  author={Sanborn, Sophia and Shewmake, Christian and Olshausen, Bruno and Hillar, Christopher},
  booktitle={International Conference on Learning Representations (ICLR)},
  year={2023}
}

@InProceedings{cohenc16,
  title = 	 {Group Equivariant Convolutional Networks},
  author = 	 {Cohen, Taco and Welling, Max},
  booktitle = 	 {Proceedings of The 33rd International Conference on Machine Learning},
  pages = 	 {2990--2999},
  year = 	 {2016},
  editor = 	 {Balcan, Maria Florina and Weinberger, Kilian Q.},
  volume = 	 {48},
  series = 	 {Proceedings of Machine Learning Research},
  address = 	 {New York, New York, USA},
  month = 	 {20--22 Jun},
  publisher =    {PMLR},
  url = 	 {https://proceedings.mlr.press/v48/cohenc16.html}
}

@inproceedings{Kakarala2009CompletenessOB,
  title={{Completeness of bispectrum on compact groups}},
  author={Kakarala, Ramakrishna},
  booktitle={arXiv preprint arXiv:0902.0196},
  year={2009},
  url={https://api.semanticscholar.org/CorpusID:18425284}
}

@article{giannakis1989signal,
    author = {Georgios B. Giannakis},
    journal = {J. Opt. Soc. Am. A},
    number = {5},
    pages = {682--697},
    publisher = {Optica Publishing Group},
    title = {Signal reconstruction from multiple correlations: frequency- and time-domain approaches},
    volume = {6},
    month = {May},
    year = {1989},
    url = {https://opg.optica.org/josaa/abstract.cfm?URI=josaa-6-5-682},
    doi = {10.1364/JOSAA.6.000682}
}

@article{chen2018spectral,
  title={A spectral method for stable bispectrum inversion with application to multireference alignment},
  author={Chen, Hua and Zehni, Mona and Zhao, Zhizhen},
  journal={IEEE Signal Processing Letters},
  volume={25},
  number={7},
  pages={911--915},
  year={2018},
  publisher={IEEE}
}

@inproceedings{weiler2021general,
author = {Weiler, Maurice and Cesa, Gabriele},
title = {General {E}(2)-equivariant steerable {CNN}s},
year = {2019},
publisher = {Curran Associates Inc.},
address = {Red Hook, NY, USA},
booktitle = {Proceedings of the 33rd International Conference on Neural Information Processing Systems},
articleno = {1286},
numpages = {12}
}

@article{lecun2010mnist,
  author = {LeCun, Yann and Cortes, Corinna},
  howpublished = {http://yann.lecun.com/exdb/mnist/},
  lastchecked = {2016-01-14 14:24:11},
  title = {{MNIST} handwritten digit database},
  journal = {ATT Labs},
  volume = {2},
  url = {http://yann.lecun.com/exdb/mnist/},
  username = {mhwombat},
  year = 2010
}

@article{sundaramoorthy1990,
  author    = {G. Sundaramoorthy and M. R. Raghuveer and S. A. Dianat},
  title     = {Bispectral reconstruction of signals in noise: amplitude reconstruction issues},
  journal   = {IEEE Transactions on Acoustics, Speech, and Signal Processing},
  volume    = {38},
  number    = {7},
  pages     = {1297--1306},
  year      = {1990},
  doi       = {10.1109/29.57500}
}

@inproceedings{pinilla2023,
  author    = {L. Pinilla and A. Mishra and B. M. Sadler},
  title     = {Unique Bispectrum Inversion for Signals with Finite Spectral/Temporal Support},
  booktitle = {Proceedings of ICASSP 2023},
  year      = {2023},
  publisher = {IEEE},
  doi       = {10.1109/ICASSP49357.2023.10095922}
}

@article{bendory2018,
  author    = {Bendory, Tamir and Boumal, Nicolas and Ma, Chao and Zhao, Zhizhen and Singer, Amit},
  title     = {Bispectrum Inversion with Application to Multireference Alignment},
  journal   = {IEEE Transactions on Signal Processing},
  volume    = {66},
  number    = {4},
  pages     = {1037--1050},
  year      = {2018},
  doi       = {10.1109/TSP.2017.2775591}
}

@article{edidin2025,
author = {Edidin, Dan and Katz, Joshua},
year = {2025},
month = {12},
pages = {},
title = {The Reflection-Invariant Bispectrum: Signal Recovery in the Dihedral Model: The Reflection-Invariant Bispectrum: Signal Recovery in the Dihedral ModelD. Edidin, J. Katz},
volume = {32},
journal = {Journal of Fourier Analysis and Applications},
doi = {10.1007/s00041-025-10209-z}
}

@article{kakarala2012,
  author    = {R. Kakarala},
  title     = {The bispectrum as a source of phase-sensitive invariants for Fourier descriptors: a group-theoretic approach},
  journal   = {Journal of Mathematical Imaging and Vision},
  volume    = {44},
  number    = {3},
  pages     = {341--353},
  year      = {2012},
  doi       = {10.1007/s10851-012-0334-6}
}

@inproceedings{kakarala1993,
  author    = {R. Kakarala and B. M. Bennett and M. D’Zmura and G. J. Iverson},
  title     = {Bispectral techniques for spherical functions},
  booktitle = {Proceedings of ICASSP '93},
  year      = {1993},
  pages     = {237--240},
  address   = {Minneapolis, MN},
  publisher = {IEEE},
  doi       = {10.1109/ICASSP.1993.319593}
}

@article{herring2019gauss,
  title        = {Gauss--Newton Optimization for Phase Recovery from the Bispectrum},
  author       = {Herring, James and Nagy, James and Ruthotto, Lars},
  journal      = {IEEE Transactions on Computational Imaging},
  year         = {2019},
  pages        = {1--1},
  doi          = {10.1109/TCI.2019.2948784}
}

@article{article,
author = {Kakarala, Ramakrishna},
year = {2009},
month = {02},
pages = {},
title = {The Bispectrum as a Source of Phase-Sensitive Invariants for Fourier Descriptors: A Group-Theoretic Approach},
volume = {44},
journal = {Journal of Mathematical Imaging and Vision},
doi = {10.1007/s10851-012-0330-6}
}

@article{reddy1996fft,
  title={An FFT-based technique for translation, rotation, and scale-invariant image registration},
  author={Reddy, B. S. and Chatterji, B. N.},
  journal={IEEE Transactions on Image Processing},
  volume={5},
  number={8},
  pages={1266--1271},
  year={1996}
}

@article{lowe2004sift,
  title={Distinctive Image Features from Scale-Invariant Keypoints},
  author={Lowe, David G.},
  journal={International Journal of Computer Vision},
  volume={60},
  number={2},
  pages={91--110},
  year={2004}
}

@inproceedings{bay2006surf,
  title={SURF: Speeded Up Robust Features},
  author={Bay, Herbert and Tuytelaars, Tinne and Van Gool, Luc},
  booktitle={European Conference on Computer Vision (ECCV)},
  pages={404--417},
  year={2006},
  publisher={Springer}
}

@inproceedings{rublee2011orb,
  title={ORB: An efficient alternative to SIFT or SURF},
  author={Rublee, Ethan and Rabaud, Vincent and Konolige, Kurt and Bradski, Gary},
  booktitle={Proceedings of the IEEE International Conference on Computer Vision (ICCV)},
  pages={2564--2571},
  year={2011}
}

@inproceedings{alcantarilla2012kaze,
  title={KAZE Features},
  author={Alcantarilla, Pablo F. and Bartoli, Adrien and Davison, Andrew J.},
  booktitle={European Conference on Computer Vision (ECCV)},
  pages={214--227},
  year={2012},
  publisher={Springer}
}

@article{hu1962visual,
  title={Visual Pattern Recognition by Moment Invariants},
  author={Hu, M.K.},
  journal={IRE Transactions on Information Theory},
  volume={8},
  number={2},
  pages={179--187},
  year={1962}
}

@article{teague1980zernike,
  title={Image analysis via the general theory of moments},
  author={Teague, M. R.},
  journal={JOSA},
  volume={70},
  number={8},
  pages={920--930},
  year={1980}
}

@inproceedings{thomas2018tfns,
  title={Tensor Field Networks: Rotation- and Translation-Equivariant Neural Networks for 3D Point Clouds},
  author={Thomas, Nathaniel and Smidt, T. and Kearnes, S. and Yang, L. and Li, L. and Kohlhoff, K. and Riley, P. and Kohlhoff, K.},
  booktitle={NeurIPS},
  year={2018}
}

@article{cohen2018spherical,
author = {Cohen, Taco and Geiger, Mario and Koehler, Jonas and Welling, Max},
year = {2018},
month = {01},
pages = {},
title = {{Spherical CNNs}},
journal = {International Conference on Learning Representations (ICLR)},
doi = {10.48550/arXiv.1801.10130}
}

@inproceedings{sabour2017dynamic,
  title={Dynamic Routing Between Capsules},
  author={Sabour, Sara and Frosst, Nicholas and Hinton, Geoffrey E.},
  booktitle={Advances in Neural Information Processing Systems (NeurIPS)},
  pages={3856--3866},
  year={2017}
}

@article{Verrall_Kakarala_1998, title={Disk-harmonic coefficients for invariant pattern recognition}, volume={15}, rights={https://doi.org/10.1364/OA_License_v1#VOR}, ISSN={1084-7529, 1520-8532}, DOI={10.1364/JOSAA.15.000389}, number={2}, journal={Journal of the Optical Society of America A}, author={Verrall, Steven C. and Kakarala, Ramakrishna}, year={1998}, month=feb, pages={389}, language={en} }

@article{donoho1994ideal,
  author  = {Donoho, David L. and Johnstone, Iain M.},
  title   = {Ideal spatial adaptation by wavelet shrinkage},
  journal = {Biometrika},
  volume  = {81},
  number  = {3},
  pages   = {425--455},
  year    = {1994},
  doi     = {10.1093/biomet/81.3.425}
}

@article{matson1991weighted,
  author  = {Matson, Charles L.},
  title   = {Weighted-least-squares phase reconstruction from the bispectrum},
  journal = {Journal of the Optical Society of America A},
  volume  = {8},
  number  = {12},
  pages   = {1905--1913},
  year    = {1991},
  doi     = {10.1364/JOSAA.8.001905}
}

@book{tikhonov1977solutions,
  author    = {Tikhonov, Andrey N. and Arsenin, Vasiliy Y.},
  title     = {Solutions of Ill-Posed Problems},
  series    = {Scripta Series in Mathematics},
  publisher = {V. H. Winston \& Sons},
  address   = {Washington, DC},
  note      = {Translation editor Fritz John; distributed by Halsted Press (Wiley), New York},
  year      = {1977}
}

@book{wiener1949extrapolation,
  author    = {Wiener, Norbert},
  title     = {Extrapolation, Interpolation, and Smoothing of Stationary Time Series, with Engineering Applications},
  publisher = {Technology Press of the Massachusetts Institute of Technology and John Wiley \& Sons},
  address   = {Cambridge, MA and New York},
  year      = {1949}
}

@inproceedings{worrall2017harmonic,
  author    = {Worrall, Daniel E. and Garbin, Stephan J. and Turmukhambetov, Daniyar and Brostow, Gabriel J.},
  title     = {Harmonic Networks: Deep Translation and Rotation Equivariance},
  booktitle = {Proceedings of the IEEE Conference on Computer Vision and Pattern Recognition (CVPR)},
  pages     = {7168--7177},
  year      = {2017},
  doi       = {10.1109/CVPR.2017.758}
}

@inproceedings{li2023using,
  author    = {Li, Wanxin and Prasad, Ashok and Miolane, Nina and Dao Duc, Khanh},
  title     = {Using a {R}iemannian Elastic Metric for Statistical Analysis of Tumor Cell Shape Heterogeneity},
  booktitle = {Geometric Science of Information (GSI 2023)},
  series    = {Lecture Notes in Computer Science},
  editor    = {Nielsen, Frank and Barbaresco, Fr{\'e}d{\'e}ric},
  publisher = {Springer Nature Switzerland},
  pages     = {583--592},
  year      = {2023},
  doi       = {10.1007/978-3-031-38299-4_60}
}

@article{alizadeh2019tismorph,
  author  = {Alizadeh, Elaheh and Xu, Wenlong and Castle, Jordan and Foss, Jacqueline and Prasad, Ashok},
  title   = {{TISMorph}: A tool to quantify texture, irregularity and spreading of single cells},
  journal = {PLOS ONE},
  volume  = {14},
  number  = {6},
  pages   = {e0217346},
  year    = {2019},
  doi     = {10.1371/journal.pone.0217346}
}

@misc{fashionmnist,
  author        = {Xiao, Han and Rasul, Kashif and Vollgraf, Roland},
  title         = {{Fashion-MNIST}: a Novel Image Dataset for Benchmarking Machine Learning Algorithms},
  year          = {2017},
  eprint        = {1708.07747},
  archivePrefix = {arXiv},
  primaryClass  = {cs.LG},
  url           = {https://arxiv.org/abs/1708.07747}
}

@article{marron1990unwrapping,
  author  = {Marron, J. C. and Sanchez, P. P. and Sullivan, R. C.},
  title   = {Unwrapping algorithm for least-squares phase recovery from the modulo $2\pi$ bispectrum phase},
  journal = {Journal of the Optical Society of America A},
  volume  = {7},
  number  = {1},
  pages   = {14--20},
  year    = {1990},
  doi     = {10.1364/JOSAA.7.000014}
}
}


\appendix

\section{Formal Definitions}
\label{app:definitions}

\subsection{Groups and Group Action}
A group $\mathcal{G}$ is a set $\{a,b,c,...\} \in G$ equipped with an operation $\circ: G \times G \to G$  that combines two elements of the set to produce a third element within the same set. The operation must be associative, and have an identity element and inverse. A group can also act on an external set $S$ via the group action $g: G \times S \to S$. For example, if $G = (\mathbb{Z}, +)$, group action could be addition acting on $\mathbb{R}$.


\subsection{Equivariance, Invariance, and Irreducible Representations}

We adopt \citep{sanborn2023bispectral}'s description. Consider a group $G$ acting on a space $X$. Next, consider a function $\phi: X \to Y$. $\phi$ is \textit{G-equivariant} if $\phi(g \circ x) = g' \circ \phi(x)$ for all $g \in G, x\in X$, where $g'\in G'$ is homomorphic to $G$. $\phi$ is \textit{G-invariant} if $\phi(g \circ x) = \phi(x) $ for all $g\in G, x\in X$. $\phi$ is a \textit{complete invariant} if $\phi(x_1) = \phi(x_2) \iff x_2 = g\circ x_1$, guaranteeing that $\phi$ maps all elements of an \textit{orbit} $\{\phi(gx): g \in G \}$ to a unique point in $Y$.  A \textit{representation} of a group is a map $\rho : G \to GL(V)$ that assigns elements of $G$ to elements of the group of linear transformations (e.g. matrices) over a vector space $V$ such that $\rho(g_1 g_2) = \rho(g_1)\rho(g_2)$. A representation is \textit{reducible} if a change of basis can decompose the representation into a direct sum of other representations. An \textit{irreducible representation} is one that cannot be further reduced in this manner. Put simply, \textit{representation theory} describes group elements as matrices so that group action can be described by matrix multiplication.

\subsection{Fourier Transform, Power Spectrum, and Translation-Invariant Bispectrum}

The Fourier transform provides a canonical example of an equivariant representation. A signal $f: \mathbb{Z}/n\mathbb{Z} \to \mathbb{C}$ can be expanded in the Fourier basis as 
\begin{equation} \label{eq:ft-expansion}
    f(x) = \sum^{n-1}_{k=0} F_k e^{i 2 \pi k x /n}.
\end{equation}
Using shorthand $\nu = 2 \pi k /n$,
\begin{equation}
    F_\nu = |F_\nu| e^{-i \nu \phi_\nu} = \frac{1}{n}\sum^{n-1}_{x=0} f(x) e^{-i \nu x}
\end{equation}
Respectively, $|F_\nu|$ and $\phi_\nu$ are the magnitude and phase of $F_\nu$, and both are crucial for signal reconstruction from Fourier coefficients using Eq. \ref{eq:ft-expansion}. From a group theoretic perspective, a Fourier transform is a projection of a signal onto the irreducible representations of $(\mathbb{Z}/n\mathbb{Z}, +)$, indexed by integer frequencies $k$. Under translation $g \in \mathbb{Z}/n\mathbb{Z}$, $f$ transforms as $(g \circ f)(x) = f(x - g) = \tilde{f}(x)$ and the Fourier coefficients transform as
\begin{equation}
    \tilde{F}_\nu = e^{-i \nu g} F_\nu = |F_\nu| e^{-i \nu (\phi_\nu + g)},
\end{equation}
 making them equivariant to $(\mathbb{Z}/n\mathbb{Z}, +)$. 

Next, invariant quantities can be constructed from equivariant ones. For example, the power spectrum
\begin{equation}
    P_\nu  = F_\nu F^*_\nu = |F_\nu| e^{-i \nu \phi_{\nu}} |F_\nu| e^{+i \nu \phi_{\nu}} = |F_\nu|^2
\end{equation}
does not change under translation $g \in \mathbb{Z}/n\mathbb{Z}$, since the phase cancels entirely. However, without $\phi_\nu$, signal reconstruction is impossible.

The \textit{translation-invariant bispectrum} \citep{tukey1953spectral, brillinger1991some, nikias1993signal}, developed in the mid-20th century for signal processing, resolves this limitation by constructing invariants that preserve relative phase information while disregarding global phase information altered by $(g \circ f)$. It is defined as
\begin{equation}
B_{\nu_1,\nu_2} = F_{\nu_1} F_{\nu_2} F^*_{\nu_1+\nu_2}.
\end{equation}
Under translation,
\begin{equation}
\tilde{B}_{\nu_1,\nu_2} \rightarrow
e^{-i \nu_1 g} F_{\nu_1}
\cdot
e^{-i \nu_2 g} F_{\nu_2}
\cdot
e^{i (\nu_1+\nu_2) g} F^*_{\nu_1+\nu_2} = B_{\nu_1,\nu_2}.
\end{equation}
The key observation is that the bispectrum combines Fourier coefficients in a way that cancels global phase, while preserving relative phase relationships $\phi_\nu$ between coefficients $F_\nu$. As a result, it is invariant to translation but retains sufficient information for reconstruction under mild conditions, as shown in \citep{giannakis1989signal}. For this reason, it is referred to as a \emph{complete invariant}. This construction illustrates a general principle: \emph{invariants can be built by combining equivariant features that transform according to irreducible representations}.

\subsection{Bispectrum Inversion}
\label{app:classical-inversion}

To reconstruct the original signal from the bispectrum, Fourier coefficients can be reconstructed from bispectral coefficients according to the algorithm
\cite{giannakis1989signal}:
\begin{equation}
    1) |F_0| = |b_{00}|^{1/3}, \space 
    arg(F_0) = arg(b_{00}); \space2) \space |F_1| = \sqrt{\frac{b_{10}}{F_0}}; \space3) \space F_{i >1} = \left( \frac{b_{1(i-1)}}{F_1F_{i-1}}\right)^*.
\end{equation}
The global phase of the reconstructed signal is set by the phase of $F_1$. \cite{giannakis1989signal} also developed an inversion algorithm for the 2D translation-invariant bispectrum, which is constructed using 2D Fourier coefficients and can therefore be defined directly on images. Because the signal can be fully reconstructed (up to a global translation) from its bispectrum coefficients, the bispectrum is referred to as a \emph{complete invariant}. This also means that two signals, identical up to a global translation, will map to the same set of bispectrum coefficients.

 Beyond \citep{giannakis1989signal}'s 1D translation bispectrum inversion,  \citep{sundaramoorthy1990} developed another analytical inversion algorithm that is more robust to noise, \citep{pinilla2023} revisited this inversion problem, providing a gradient-based inversion procedure, and \citep{bendory2018} derived an inversion optimized for multi-reference alignment. \citep{giannakis1989signal} also extended his 1D inversion scheme to invert the 2D translation-invariant bispectrum, invariant to the product of cyclic groups $C_n \times C_n$.  \cite{mataigne2024selective} generalized that proof for the product of any number of cyclic groups, \textit{i.e.}, to any $G$-bispectrum, for signals defined over a finite and commutative group $G$. A reflection-invariant bispectrum was also proposed by \citep{edidin2025}, who formulated a gradient descent optimization for its inversion. Methods to invert the $G$-bispectrum of signals defined on a (continuous) Lie group $G$ have been proposed in \citep{kakarala2012,kakarala1993}. 

\section{Example Disk Harmonic Coefficient Truncation}
\label{app:dh-truncation}
Table~\ref{tab:indices} shows an example of this correspondence between $j$ and $(n,k)$ for an $8 \times 8$ image expanded in the truncated DH basis. In the disk bispectrum definition, $-N_m \leq n_{j_1}+n_{j_2} \leq N_m \text{, and } k_3 \le K_{n_{j_1} + n_{j_2}}$.


\begin{table*}[t]
    \centering
    \scriptsize
    \resizebox{\textwidth}{!}{
    \begin{tabular}{c|ccccccccccccccccccccc}
    \hline
    $n \rightarrow$ & -10 & -9 & -8 & -7 & -6 & -5 & -4 & -3 & -2 & -1 & 0 & 1 & 2 & 3 & 4 & 5 & 6 & 7 & 8 & 9 & 10 \\
    $k \downarrow$ &  &  &  &  &  &  &  &  &  &  &  &  &  &  &  &  &  &  &  &  &  \\
    \hline
    1 & 44 & 38 & 30 & 25 & 19 & 15 & 10 & 6 & 3 & 1 & 0 & 2 & 4 & 7 & 11 & 16 & 20 & 26 & 31 & 39 & 45 \\
    2 & - & - & - & 48 & 40 & 32 & 23 & 17 & 12 & 8 & 5 & 9 & 13 & 18 & 24 & 33 & 41 & 49 & - & - & - \\
    3 & - & - & - & - & - & - & 42 & 34 & 27 & 21 & 14 & 22 & 28 & 35 & 43 & - & - & - & - & - & - \\
    4 & - & - & - & - & - & - & - & - & 46 & 36 & 29 & 37 & 47 & - & - & - & - & - & - & - & - \\
    \hline
    \end{tabular}
    }
      \caption{\textbf{The SDB keeps one coefficient for each DH coefficient with $n \geq 0$, the right half of this index map, so its size is $(m+K_0)/2$.} DH coefficient index map $ j \leftrightarrow(n,k)$, for an $8 \times 8$ image, truncated to the $\lfloor \pi L^2/4 \rfloor = 50$ lowest frequencies, i.e. one basis function per pixel inside the disk, the maximum truncation of \citet{fast_dhcs_2023}. This is a larger truncation than the library default bandlimit used in Fig.~\ref{fig:spatial-complexity}, which keeps $m=15$ coefficients at $L=8$. Columns and rows are different values of $n$ and $k$ respectively, and table entries are $j$. Here, $m=50$, $N_m = 10$, and $K_{\max} = 4$.}
      \label{tab:indices}
    \end{table*}

\section{Proofs for Theorems on the Disk Bispectrum}

\begin{proposition}[Complexity of the disk bispectrum]
    Denote $L$ the size of the image, $m$ the number of disk harmonic coefficients, $N_m$ the maximum angular frequency. The space complexity of the disk bispectrum is $O(m^3/N_m)$. The time complexity of the disk bispectrum is  $O(L^3 + m^3/N_m)$.
\end{proposition}

\begin{proof}
    \textbf{Space complexity:} The space complexity is equal to the number of bispectral coefficients $b_{j_1, j_2, k_3}$ for $j_1\in \{1, ..., m\}$, $j_2 \in \{ j \in \{1, ..., m\}~|~-N_m \leq n_{j_1}+n_{j_2} \leq N_m\}$ and $k_3 \in \{1, ..., K_{n_{j_1}+n_{j_2}}\}$:
    \begin{align*}
    N &= 
    \sum_{j_1=1}^m 
    \sum_{\substack{j_2=1 \\ -N_m \leq n_{j_1}+n_{j_2} \leq N_m}}^m 
    \sum_{k_3=1}^{K_{n_{j_1}+n_{j_2}}} 1 \\
    &= O(m^2.K_{\max}) 
    = O(m^3/ N_m),
    \end{align*}
    where $K_{\max} = \max \{K_n~|~n \in \{-N_m, ..., N_m\}\}$.
    
    \textbf{Time complexity:} The time complexity is the number of operations required to compute the disk harmonic transform, and then 3 operations per bispectrum coefficient. The time complexity required to compute the disk harmonic transform is $O(L^3)$ \citep{zhao2014rotationally}, hence the time complexity is $O(L^3+m^3/ N_m)$.
\end{proof}



\section{Proofs for Theorems on the Selective Disk Bispectrum}
\label{app:sdb-proofs}

\begin{theorem}[Inversion] \label{th:inversion-app}
    Let $f: \mathbb{D} \to \mathbb{R}$ be a real-valued square-integrable function and let $a_{n, k}$ be its disk harmonic coefficients, bandlimited by $\lambda$. Assume that $a_{n, 1} \neq 0$ for $n \in \{0, ..., N_m-1\}$. Then, its disk bispectrum $b$ can be inverted, that is $f$ can be reconstructed from $b$ up to a planar disk rotation and frequency $\lambda$. The time complexity of the inversion is $O(m+L^3)$.
\end{theorem}

\begin{proof}

We show how to recover the DH coefficients of an image $f$, up to a global phase shift, from its disk bispectrum $b$ coefficients. Afterwards, the disk harmonic expansion \cite{Verrall_Kakarala_1998} in Sec. \ref{sec:background} can be used to reconstruct $f$, up to a global orientation shift and bandlimit.

\textbf{Column $n=0$} We first reconstruct the disk coefficients with Bessel order $n=0$. We have:
\begin{equation}
        b_{0, 0, 1} 
        = a_{n_{0},k_{0}}.a_{n_{0},k_{0}}.a_{n_{0}+n_{0},1}^* 
        = a_{0, 1}.a_{0, 1}.a_{0, 1}^*
        = a_{0, 1}.\|a_{0, 1}\|^2
\end{equation}
Hence: $\|a_{0, 1}\| = \|b_{0, 0, 1} \|^{1/3}$ and $\arg(a_{0,1}) = \arg(b_{0, 0, 1})$. For the rest of this column, i.e., for $k_3 \le K_{n_{j_1} + n_{j_2}} = K_0$:
\begin{equation}
        b_{0, 0, k_3} 
        = a_{n_{0},k_{0}}.a_{n_{0},k_{0}}.a_{n_{0}+n_{0},k_3}^* 
        = a_{0, 1}.a_{0, 1}.a_{0, k_3}^*
        = a_{0, 1}^2.a_{0, k_3}^*
\end{equation}
Hence, $a_{0, k_3} = \left(\frac{b_{0, 0, k_3}}{a_{0, 1}^2}\right)^*$ for $k_3 \in \{1, ..., K_0\}$.

\textbf{Column $n=1$} 

Next, we reconstruct the disk coefficients with Bessel order $n=1$. We have:
\begin{equation}
        b_{2, 0, 1} = a_{n_{2},k_{2}}.a_{n_{0},k_{0}}.a_{n_{0}+n_{2},1}^*
        = a_{1, 1}.a_{0, 1}.a_{1, 1}^*
        = a_{0, 1}.\|a_{1, 1}\|^2
\end{equation}
Hence: $\|a_{1, 1}\| = \|\frac{b_{2, 0, 1}}{a_{0, 1}} \|^{1/2}$ and we fix its phase argument to $0$, which fixes the orientation of the reconstructed image to something arbitrary: $\arg(a_{1, 1}) = 0.$ For the rest of this column, i.e., for every $k_3 \le K_{n_{j_1} + n_{j_2}} = K_1$:
\begin{equation}
        b_{2, 0, k_3} 
        = a_{n_{2},k_{2}}.a_{n_{0},k_{0}}.a_{n_{0}+n_{2},k_3}^*
        = a_{1, 1}.a_{0, 1}.a_{1, k_3}^*
\end{equation}
Hence $a_{1, k_3} = \left(\frac{b_{2, 0, k_3} }{a_{0, 1}.a_{1, 1}}\right)^*$ for $k_3 \in \{1, ..., K_1\}$.

\textbf{Column $n$ from column $n-1$} 

Assume that we have recovered the disk harmonic coefficients up to $n-1$, i.e., all $a_{n', k'}$ for $n' \in \{0, ..., n-1\}$ and $k' \in \{1, ..., K_{n'}\}$. We show that we can recover the nth column. We have, for every $k_3 \le K_{n_{j_1} + n_{j_2}} \equiv K_n$:
\begin{equation}
        b_{2, i(n-1, 1), k} 
        = a_{n_{2},k_{2}}.a_{n_{i(n-1, 1)},k_{i(n-1, 1)}}.a_{n_{2}+n_{i(n-1, 1)},k}^* 
        = a_{1, 1}.a_{n-1, 1}.a_{n, k}^*,
\end{equation}
where $i:(n,k) \to j$. Hence $a_{n, k} = \left(\frac{b_{2, i(n-1, 1), k} }{a_{1, 1}.a_{n-1, 1}}\right)^*$ for $k \in \{1, ..., K_{n}\}$. With this, we can recover all disk harmonics up to the predefined bandlimit $\lambda$, assuming $a_{n, 1} \neq 0$, which is standard in bispectrum inversions and satisfied with any amount of noise on the images. Choosing an arbitrary phase argument for $a_{1, 1}$ means the disk harmonic expansion will reconstruct $f$ up to a global rotation.

We note that reconstructing each disk harmonic coefficient requires exactly one disk bispectrum coefficient. Hence, the time complexity of this step is $O(m)$. Second, the time complexity of the inverse disk harmonic transform is $O(L^3)$, hence the final complexity of $O(m+L^3)$.
\end{proof}

\begin{proposition}[Complexity of the selective disk bispectrum]
    Denote $L$ the size of the image, $m$ the number of disk harmonic coefficients, $N_m$ the maximum angular frequency. The number of selective bispectrum coefficients is $N = \sum_{n= 0}^{N_m} K_n$.
    The space complexity of the selective disk bispectrum is $O(m)$. Its time complexity is $O(L^3 + m)$.
\end{proposition}

\begin{proof}
    The number of selective bispectrum coefficients is:
    \begin{equation}
        N = K_0 + \sum_{n_+=0}^{N_m-1} K_{n_+ +1} = K_0 + \sum_{n=1}^{N_m}K_{n} = \sum_{n= 0}^{N_m} K_n.
    \end{equation}
    \textbf{Space complexity:} We have $N = O(N_m.K_{\max}) = O(m)$ for space complexity.

    \textbf{Time complexity:} The time complexity is the number of operations required to compute the disk harmonic transform, and then 3 operations per selective bispectrum coefficient. The time complexity required to compute the disk harmonic transform is $O(L^3)$ \citep{zhao2014rotationally}, hence the time complexity is $O(L^3+m)$.
\end{proof}
\begin{proposition}[Time complexity of the approximated selective disk bispectrum]
    Denote $L \times L$ the size of the image, $m$ the number of DH coefficients. The time complexity of the approximate selective disk bispectrum, using DH coefficients from \cite{fast_dhcs_2023}, is $O(L^2 \log L + m)$.
\end{proposition}

\begin{theorem}[Complete Invariance]
Let \( f \) and \( f' \) be a pair of real-valued, square-integrable functions on the disk, and let the bispectrum be defined as in Selective Disk Bispectrum Definition. Assume that the disk harmonic coefficients \( a_{n,1} \) of \( f \) are non-zero for all \( n \in \{0, \dots, N_m - 1\} \). Then \( f' = f \circ R_\phi \) for some 2D rotation \( \phi \in SO(2) \) if and only if $b^f = b^{f'}$. Thus, the selective bispectrum is a complete invariant.
\end{theorem}

\begin{proof}
    \textbf{If $f' = f \circ R_\phi$ then $b^f = b^{f'}$.} We first prove that if $f' = f \circ R_\phi$ then $b^f = b^{f'}$, i.e., prove the invariance of the selective bispectrum. We compute the bispectral coefficients of $f'$, as:
    \begin{align*}
        b^{f'}_{j_1, j_2, k_3}
        &= a^{f'}_{n_{j_1},k_{j_1}} \cdot a^{f'}_{n_{j_2},k_{j_2}}\cdot {a^{f'}_{n_{j_1}+n_{j_2},k_3}}^*\\
        &= e^{in_{j_1}\phi} \cdot a^{f}_{n_{j_1},k_{j_1}} \cdot e^{in_{j_2}\phi} \cdot a^{f}_{n_{j_2},k_{j_2}}\cdot \\
        &\times e^{-i(n_{j_1}+n_{j_2})\phi}\cdot {a^{f}_{n_{j_1}+n_{j_2},k_3}}^*\\
        &= a^{f}_{n_{j_1},k_{j_1}} \cdot a^{f}_{n_{j_2},k_{j_2}}\cdot {a^{f}_{n_{j_1}+n_{j_2},k_3}}^*\\
        &=b^{f}_{j_1, j_2, k_3}.
    \end{align*}
    Hence, two functions that differ from a rotation have the same full disk bispectrum and thus the same selective disk bispectrum.
    
    \textbf{If $b^f = b^{f'}$ then there exists $\phi \in SO(2)$ such that $f' = f \circ R_\phi$}. Let $\tilde{f}$ be the function reconstructed from our inversion algorithms from $b^f= b^{f'}$. We have that there exists $\phi, \phi'$ such that $\tilde{f} = f \circ R_\phi$ and $\tilde{f} = f' \circ R_{\phi'}$. Hence, $f' = f \circ R_{\phi - \phi'}$, where $\phi - \phi'$ is itself a rotation in $SO(2)$, completing the proof that $f'$ and $f$ differ by a planar rotation.
\end{proof}

\section{Proofs for Theorems on Approximating the Selective Disk Bispectrum}
\label{app:approx-proofs}
To further reduce the time complexity of the selective bispectrum, we propose leveraging a recent numerical approximation of the DH coefficients \citep{fast_dhcs_2023} which computes the disk harmonic transform in $O(L^2 \log L)$. We prove that using this DH coefficient approximation, we still have precise guarantees on the accuracy of the SDB coefficient approximations, the SDB rotation invariance, and reconstruction of the original image from SDB coefficients.

\begin{theorem}[Accuracy Guarantee]
\label{th:accuracy-app}
Let \( 0 < \varepsilon \leq 1 \) be a target accuracy, and assume that \( \lambda \leq \sqrt{\pi p} \) and \( |\log \varepsilon| \leq \sqrt{p} \), where $\lambda$ is the bandlimit for the disk harmonic coefficients and \( p = L^2 \) the number of pixels in the image $f$ of size $L$. Let \( b^f \) and \( \tilde{b}^f \) denote the exact and approximate selective disk bispectra of \( f \), where \( \tilde{b}^f \) is computed using the approximate disk harmonic transform of \citet{fast_dhcs_2023}. Then,
\[
\| \tilde{b}^f - b^f \|_{\infty} \leq C\varepsilon \| f \|_{1},
\]
\end{theorem}
\begin{proof}
    Consider one coefficient of the selective disk bispectrum $b$ and its approximate version $\tilde{b}$ as:
    \begin{equation}
        b = a_1 \cdot a_2 \cdot a_3,
        \quad
        \text{and:}
        \quad
        \tilde{b} = \tilde{a}_1 \cdot \tilde{a}_2 \cdot \tilde{a}_3,
    \end{equation}
where we index the disk harmonic coefficients by $1, 2, 3$ for simplicity, without loss of generality.

The approximate disk harmonic coefficients, $\tilde{a}$, are given by \citet{fast_dhcs_2023}. Using their Theorem 4.1., we know that:
\begin{equation}
    \|\tilde{a} - a\|_{\ell^\infty} \leq \varepsilon \|f\|_{\ell^1}.
\end{equation}

Hence, we can compute:
\begin{align*}
    & |\tilde{b} - b| \\
    &=
    |\tilde{a}_1 
    \cdot \tilde{a}_2 \cdot \tilde{a}_3
    -
    a_1 
    \cdot a_2 \cdot a_3|\\
    &=
    |\tilde{a}_1 
    \cdot \tilde{a}_2 \cdot \tilde{a}_3
    -
    a_1 
    \cdot \tilde{a}_2 \cdot \tilde{a}_3
    +
    a_1 
    \cdot \tilde{a}_2 \cdot \tilde{a}_3
    -
    a_1 
    \cdot a_2 \cdot \tilde{a}_3 \\
    &\quad +
    a_1 
    \cdot a_2 \cdot \tilde{a}_3
    -
    a_1 
    \cdot a_2 \cdot a_3|\\
    &=
    |(\tilde{a}_1 - a_1)\cdot \tilde{a}_2 \cdot \tilde{a}_3 
    + 
    a_1 \cdot (\tilde{a}_2 - a_2) \cdot \tilde{a}_3 \\
    &\quad + a_1 \cdot a_2 \cdot (\tilde{a}_3 - a_3)|\\
    &\leq 
    |\tilde{a}_1 - a_1|\cdot |\tilde{a}_2| \cdot |\tilde{a}_3 |
    + 
    |a_1| \cdot |\tilde{a}_2 - a_2| \cdot |\tilde{a}_3| \\
    & \quad + |a_1 |\cdot |a_2| \cdot |\tilde{a}_3 - a_3|
    ~\text{(triangle inequality)}\\
    &\leq 
    \varepsilon \|f\|_{\ell^1}\cdot |\tilde{a}_2| \cdot |\tilde{a}_3 |
    + 
    |a_1| \cdot \varepsilon \|f\|_{\ell^1} \cdot |\tilde{a}_3| \\
    &\quad + |a_1 |\cdot |a_2| \cdot \varepsilon \|f\|_{\ell^1}
    ~\text{(\cite{fast_dhcs_2023} Th. 4.1)}\\
    &=\varepsilon \|f\|_{\ell^1}\cdot \left( 
    |\tilde{a}_2| \cdot |\tilde{a}_3 |
    + 
    |a_1| \cdot |\tilde{a}_3|
    + |a_1 |\cdot |a_2| \right) \\
    &\quad ~\text{(reorganizing terms)}
\end{align*}

To conclude, we must bound $|a_j|$ and $|\tilde{a}_j|$. We start with the exact disk harmonic coefficient. By definition:
\begin{equation}
    a_j = \int_\mathbb{D} f(x).\psi_j^*(x)dx = \langle f, \psi_j \rangle,
\end{equation}
where $\psi_j$ is the disk harmonic of index $j$. Applying the Cauchy--Schwarz inequality, we get:
\begin{equation}
    |a_j| \leq \|f\|_2\cdot \|\psi_j\|_2 = \|f\|_2,
\end{equation}
where we use the fact that the disk harmonics form an orthonormal basis and thus have norm $1$.

Then, we have:
\begin{equation}
    |\tilde{a}_j| = |\tilde{a}_j - a_j + a_j| \leq |a_j| + |\tilde{a}_j - a_j| \leq \|f\|_2 + \varepsilon \|f\|_1 
\end{equation}

We plug this in the inequality derivations above.

\begin{align*}
    |\tilde{b} - b|
    &\leq \varepsilon \|f\|_{\ell^1}\cdot \left( 
    |\tilde{a}_2| \cdot |\tilde{a}_3 |
    + 
    |a_1| \cdot |\tilde{a}_3|
    + |a_1 |\cdot |a_2| \right)\\
    &\leq \varepsilon \|f\|_{\ell^1}\cdot \{ 
    (\|f\|_2 + \varepsilon \|f\|_1 )^2 \\
    & \quad+ 
    \|f\|_2 \cdot (\|f\|_2 + \varepsilon \|f\|_1 )
    + \|f\|_2\cdot \|f\|_2 \}.
\end{align*}
Since the disk is a compact domain, $f$ can be bounded from above by a constant, and hence its $L_1$ and $L_2$ norms are also bounded from above. Hence, we can bound the terms in the parenthesis by a constant $C$.
\begin{align*}
    |\tilde{b} - b|
    &\leq \varepsilon \|f\|_{\ell^1}\cdot \left( 
    |\tilde{a}_2| \cdot |\tilde{a}_3 |
    + 
    |a_1| \cdot |\tilde{a}_3|
    + |a_1 |\cdot |a_2| \right)\\
    &\leq \varepsilon \|f\|_{\ell^1}\cdot C
\end{align*}

This concludes the proof.
\end{proof}

As a corollary, we find that the bispectrum coefficients are approximately invariant with controlled guarantees.

\begin{corollary}[Invariance Guarantee]
    Let $f, f'$ be a pair of real-valued square-integrable functions on the disk such that there exists a $\phi \in SO(2)$ such that $f' = f \circ R_\phi$. Under the notations and assumptions of Theorem~\ref{th:accuracy}, we have:
   \[
\| \tilde{b}^f - \tilde{b}^{f'} \|_{\infty} \leq 2C\varepsilon \| f \|_{1},
\]
    for a constant \( C \) depending only on a bound of \( f \) on its compact disk support.
\end{corollary}

\begin{proof}
    Consider one coefficient of $\tilde b^{f}$ and $\tilde b^{f'}$, respectively. We have
    \begin{align*}
        | \tilde b^{f} - \tilde b^{f'}|
        & =| \tilde b^{f} - b^f + b^f - b^{f'} + b^{f'} - \tilde b^{f'}|\\
        &\leq | \tilde b^{f} - b^f| + |b^f - b^{f'}| + |b^{f'} - \tilde b^{f'}|\\
        &\quad ~\text{(triangle inequality)}\\
        &=| \tilde b^{f} - b^f| + |b^{f'} - \tilde b^{f'}|\\
        &\quad ~\text{(invariance of bispectrum)}\\
        &\leq 2C\varepsilon \|f\|_1\\
        &\quad ~\text{($f$ and $f'$ have the same bound on the disk).}
    \end{align*}
    If all coefficients are bounded, then their $L_\infty$ norm is also bounded. This concludes the proof.
\end{proof}
Linking the accuracy to $f$ is useful, as it allows us to control accuracy through number of pixels $p$. If $|f|$ and $C$ are large, but we still want to reach a given accuracy, then we can choose $p$ accordingly such that $|\log \varepsilon| \leq \sqrt{p}$ to match the desired accuracy. 

\section{Details of Classification MLP and Hyperparameters}
\label{app:mlp-details}
We vary MLP parameter counts $p \in \{10k, 50k, 500k  \}$ and image resolutions $L\times L \in \{28\times 28, 56\times 56, 112\times 112  \}$. The MLP architecture (0–2 hidden layers) is determined by the input feature size and parameter budget. We train for 15 epochs with learning rate $0.005$, dropout $0.1$, and batch size $512$. We use the classic train/test split of MNIST and FashionMNIST. 20 percent of the train dataset is used for validation.

\begin{figure}[t]
  \centering
  \includegraphics[width=\textwidth]{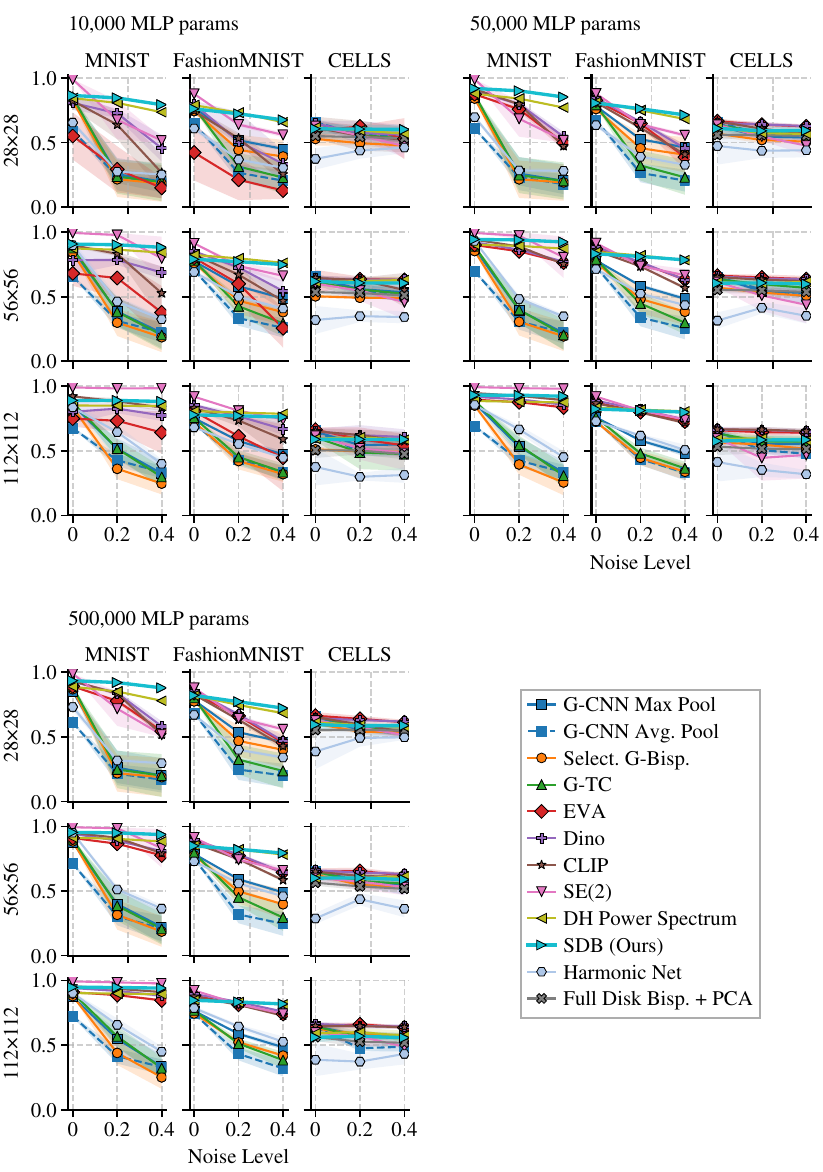}
  \caption{\textbf{The SDB's advantage over convolutional descriptors under noise holds at every image size and MLP budget.} Within each block, rows are image sizes $L \in \{28, 56, 112\}$ and columns are datasets (MNIST, FashionMNIST, and osteosarcoma CELLS). The three blocks correspond to MLP budgets of roughly $10{,}000$, $50{,}000$ and $500{,}000$ parameters. Every panel plots test accuracy against the Gaussian noise level $\sigma \in \{0, 0.2, 0.4\}$ on a shared accuracy scale, as mean over seeds with $\pm 1$ standard deviation shaded. The protocol matches Fig.~\ref{fig:classification}. The main-text figure is the $L = 28$, $500{,}000$-parameter slice of this grid.}
  \label{fig:classification-appendix}
\end{figure}

\section{Stable MRA Under Noise}
\label{sec:damping}

\paragraph{Setting.} We observe $N$ copies $x_i = R_{\theta_i} f + \eta_i$, $i=1,\dots,N$, of a template $f$ with unknown rotations $\theta_i \sim \mathrm{Unif}[0,2\pi)$ and white pixel noise $\eta_i \sim \mathcal{N}(0,\sigma^{2} I)$. Since the SDB is rotation-invariant, the SDB of each observation approximates the SDB of $f$. Without inversion damping, we would average them, subtract the bias of Th.~\ref{th:expectation-value-main-text}, and invert the result $\hat b$ with the recursion of Th.~\ref{th:inversion}. The template is recovered only up to a global rotation, so reconstruction error is measured after minimization over $SO(2)$.

\paragraph{Why bias correction is not enough.} After the bias is subtracted, $\hat b$ is an unbiased estimate of $b$, so $\hat b \to b$ as $N \to \infty$ and the recursion recovers $f$ in the limit. At any finite $N$, however, $\hat b$ differs from $b$ by a random fluctuation (Th.~\ref{th:variance}), and the recursion divides by estimated quantities. Bias correction fixes where $\hat b$ is centred; it does not prevent a divisor from being a random number close to zero. That is the problem addressed here.

\paragraph{The recursion under noise.} For $n \ge 2$ the recursion recovers
\begin{equation}
  \hat a_{n,k} = \left( \frac{\hat b_{2,n-1,k}}{\hat d_n} \right)^*,
  \qquad
  \hat d_n = \hat a_{1,1}\,\hat a_{n-1,1},
  \label{eq:cascade}
\end{equation}
with $\hat a_{1,1}$ and $\hat a_{n-1,1}$ recovered at earlier steps. With exact inputs, $b_{2,n-1,k} = a_{1,1}a_{n-1,1}a_{n,k}^*$ contains the divisor $d_n = a_{1,1}a_{n-1,1}$ as a factor, so it cancels and \eqref{eq:cascade} returns $a_{n,k}$ without issue, however small $d_n$ is. However, if the estimations of $b_{2,n-1,k}$ and $d_n$ are corrupted by noise after averaging many noisy signals, the appropriate terms do not cancel exactly. If $d_n$ is close to zero and does not cancel, this causes instabilities that propagate through the recursive inversion.

\paragraph{Noise floor.} Each observation is the rotated template plus pixel noise of standard deviation $\sigma$. Because the DH basis is orthonormal, the same is true of its coefficients. Every $a_{n,k}$ is the template coefficient, rotated, plus noise of standard deviation $\sigma$. For the SDB coefficieint $b_{2,n-1,k} = a_{1,1}\,a_{n-1,1}\,a_{n,k}^*$ with $n \ge 2$, Th.~\ref{th:variance-app} gives,
\begin{equation}
  \begin{aligned}
  \mathrm{Var}(b_{2,n-1,k})
  &= \sigma^{2}\big(\lvert a_{1,1}\rvert^{2}\lvert a_{n-1,1}\rvert^{2} + \lvert a_{1,1}\rvert^{2}\lvert a_{n,k}\rvert^{2} + \lvert a_{n-1,1}\rvert^{2}\lvert a_{n,k}\rvert^{2}\big) \\
  &\quad + \sigma^{4}\big(\lvert a_{1,1}\rvert^{2} + \lvert a_{n-1,1}\rvert^{2} + \lvert a_{n,k}\rvert^{2}\big)
  + \sigma^{6},
  \end{aligned}
  \label{eq:triple-variance}
\end{equation}
and the average over $N$ independent observations has variance $\mathrm{Var}(b_{2,n-1,k})/N$. The $\sigma^{4}$ and $\sigma^{6}$ terms are negligible against the $\sigma^{2}$ term when $\sigma \ll \lvert a_{1,1}\rvert$ and $\sigma \ll \lvert a_{n-1,1}\rvert$ (a condition on ratios, since $\sigma$ and the coefficients have the same units). We assume this in what follows.

The recursion is seeded by $\hat a_{0,1} = \hat b_{001}^{1/3}$ and $\lvert \hat a_{1,1}\rvert = (\hat b_{201}/\hat a_{0,1})^{1/2}$ (Th.~\ref{th:inversion}). By \eqref{eq:variance-001}, the averaged $\hat b_{001}$ has standard deviation $\sqrt{9\sigma^{2}a_{0,1}^{4}/N}$ at leading order, so, to first order, $\hat a_{0,1}$ has standard deviation $\sqrt{9\sigma^{2}a_{0,1}^{4}/N}\big/(3a_{0,1}^{2}) = \sigma/\sqrt{N}$. This error only changes the mean brightness of the reconstruction, and since $a_{0,1}$ is the largest coefficient it is negligible in relative terms, so we take $\hat a_{0,1} = a_{0,1}$ in what follows. The coefficient that matters is $a_{1,1}$, which enters every divisor $d_n$. By \eqref{eq:variance-201}, the averaged $\hat b_{201}$ has standard deviation $\sqrt{\sigma^{2}(\lvert a_{1,1}\rvert^{4} + 2a_{0,1}^{2}\lvert a_{1,1}\rvert^{2})/N}$ at leading order. Since $\lvert \hat a_{1,1}\rvert^{2} = \hat b_{201}/a_{0,1}$ with $a_{0,1}$ fixed, $\lvert \hat a_{1,1}\rvert^{2}$ has standard deviation $\operatorname{std}(\hat b_{201})/a_{0,1}$; and since $\lvert \hat a_{1,1}\rvert = (\lvert \hat a_{1,1}\rvert^{2})^{1/2}$ with $\frac{d}{dy}\,y^{1/2} = 1/(2y^{1/2})$, to first order $\lvert \hat a_{1,1}\rvert$ has standard deviation $\operatorname{std}(\lvert \hat a_{1,1}\rvert^{2})/(2\lvert a_{1,1}\rvert)$. Together,
\begin{equation}
  \begin{aligned}
  \operatorname{std}(\lvert \hat a_{1,1}\rvert)
  &= \frac{\operatorname{std}(\hat b_{201})}{2a_{0,1}\lvert a_{1,1}\rvert}
  = \frac{\sqrt{\sigma^{2}(\lvert a_{1,1}\rvert^{4} + 2a_{0,1}^{2}\lvert a_{1,1}\rvert^{2})/N}}{2a_{0,1}\lvert a_{1,1}\rvert} \\
  &= \frac{\sigma}{\sqrt{N}}\sqrt{\frac{1}{2} + \frac{\lvert a_{1,1}\rvert^{2}}{4a_{0,1}^{2}}}
  \approx \frac{\sigma}{\sqrt{2N}},
  \end{aligned}
  \label{eq:a11-error}
\end{equation}
where the last step follows since $a_{0,1}$ is the largest coefficient. The phase of $a_{1,1}$ is fixed to zero, which sets the global rotation of the reconstruction.

For a general step, write $e_b = \hat b_{2,n-1,k} - b_{2,n-1,k}$ for the error of the averaged bispectrum. If the divisor were exact, the recursion \eqref{eq:cascade} would pass this error through as $e_b/d_n$, of standard deviation
\begin{equation}
  \frac{\sqrt{\mathrm{Var}(b_{2,n-1,k})/N}}{\lvert a_{1,1}\rvert\lvert a_{n-1,1}\rvert}
  \approx \frac{\sigma}{\sqrt{N}}
    \sqrt{1 + \frac{\lvert a_{n,k}\rvert^{2}}{\lvert a_{n-1,1}\rvert^{2}} + \frac{\lvert a_{n,k}\rvert^{2}}{\lvert a_{1,1}\rvert^{2}}} .
  \label{eq:error-derivation}
\end{equation}
The divisor is not exact, but its error is built from the same noise as $e_b$ and, to first order, cancels part of \eqref{eq:error-derivation} rather than adding to it, just as $d_n$ cancels exactly in the noiseless case; \eqref{eq:error-derivation} is therefore a conservative estimate of the error in each recovered coefficient. The square root is of order one, since the coefficients decay with $n$ and $a_{1,1}$, $a_{n-1,1}$ are leading radial modes. Up to the global rotation, which the error metric removes, each recovered coefficient therefore carries an error of order
\begin{equation}
  s = \frac{\sigma}{\sqrt{N}},
  \label{eq:noise-floor}
\end{equation}
the usual $1/\sqrt{N}$ of averaging $N$ measurements. We call $s$ the noise floor: a recovered coefficient smaller than $s$ cannot be distinguished from noise.

\paragraph{Why the recursion fails.} This holds only while the divisors are well above the noise. Because the coefficients of any template decay with $n$ while $s$ does not, there is always a frequency beyond which $\lvert a_{n-1,1}\rvert \lesssim s$. There $\hat a_{n-1,1}$ is mostly noise, so $\hat d_n$ is a random number of scale $\lvert a_{1,1}\rvert s$ centred near zero, and the next step divides $\hat b_{2,n-1,k}$, whose own noise does not cancel against it, by this number. The result can be arbitrarily large, and through the chain it corrupts every higher frequency. More observations lower $s$ as $1/\sqrt{N}$, which moves this frequency up but does not remove it. This failure mode is known. For the analogous recursion in one-dimensional multi-reference alignment, called frequency marching, \citet{bendory2018} note that it is exact without noise but that, because of its recursive nature, error in the low frequencies propagates to the high frequencies; in bispectral phase recovery for speckle imaging the recursion was replaced by weighted least squares over all bispectrum coefficients for the same reason~\cite{marron1990unwrapping, matson1991weighted}, and \citet{bendory2018} likewise turn to least-squares and semidefinite formulations. We keep the recursion and regularize the division.

\paragraph{Damped inversion.} We replace the division in \eqref{eq:cascade} by
\begin{equation}
  \hat a^{\varepsilon}_{n,k}
  = \left( \frac{\hat b_{2,n-1,k}\,\hat d_n^{*}}{\lvert \hat d_n\rvert^{2} + \varepsilon^{2}} \right)^*
  = \frac{\lvert \hat d_n\rvert^{2}}{\lvert \hat d_n\rvert^{2} + \varepsilon^{2}}\;\hat a_{n,k},
  \label{eq:damped-division}
\end{equation}
which is the exact recovery multiplied by a real factor between $0$ and $1$. This is the Tikhonov-regularized inverse of $\hat d_n$~\cite{tikhonov1977solutions}. When $\lvert \hat d_n\rvert \gg \varepsilon$ the factor is one and nothing changes. When $\lvert \hat d_n\rvert \ll \varepsilon$ the factor tends to zero, and the coefficient is set to zero rather than to the arbitrarily large value of the exact recursion. For every $\hat d_n$, the damped quotient has magnitude $\lvert \hat b_{2,n-1,k}\rvert\,\lvert \hat d_n\rvert/(\lvert \hat d_n\rvert^{2} + \varepsilon^{2}) \le \lvert \hat b_{2,n-1,k}\rvert/(2\varepsilon)$. With $\varepsilon$ at the noise scale of $\hat d_n$, the coefficients suppressed are exactly those whose divisor cannot be distinguished from noise, where nothing was recoverable in the first place.

\paragraph{Choice of $\varepsilon$.} The noise in $\hat d_n = \hat a_{1,1}\hat a_{n-1,1}$ is mainly that of $\hat a_{n-1,1}$, of standard deviation $s$, multiplied by $\lvert a_{1,1}\rvert$, so the natural scale is $\varepsilon = \lvert a_{1,1}\rvert s$. At exactly this scale a divisor at the noise level is only halved, whereas a predecessor far below the noise should be removed almost entirely. We therefore enlarge the scale by the noise-to-signal ratio of the predecessor, as a Wiener filter does~\cite{wiener1949extrapolation},
\begin{equation}
  \varepsilon_n^{2}
  = \lvert \hat a_{1,1}\rvert^{2}\, s^{2}
    \left(1 + \frac{s^{2}}{\lvert a_{n-1,1}\rvert^{2}}\right),
  \label{eq:wiener-epsilon}
\end{equation}
so that the bracket is near one when $\lvert a_{n-1,1}\rvert \gg s$ and grows without bound when $\lvert a_{n-1,1}\rvert \ll s$.

\paragraph{Estimating $\sigma$ and $\lvert a_{n-1,1}\rvert^{2}$.} Both follow from the second moments of Th.~\ref{th:second-moment}. For a noisy coefficient $\tilde a_{n,k} = a_{n,k} + N_{n,k}$, $\mathbb{E}[N_{n,k}] = 0$ and $\mathbb{E}\lvert N_{n,k}\rvert^{2} = \sigma^{2}$ give $\mathbb{E}\lvert \tilde a_{n,k}\rvert^{2} = \lvert a_{n,k}\rvert^{2} + \sigma^{2}$, and a rotation of the observation multiplies $a_{n,k}$ by a unit phase, so the identity holds for every observation regardless of its unknown rotation. The mean squared norm over observations, $\hat S_{n,k} = N^{-1}\sum_{i}\lvert a_{n,k}(x_i)\rvert^{2}$, therefore estimates $\lvert a_{n,k}\rvert^{2} + \sigma^{2}$. At the highest angular frequencies the signal term is negligible, leaving $\sigma^{2}$, so we take
\begin{equation}
  \hat\sigma^{2} = \operatorname{median}_{\,n \ge \tau n_{\max}} \hat S_{n,k},
  \qquad
  \lvert a_{n-1,1}\rvert^{2} \approx \max\big(0,\; \hat S_{n-1,1} - \hat\sigma^{2}\big),
  \label{eq:sigma-estimator}
\end{equation}
with $\tau = 0.7$, the median guarding against residual signal in the same way as the noise-level estimator of \citet{donoho1994ideal}. The predecessor power is taken from $\hat S_{n-1,1}$ rather than from the recovered $\hat a_{n-1,1}$ because $\hat S_{n-1,1}$ never passes through a division, so it remains more reliable.

\begin{figure}[t]
  \centering
  \includegraphics[width=\textwidth]{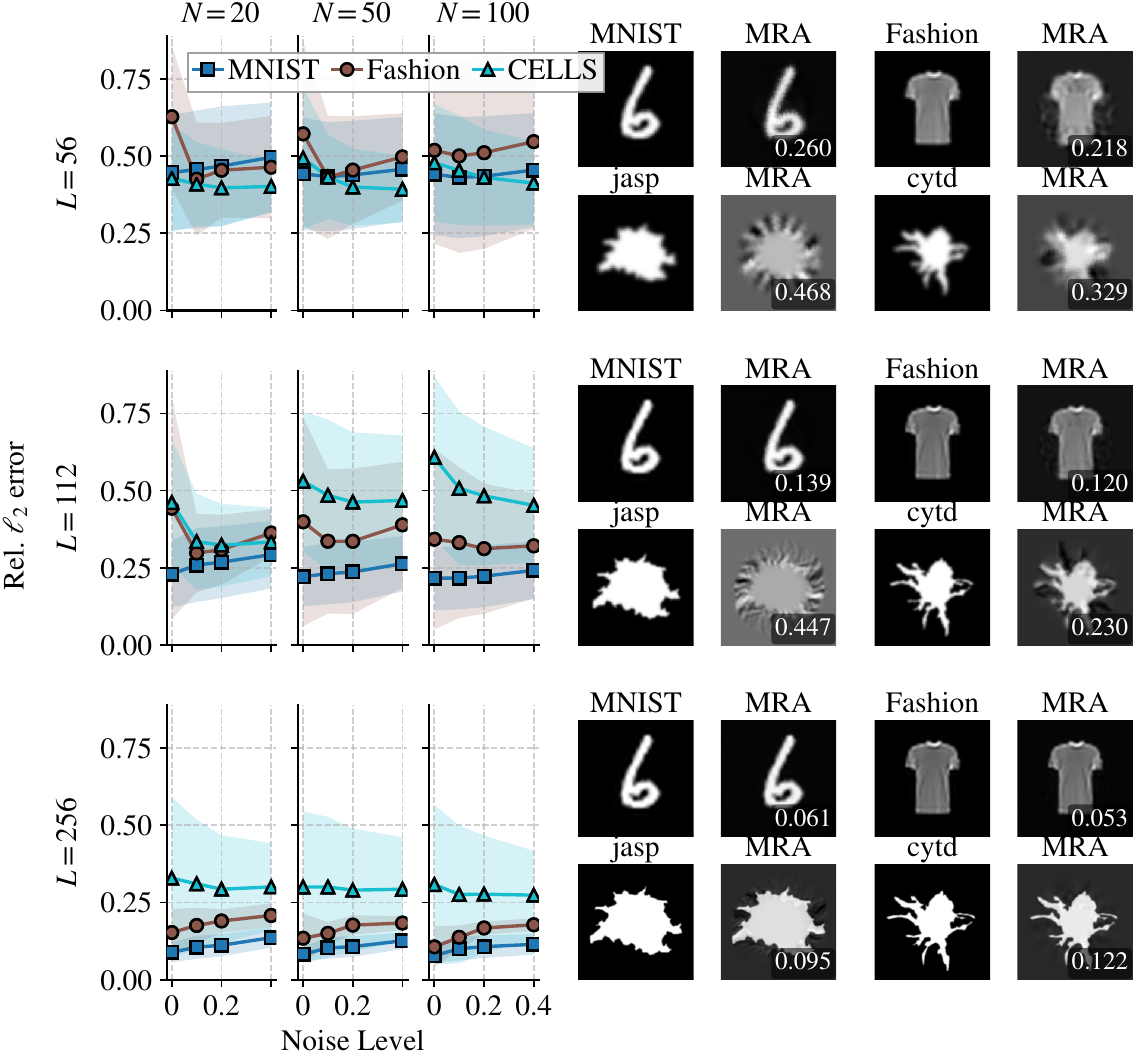}
  \caption{Multi-reference alignment across image sizes. Each row repeats the main-text experiment at a different resolution, $L \in \{56, 112, 256\}$, with bandlimits $44$, $84$ and $160$ respectively. The main-text figure is the $L = 256$ row.}
  \label{fig:mra-appendix}
\end{figure}

All multi-reference alignment results, shown in Fig. \ref{fig:mra-appendix}, use the damped recursion with $\varepsilon_n$ from \eqref{eq:wiener-epsilon}.

\section{SDB Round Trip Error}
\label{sdb-round-trip-error}

\begin{figure}[p]
  \centering
  \includegraphics[height=0.9\textheight,keepaspectratio]{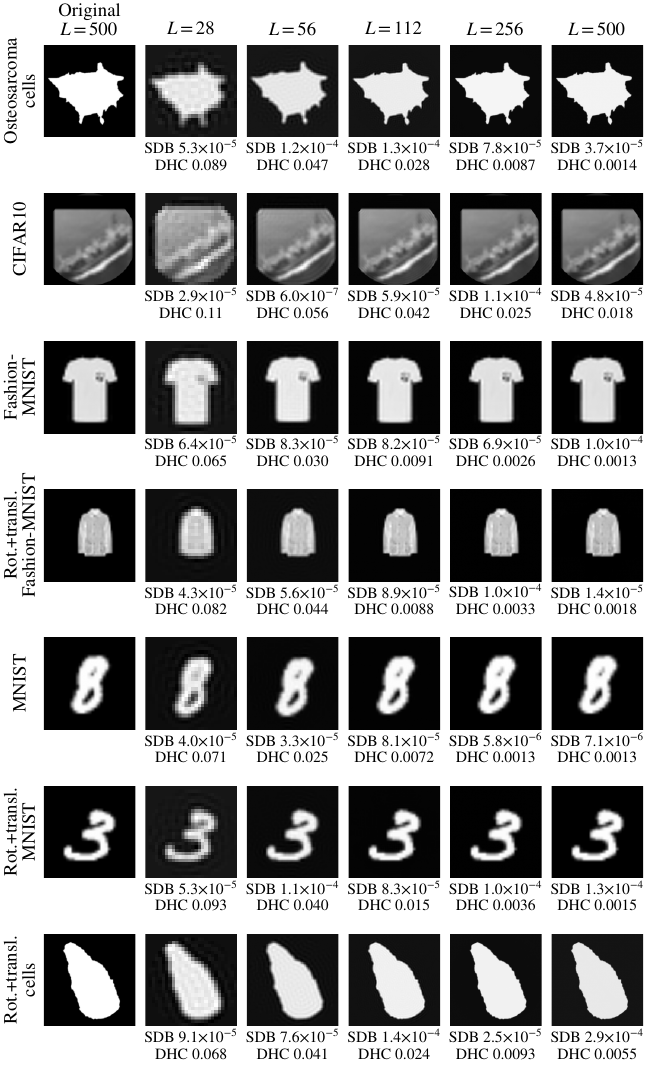}
  \caption{Inversion examples at $L\in \{28, 56, 112, 256, 500$\}, with one row per dataset. The SDB error reports the error from mapping an image's DH coefficients to the SDB and back to DH coefficients. The DH coefficient (DHC) error reports the reconstruction error from mapping an image to its DHCs and back.}
  \label{fig:reconstruction-examples}
\end{figure}

Reconstruction accuracy is tied to the accuracy of the bispectrum map. Theorem~\ref{th:accuracy} bounds the error between the exact bispectrum and the bandlimited implemented bispectrum by a term that shrinks as the number of pixels grows. We demonstrate this on full evaluation sets across six datasets at $L\in\{28,56,112,256,500\}$, using all $N{=}650$ osteosarcoma cell silhouettes \citep{alizadeh2019tismorph, li2023using} and the full $10{,}000$-image test sets for MNIST, Fashion-MNIST, rotated and translated (RT) MNIST, rotated and translated (RT) Fashion-MNIST, and CIFAR10. Errors are median relative $\ell_2$ after rotation registration. Each entry is SDB inversion error (with DHC round-trip error in parentheses). Example reconstructions for each dataset and image size can be found in Fig. \ref{fig:reconstruction-examples}.

\begin{center}
\small
\begin{tabular}{lccccc}
\hline
dataset & $L{=}28$ & $L{=}56$ & $L{=}112$ & $L{=}256$ & $L{=}500$ \\
\hline
osteosarcoma cells & 0.259 (0.051) & 0.153 (0.073) & 0.076 (0.030) & 0.086 (0.056) & 0.079 (0.038) \\
CIFAR10 & 0.281 (0.133) & 0.207 (0.160) & 0.144 (0.121) & 0.077 (0.061) & 0.055 (0.044) \\
Fashion-MNIST & 0.342 (0.123) & 0.245 (0.178) & 0.153 (0.115) & 0.059 (0.019) & 0.038 (0.009) \\
RT Fashion-MNIST & 0.360 (0.138) & 0.305 (0.242) & 0.206 (0.171) & 0.088 (0.044) & 0.053 (0.015) \\
MNIST & 0.446 (0.134) & 0.250 (0.158) & 0.136 (0.085) & 0.064 (0.017) & 0.043 (0.007) \\
RT MNIST & 0.481 (0.163) & 0.315 (0.203) & 0.163 (0.102) & 0.069 (0.022) & 0.046 (0.009) \\
\hline
\end{tabular}
\end{center}

At $L{=}500$, a common microscopy image size, every dataset has median SDB error at most $0.08$. The highest is osteosarcoma cells at $0.079$ and the lowest is Fashion-MNIST at $0.038$. Osteosarcoma cells already reach $0.076$ at $L{=}112$. At this resolution the reconstructed cell morphology is visually not distinguishable from the original. Most residual numerical error is an intensity fade, for example white turning to grey, while structure is preserved.

The key point is that SDB error stays close to DHC round-trip error at large $L$. Much of the reconstruction error is already present when we project onto disk harmonics and reconstruct the image, without ever forming the SDB. Selective inversion does not introduce a large new failure mode on top of bandlimiting.

\section{SDB Timing}
\label{sdb-timing}

Table~\ref{tab:sdb-timing} reports wall-clock time to compute the SDB (forward) and to invert it (inverse), compared with the disk-harmonic coefficient (DHC) transform of the same image. Times are mean$\pm$s.d.\ over five seeds of five timed repetitions on a single image, at the library default bandlimit $\lambda = \max(20, L/4)$, matching the round-trip errors of Sec.~\ref{sdb-round-trip-error}.

\begin{table}[t]
\centering
\small
\begin{tabular}{llcccc}
\hline
$L$ & $\lambda$ & SDB fwd (s) & SDB inv (s) & DHC fwd (s) & DHC inv (s) \\
\hline
28 & 20 & $0.104{\pm}0.004$ & $0.091{\pm}0.004$ & $0.075{\pm}0.001$ & $0.077{\pm}0.004$ \\
56 & 20 & $0.104{\pm}0.002$ & $0.122{\pm}0.008$ & $0.075{\pm}0.001$ & $0.099{\pm}0.002$ \\
112 & 28 & $0.149{\pm}0.035$ & $0.146{\pm}0.024$ & $0.121{\pm}0.032$ & $0.107{\pm}0.006$ \\
256 & 64 & $0.122{\pm}0.003$ & $0.422{\pm}0.003$ & $0.095{\pm}0.003$ & $0.273{\pm}0.002$ \\
500 & 125 & $0.167{\pm}0.002$ & $1.537{\pm}0.008$ & $0.146{\pm}0.004$ & $0.967{\pm}0.004$ \\
\hline
\end{tabular}
\caption{\textbf{The SDB forward pass stays comparable to a DHC transform at every resolution.} Forward and inverse wall-clock time versus image size. Inversion cost grows with $L$ because the sequential recovery walks every selected coefficient, reaching $1.5$ s per image at $L = 500$.}
\label{tab:sdb-timing}
\end{table}

\section{SDB Under Image Translation}
\label{sdb-translation}

SDB is rotation-invariant but not translation-invariant, so translations change the descriptor. We therefore study the effect of translations on the SDB representation. A translation is called \emph{admissible} when the object's support remains inside the computational disk after the translation (no clipping at the disk boundary).

\paragraph{Practical context.} At high SNR (for example, histopathology or clean microscopy), centering is cheap and reliable. An intensity center-of-mass or a single registration pass costs $O(N)$ in the number of pixels and is already part of most pipelines. At low SNR (for example, cryo-EM), centering can be harder, and particles may retain residual translations. SDB is suited to settings where reliable centering is available; the experiments below show that it also tolerates residual translations.

\paragraph{Experiment (i): SDB separability for translated MNIST.} We take one center-of-mass-centered prototype per digit. For each maximum translation radius $r\in\{0,1,2,4,8\}$ pixels ($0\%$, $1.8\%$, $3.6\%$, $7.1\%$, and $14.3\%$ of the disk diameter $56$), we draw twenty random admissible translations with $|\Delta_x|,|\Delta_y|\le r$ and compute pairwise $\ell_2$ distances on the real and imaginary parts of the SDB vectors. \emph{Separability} is the mean between-digit SDB $\ell_2$ divided by the mean within-digit distance. A ratio $\gg 1$ means digit identity remains recoverable; a ratio near $1$ means collapse. At $r=0$, the ratio is $\infty$.

\begin{center}
\small
\begin{tabular}{rrr}
\hline
max translation $r$ & \% of disk diameter & SDB sep.\ after translation \\
\hline
0 px & $0\%$ & $\infty$ \\
1 px & $1.8\%$ & 22.6 \\
2 px & $3.6\%$ & 12.8 \\
4 px & $7.1\%$ & 7.2 \\
8 px & $14.3\%$ & 3.7 \\
\hline
\end{tabular}
\end{center}

Even without perfect centering, same-digit SDBs stay clustered at moderate residual translations. Digit identity therefore remains recoverable from $\ell_2$ distances in SDB space.

\paragraph{Experiment (ii): classification of translated and rotated MNIST.} We evaluate classification on images that are both rotated and translated with an admissible translation, and include SE(2) and Harmonic baselines. We use $L\in\{28,56,112\}$ and an MLP sized $\approx 50{,}000$ parameters. We report mean$\pm$std over seeds (typically $n{=}5$; a few cells use $n{=}4$), for $L$ and $\sigma\in\{0,0.2,0.4\}$.

\begin{center}
\footnotesize
\setlength{\tabcolsep}{3pt}
\resizebox{\linewidth}{!}{%
\begin{tabular}{llccccccccc}
\hline
$L$ & $\sigma$ & SDB & G-CNN (max) & \texttt{select\_gbisp} & \texttt{tc} & SE(2) & Harmonic & DINOv3 & EVA & OpenCLIP \\
\hline
28  & 0   & $.837{\pm}.002$ & $.720{\pm}.006$ & $.712{\pm}.005$ & $.761{\pm}.002$ & $.969{\pm}.000$ & $.681{\pm}.091$ & $.887{\pm}.002$ & $.865{\pm}.005$ & $.904{\pm}.003$ \\
    & 0.2 & $.743{\pm}.039$ & $.225{\pm}.080$ & $.199{\pm}.094$ & $.217{\pm}.093$ & $.613{\pm}.203$ & $.196{\pm}.017$ & $.762{\pm}.009$ & $.722{\pm}.011$ & $.764{\pm}.006$ \\
    & 0.4 & $.551{\pm}.072$ & $.177{\pm}.075$ & $.163{\pm}.085$ & $.169{\pm}.087$ & $.171{\pm}.107$ & $.145{\pm}.018$ & $.483{\pm}.030$ & $.440{\pm}.029$ & $.442{\pm}.019$ \\
\cmidrule(lr){1-11}
56  & 0   & $.850{\pm}.001$ & $.740{\pm}.003$ & $.732{\pm}.003$ & $.788{\pm}.003$ & $.978{\pm}.001$ & $.897{\pm}.045$ & $.916{\pm}.003$ & $.896{\pm}.007$ & $.938{\pm}.001$ \\
    & 0.2 & $.812{\pm}.021$ & $.278{\pm}.074$ & $.233{\pm}.078$ & $.275{\pm}.073$ & $.903{\pm}.061$ & $.283{\pm}.047$ & $.862{\pm}.008$ & $.851{\pm}.008$ & $.896{\pm}.004$ \\
    & 0.4 & $.735{\pm}.042$ & $.175{\pm}.064$ & $.158{\pm}.062$ & $.165{\pm}.065$ & $.659{\pm}.296$ & $.188{\pm}.025$ & $.744{\pm}.006$ & $.727{\pm}.006$ & $.728{\pm}.006$ \\
\cmidrule(lr){1-11}
112 & 0   & $.808{\pm}.002$ & $.724{\pm}.006$ & $.693{\pm}.004$ & $.757{\pm}.004$ & $.981{\pm}.000$ & $.872{\pm}.016$ & $.914{\pm}.002$ & $.901{\pm}.003$ & $.944{\pm}.002$ \\
    & 0.2 & $.772{\pm}.034$ & $.399{\pm}.060$ & $.332{\pm}.057$ & $.398{\pm}.064$ & $.919{\pm}.048$ & $.562{\pm}.035$ & $.896{\pm}.006$ & $.870{\pm}.004$ & $.928{\pm}.005$ \\
    & 0.4 & $.697{\pm}.032$ & $.243{\pm}.055$ & $.201{\pm}.048$ & $.224{\pm}.051$ & $.907{\pm}.053$ & $.273{\pm}.052$ & $.848{\pm}.008$ & $.827{\pm}.006$ & $.891{\pm}.002$ \\
\hline
\end{tabular}%
}
\end{center}

Under translated and rotated MNIST, SDB stays far above G-CNN, \texttt{select\_gbisp}, \texttt{tc}, and Harmonic as noise grows. SE(2) is strongest at larger $L$ across noise levels; at $L{=}28$ it is strong when noiseless but drops sharply as $\sigma$ grows. Among pretrained encoders, OpenCLIP and DINOv3 lead at low $\sigma$, while SDB is competitive at higher noise, especially at $L{=}56$.

\paragraph{Translation equivariance.} SDB is not translation-invariant. More precisely, under an in-disk support condition, it is translation-equivariant at the disk-harmonic and SDB level. This matters because the change under small admissible translations is structured and continuous in the translation, which explains the clustering in Experiment~(i).

Take the unit disk $D=\{x:|x|\le 1\}$ with orthonormal basis $\{\psi_{nk}\}$ and coefficients $a(f)=(a_{nk}(f))$, where $a_{nk}(f)=\langle f,\psi_{nk}\rangle$. Let the object be supported in a smaller disk $K_0=\{x:|x|\le\rho\}$ with $\rho<1$, and zero outside $K_0$. An admissible translation $\tau$ satisfies $|\tau|+\rho\le 1$, so the translated support $K_\tau$ stays inside $D$. Define the translated image by $(T_\tau f)(x)=f(x-\tau)$ on $K_\tau$ and $0$ elsewhere. On this class of images, both $a$ and $T_\tau$ are linear. Expanding $f$ in the basis and using linearity gives $a(T_\tau f)=U_\tau a(f)$, where $U_\tau$ is a fixed matrix depending on $\tau$ and $D$ (but not on $f$), with entries given by inner products of translated basis functions against the original basis. Each SDB coefficient is a fixed triple product of disk-harmonic coefficients, so applying SDB after translation is the same as applying the triple-product map to $U_\tau a(f)$. Thus SDB is equivariant under admissible translations, not invariant. Center rotations cancel in the triple products, yielding rotation invariance. As $\tau\to 0$, $U_\tau$ approaches the identity, so slightly translated SDBs stay near the centered SDB and Experiment~(i) separability remains $\gg 1$ at small $r$.

\section{Proofs for the Selective Disk Bispectrum Distribution Under Gaussian Noise}
\label{app:noise-proofs}

\begin{theorem}[SDB Expectation]\label{th:expectation-value}
Consider a noiseless signal $f$ with bispectrum $b$ and DH coefficients $a_{n,k}$. Upon corruption with Gaussian noise, $\tilde{f} = f + n$ for $n \in\mathcal{N}(0, \sigma^2)$. The expectation of $\tilde{b}$ is given by
\begin{equation}
\mathbb{E}[\tilde{b}_{00k}] = b_{00k} + \sigma^2 a^*_{0,k} + \delta_{k,1}2 \sigma^2 a_{0,1}; \quad \mathbb{E}[\tilde{b}_{201}] = b_{201} + \sigma^2 a_{0,1}; \quad \text{Else, } \mathbb{E}[\tilde{b}_{2nk}] = b_{2nk},
\end{equation}
which can be re-written as
\begin{equation}
\left\{
\begin{aligned}
    & \mathbb{E}[\tilde{b}_{00(k =1)}] = b_{001} + \sigma^2 a^*_{0,1} + 2 \sigma^2 a_{0,1} \\
    & \mathbb{E}[\tilde{b}_{00(k\neq1)}] = b_{00k} + \sigma^2 a^*_{0,k} \\
    & \mathbb{E}[\tilde{b}_{201}] = b_{201} + \sigma^2 a_{0,1} \\
    & \text{Else, no correction: } \mathbb{E}[\tilde{b}_{2nk}] = b_{2nk}.
\end{aligned}
\right.
\end{equation}
\end{theorem}

\begin{proof}
Consider a signal $f$ corrupted with Gaussian noise such that $\tilde{f} = f + n$, and $n \sim \mathcal{N}(0, \sigma^2)$. The DHCs of the noiseless signal $f$ are given by 
\begin{equation}
    a_{n_j k_j} = \int_0^{2\pi} \int_0^1 c_{nk} f(r,\theta) J_{n}(\lambda_{nk}r) e^{-i n \theta}r dr d\theta.
\end{equation}
The DHCs of the noisy signal $\tilde{f}$ are given by 
\begin{align}
    \tilde{a}_{n_j k_j} &= \int_0^{2\pi} \int_0^1 c_{nk} [f(r,\theta) + n(r, \theta)] J_{n}(\lambda_{nk}r) e^{-i n \theta}r dr d\theta \\
    & = a_{n_j, k_j} + N_{n_j, k_j}
\end{align}
where we denote $N_{n_j, k_j}$ as the DHC of the noise. The SDB coefficients are thus given by 
\begin{equation}
    \tilde{b}_{j_1 j_2 k_3} = (a_{n_{j_1}, k_{j_1}} + N_{n_{j_1}, k_{j_1}})(a_{n_{j_2}, k_{j_2}} + N_{n_{j_2}, k_{j_2}})(a^*_{n_{j_1}+ n_{j_2}, k_{3}} + N^*_{n_{j_1}+ n_{j_2}, k_{3}}).
\end{equation}
Upon expansion, we see a polynomial of the form
\begin{align}
    & \tilde{b}_{j_1 j_2 k_3} = a_{n_{j_1}, k_{j_1}} a_{n_{j_2}, k_{j_2}} a^*_{n_{j_1}+n_{j_2}, k_3}  \\
    & +\{2 \space a \text{ term}\}\{1 \space N\text{ term}\} + \{1 \space a \text{ term}\}\{2 \space N\text{ term}\} \\
    & + N_{n_{j_1}, k_{j_1}} N_{n_{j_2}, k_{j_2}} N^*_{n_{j_1}+n_{j_2}, k_3}.
\end{align}
Taking the expectation
\begin{align}
    & \mathbb{E}[\tilde{b}_{j_1 j_2 k_3}]   \\
    & = b_{j_1 j_2 k_3} + \mathbb{E}[N_{n_{j_1}, k_{j_1}} N_{n_{j_2}, k_{j_2}} N^*_{n_{j_1}+n_{j_2}, k_3}] \\
    & + \{2 \space a \text{ term}\} \mathbb{E}[\{1 \space N\text{ term}\}] + \{1 \space a \text{ term}\}\mathbb{E}[\{2 \space N\text{ term}\}] \\
& =  b_{j_1 j_2 k_3} + \{1 \space a \text{ term}\}\mathbb{E}[\{2 \space N\text{ term}\}]
\end{align}
where $b_{123}$ denotes the bispectrum of $f$. Prop. \ref{th:isserlis-dhc} proves that Isserlis' Theorem \ref{th:isserlis} applies to $\mathbb{E}[N_1, ..., N_n]$. Therefore, the odd-order $N$ moments disappear, leaving only even-order $N$ moments. 

Writing out such even moment terms, and using shorthand $\tilde{b}_{123} = \tilde{a}_{n_1, k_1}\tilde{a}_{n_2, k_2} \tilde{a}^*_{n_3, k_3}$ in place of $\tilde{b}_{j_1 j_2 k_3} = \tilde{a}_{n_{j_1}, k_{j_1}} \tilde{a}_{n_{j_2}, k_{j_2}} \tilde{a}^*_{n_{j_1}+n_{j_2}, k_3}$, we find

\begin{align}
    & \mathbb{E}[\tilde{b}_{123}]  = b_{123} + a^*_{n_3,k_3} \mathbb{E} [N_{n_1,k_1}N_{n_2,k_2}] \\
    & + a_{n_2,k_2} \mathbb{E} [N_{n_1,k_1}N^*_{n_3,k_3}] + a_{n_1,k_1} \mathbb{E} [N_{n_2,k_2}N^*_{n_3,k_3}].
\end{align}
Next, we substitute the second-order moment results of Th. \ref{th:second-moment}
\begin{align*}
        &\mathbb{E}[N_{nk} N_{n'k'}] = \sigma^2 \delta_{n,-n'} \delta_{k', k} \gamma_{n, k'}  \\
        &\mathbb{E}[N_{nk} N^*_{n'k'}] = \sigma^2   \delta_{k, k'}\delta_{n,n'},
    \end{align*}
where $\gamma_{nk'} \equiv (-1)^n \frac{|J_{1+n}(\lambda_{nk'})|}{|J_{1-n}(\lambda_{nk'})|} $, to find
 \begin{align}
    & \mathbb{E}[\tilde{b}_{123}]  = b_{123} \\
    & +\sigma^2 \{a^*_{n_3,k_3}  \delta_{n_1,-n_2} \delta_{k_2, k_1} \gamma_{n_1, k_2} \\
    &+ a_{n_2,k_2} \delta_{k_1, k_3}\delta_{n_1,n_3} + a_{n_1,k_1} \delta_{k_2, k_3}\delta_{n_2,n_3} \}.
\end{align}
Removing the shorthand,
\begin{align}
    & \mathbb{E}[\tilde{b}_{j_1 j_2 k_3}]   \\
    & = b_{j_1 j_2 k_3} +\sigma^2 \{a^*_{n_{j_1} + n_{j_2},k_3}  \delta_{n_{j_1},-n_{j_2}} \delta_{k_{j_2}, k_{j_1}} \gamma_{n_{j_1}, k_{j_2}} \\
    &+ a_{n_{j_2},k_{j_2}} \delta_{k_{j_1}, k_3}\delta_{n_{j_1},n_{j_1} + n_{j_2}} + a_{n_{j_1},k_{j_1}} \delta_{k_{j_2}, k_3}\delta_{n_{j_2},n_{j_1} + n_{j_2}} \} \\
  &  = b_{j_1 j_2 k_3} +\sigma^2 \{a^*_{0,k_3}  \delta_{n_{j_1},-n_{j_2}} \delta_{k_{j_2}, k_{j_1}} \gamma_{n_{j_1}, k_{j_2}} \\
    &+ a_{0,k_{j_2}} \delta_{k_{j_1}, k_3}\delta_{n_{j_2},0} + a_{0,k_{j_1}} \delta_{k_{j_2}, k_3}\delta_{n_{j_1},0} \}.
\end{align}
This is the expectation of the full disk bispectrum of a signal corrupted by zero-mean gaussian noise.

For the selective disk bispectrum, we consider 
\begin{equation}
\left\{
\begin{aligned}
    b_{0, 0, k} 
    &= a_{0, 1}^2 \cdot a_{0, k}^*, \\
    & \text{for } k \in \{1, \dots, K_{0}\},\\
    b_{2, n, k}  
    &= a_{1, 1} \cdot a_{n,1} \cdot a_{n+1,k}^*, \\
    & \text{for } n \in \{0, N_m-1\},~\text{and}~ k \in \{1, \dots, K_{n+1}\},
\end{aligned}
\right.
\end{equation}
\label{eq:selective_disk_bispectrum-app}
where the selective bispectrum coefficients are indexed by $j_1,n_2,k_3$ (limiting $j_1\in\{0,2\}$), as opposed to $j_1, j_2, k_3$.  This means we only consider cases where $n_{j_1} = n_{j_2} = 0, k_{j_1} = k_{j_2} = 1$ (first set of terms) and $n_{j_1} = 1, k_{j_1} = k_{j_2} = 1$ (second set of terms).

With this, we can simplify $\mathbb{E}[\tilde{b}_{j_1 j_2 k_3}]$ in the SDB case as

\begin{align}
& \mathbb{E}[\tilde{b}_{j_1 j_2 k_3}] \\
    &  = b_{j_1 j_2 k_3} +\sigma^2 \{a^*_{0,k_3}  \delta_{n_{j_1},0}\delta_{n_{j_2},0} \gamma_{0, 1} \\
    &+ a_{0,1} \delta_{1, k_3}\delta_{n_{j_2},0} + a_{0,1} \delta_{1, k_3}\delta_{n_{j_1},0} \},
\end{align}
noting $\gamma_{0, k} = 1$. Thus, bias terms only appear for the following terms
\begin{equation}
\left\{
\begin{aligned}
    & \mathbb{E}[\tilde{b}_{00(k =1)}] = b_{001} + \sigma^2 a^*_{0,1} + 2 \sigma^2 a_{0,1} \\
    & \mathbb{E}[\tilde{b}_{00(k\neq1)}] = b_{00k} + \sigma^2 a^*_{0,k} \\
    & \mathbb{E}[\tilde{b}_{201}] = b_{201} + \sigma^2 a_{0,1} \\
    & \text{Else, no correction: } \mathbb{E}[\tilde{b}_{2nk}] = b_{2nk}.
\end{aligned}
\right.
\end{equation}
\end{proof}

\begin{corollary}
The only SDB coefficients that are biased in estimation are the ones for which the bispectrum has zero-valued phase. 
\end{corollary}
\begin{proof}
    According to Th. \ref{th:expectation-value}, $\mathbb{E}[\tilde{b}]$ is biased for the following terms
\begin{equation}
\left\{
\begin{aligned}
    & \mathbb{E}[\tilde{b}_{00(k =1)}] = b_{001} + \sigma^2 a^*_{0,1} + 2 \sigma^2 a_{0,1} \\
    & \mathbb{E}[\tilde{b}_{00(k\neq1)}] = b_{00k} + \sigma^2 a^*_{0,k} \\
    & \mathbb{E}[\tilde{b}_{201}] = b_{201} + \sigma^2 a_{0,1} \\
    & \text{Else, no correction: } \mathbb{E}[\tilde{b}_{2nk}] = b_{2nk}.
\end{aligned}
\right.
\end{equation}
Recall that similarly to Fourier coefficients, disk harmonic coefficients can be written as $a_{n,k} = |a_{n,k}| e^{-i n \phi_{n,k}}$, so
\begin{align}
     b_{0, 0, k} &= a_{0, 1}^2 \cdot a_{0, k}^* \\
    &= |a_{0, 1}|^2e^{-i2\cdot0\phi_{0,1}}|a_{0, k}|e^{-i0\phi_{0,k}}\\
    & = |a_{0, 1}|^2|a_{0, k}| \\
    & \text{for } k \in \{1, \dots, K_{0}\} \\
\end{align}
\begin{align}
    b_{2, 0, 1}  & = a_{1, 1} \cdot a_{0,1} \cdot a_{1,1}^* \\
    &= |a_{1,1}|^2 a_{0,1}\\
    &= |a_{1,1}|^2 |a_{0,1}|e^{-i0\phi_{0,1}} \\
    &= |a_{1,1}|^2 |a_{0,1}|.
\end{align}
\end{proof}

\begin{theorem}[Isserlis' Theorem / Wick's Theorem]\label{th:isserlis}
If $(X_1, X_2, X_3,...X_n)$ is a real or complex zero-mean multivariate normal random vector, even-order moments (even $n$) are given by 
\begin{equation}
    \mathbb{E}[X_1, X_2, X_3,...X_n] = \mathbb{E}[\prod^{n=2m}_{i=1}X_i] = \sum_{p \in P^2_n} \prod_{\{i,j\}\in p} E[X_i X_j]
\end{equation}
where $P^2_n$ is permutation ``$n$ choose two'', so the sum is over all the pairings of $\{1,…,n\}$, i.e. all distinct ways of partitioning $\{1,…,n\}$ into pairs $\{i,j\}$, and the product is over the pairs contained in $p$. Odd order moments (odd $n$) are equal to zero.
\end{theorem}

\begin{proposition}[Isserlis Theorem Applies to zero-mean Gaussian noise in DH space]
\label{th:isserlis-dhc}
    Isserlis' Theorem \ref{th:isserlis} applies to $\mathbb{E}[N_{n_1, k_1}, ..., N_{n_j, k_j}]$, where $N_{n_j, k_j}$ is the DH transform of i.i.d. zero-mean Gaussian noise.
\end{proposition}
\begin{proof}
    Isserlis' Theorem applies to zero-mean multivariate normal vectors. A random vector is said to be \textit{k}-variate normally distributed if every linear combination of its \textit{k} components has a univariate normal distribution. Here, we will prove that the vector $(N_{n_1, k_1}, N_{n_2, k_2}, ..., N_{n_j, k_j})$ found in our second-, fourth- and sixth-order moment proofs is a zero-mean multivariate normal vector.
    
In line with all our proofs, consider a signal $f$ corrupted with i.i.d. Gaussian noise such that $\tilde{f} = f + n$, and $n \sim \mathcal{N}(0, \sigma^2)$. The DHCs of the noiseless signal $f$ are given by 
\begin{equation}
    a_{n_j k_j} = \int_0^{2\pi} \int_0^1 c_{nk} f(r,\theta) J_{n}(\lambda_{nk}r) e^{-i n \theta}r dr d\theta.
\end{equation}
The DHCs of the noisy signal $\tilde{f}$ are given by 
\begin{align}
    \tilde{a}_{n_j k_j} &= \int_0^{2\pi} \int_0^1 c_{nk} [f(r,\theta) + n(r, \theta)] J_{n}(\lambda_{nk}r) e^{-i n \theta}r dr d\theta \\
    & = a_{n_j, k_j} + N_{n_j, k_j}.
\end{align}
Recall $n$ has distribution $n \sim \mathcal{N}(0, \sigma^2)$, and the DH transform is a linear operation. Therefore, $ N_{n_j, k_j}$ will also be normally distributed with mean zero for all $j$. The linear combination of such random variables will be a univariate zero-mean normal distribution. Therefore, $(N_{n_1, k_1}, N_{n_2, k_2}, ..., N_{n_j, k_j})$, is a zero-mean multivariate normal vector for all $j$.
\end{proof}

\begin{theorem}[Second Order Moments of Noise in DHC Space]\label{th:second-moment}

The second order moments found in the bispectrum expectation and variance proofs are given by
    \begin{align*}
        &\mathbb{E}[N_{nk} N_{n'k'}] = \mathbb{E}[N^*_{nk} N^*_{n'k'}] = \sigma^2 \delta_{n,-n'} \delta_{k', k} \gamma_{n, k'}  \\
        &\mathbb{E}[N_{nk} N^*_{n'k'}] = \mathbb{E}[N^*_{nk} N_{n'k'}] = \sigma^2   \delta_{k, k'}\delta_{n,n'}
    \end{align*}
where $\gamma_{nk'} \equiv (-1)^n \frac{|J_{1+n}(\lambda_{nk'})|}{|J_{1-n}(\lambda_{nk'})|} $.
\end{theorem}

\begin{proof}
    \begin{align*}
    &\mathbb{E}[N_{nk} N_{n'k'}] \\
    &= \mathbb{E}[\int_0^{2\pi} \int_0^1 c_{nk} n(r_1, \theta_1) J_{n}(\lambda_{nk}r_1) e^{-i n \theta_1}r_1 dr_1 d\theta_1 \\
    &\times \int_0^{2\pi} \int_0^1 c_{n'k'} n(r_2, \theta_2) J_{n'}(\lambda_{n'k'}r_2) e^{-i n' \theta_2}r_2 dr_2 d\theta_2] \\
    &=  \int_0^{2\pi} \int_0^1\int_0^{2\pi} \int_0^1c_{nk}c_{n'k'}\mathbb{E}[n(r_1, \theta_1)n(r_2, \theta_2)] \\
    &\times J_{n}(\lambda_{nk}r_1) e^{-i n \theta_1} J_{n'}(\lambda_{n'k'}r_2) e^{-i n' \theta_2} r_1 r_2 dr_1 dr_2 d\theta_1 d\theta_2\\
    \end{align*}

    Noise at one point is independent from noise at another point, so

    \begin{align*}
    \mathbb{E}[n(r_1, \theta_1)n(r_2, \theta_2)] =
    \begin{cases}
        \sigma^2 & \text{if } (r_1,\theta_1) = (r_2,\theta_2)\\
        \mathbb{E}[n(r_1, \theta_1)]\mathbb{E}[n(r_2, \theta_2)] =0 & \text{otherwise}.
    \end{cases}
    \end{align*}
    
    Thus, $\mathbb{E}[n(r_1, \theta_1)n(r_2, \theta_2)] = \sigma^2 \frac{1}{r_1} \delta(r_1 - r_2) \delta(\theta_1 - \theta_2)$, yielding
    
    \begin{align*}
    &\mathbb{E}[N_{nk} N_{n'k'}] \\
    &=  \int_0^{2\pi} \int_0^1 \int_0^{2\pi} \int_0^1 c_{nk}c_{n'k'}\sigma^2\delta(r_1 - r_2) \delta(\theta_1 - \theta_2)\\
    & \times e^{-i (n \theta_1 + n' \theta_2)} J_{n}(\lambda_{nk}r_1) J_{n'}(\lambda_{n'k'}r_2) \\
    &\times    r_1 r_2 dr_1 dr_2 d\theta_1 d\theta_2\\
    \end{align*}
    
    where $\delta$ is the dirac delta function. Integrating over $r_2, \theta_2$ and setting $r_1 \to r, \theta_1 \to \theta$,
    \begin{align*}
    \mathbb{E}[N_{nk} N_{n'k'}]
    &=  c_{nk}c_{n'k'}\sigma^2 \int_0^{2\pi}  e^{-i (n + n') \theta)} d\theta \\
    &\times  \int_0^1J_{n}(\lambda_{nk}r) J_{n'}(\lambda_{n'k'}r) \space r dr\\
    \end{align*}
    Note that 
    \[
    \int_0^{2\pi}  e^{-i (n + n') \theta} d\theta= 
    \begin{cases}
        2\pi & \text{if } n+n'=0 \\
        0 & \text{otherwise}
    \end{cases}
    \]
since the Fourier transform of a scalar is only nonzero at the DC component. Thus, the ``frequency'' $n+n'$ must equal zero for this Fourier transform integral to be nonzero. Using the shorthand $\delta_{n, -n'}$ for $\delta(n + n')$,
    
    \begin{align*}
    & \mathbb{E}[N_{nk} N_{n'k'}] \\
    &=  c_{nk}c_{n'k'}\sigma^2 \delta_{n,-n'} (2 \pi) \int_0^1J_{n}(\lambda_{nk}r) J_{n'}(\lambda_{n'k'}r) \space r dr\\
    &=  c_{nk}c_{-nk'}\sigma^2  \delta_{n,-n'} (2 \pi)\\
    & \quad \times \int_0^1J_{n}(\lambda_{n,k}r) J_{-n}(\lambda_{-n,k'}r) \space r dr\\
    \end{align*}

Next, we can utilize a common orthogonality relation \cite{bendory2018, Verrall_Kakarala_1998}
    
    \[
    (2 \pi)\int_0^1 c_{nk'}J_{n}(\lambda_{n,k}r) J_{n}(\lambda_{n,k'}r) \space r dr = \frac{1}{c_{nk}} \delta_{k, k'}.
    \]

Specifically, by recalling that $J_{-n}(r) = (-1)^nJ_{n}(r)$, and $\lambda_{nk} = \lambda_{-nk}$ since $J_{-n}$ has the exact same roots as $J_{n}$, we can write
    
    \begin{align*}
    \mathbb{E}[N_{nk} N_{n'k'}]
    &=  c_{-nk'}\sigma^2  \delta_{n,-n'}  \\
    & \quad \times  (-1)^n (2 \pi)\int_0^1 c_{nk}J_{n}(\lambda_{n,k}r) J_{n}(\lambda_{n,k'}r) \space r dr\\
    &= (-1)^n \sigma^2 \frac{c_{-nk'}}{c_{nk'}} \delta_{n,-n'} \delta_{k', k}    \\
    &= (-1)^n \sigma^2 \frac{|J_{1+n}(\lambda_{nk'})|}{|J_{1-n}(\lambda_{nk'})|}  \delta_{n,-n'} \delta_{k', k}  
    \end{align*}
    where the last equality comes from the fact that 
    \[
    c_{nk} = \frac{1}{\sqrt{\pi}|J_{n+1}(\lambda_{nk})|}
    \]
    from Eq. 2.2 in \cite{bendory2018}, along with the fact that $\lambda_{n,k} = \lambda_{-n,k}$ as mentioned previously.

    Finally, naming $\gamma_{nk'} \equiv (-1)^n \frac{|J_{1+n}(\lambda_{nk'})|}{|J_{1-n}(\lambda_{nk'})|} $, we have

    \begin{equation}
        \mathbb{E}[N_{nk} N_{n'k'}] = \sigma^2 \delta_{n,-n'} \delta_{k', k} \gamma_{n, k'}
    \end{equation}
    
    The proof for $\mathbb{E}[N_{nk} N^*_{n'k'}]$ is nearly identical, so we will outline it in less detail, only highlighting key points of difference.
    \begin{align*}
    & \mathbb{E}[N_{nk} N^*_{n'k'}] \\
    &= \mathbb{E}[\int_0^{2\pi} \int_0^1 c_{nk} n(r_1, \theta_1) J_{n}(\lambda_{nk}r_1) e^{-i n \theta_1}r_1 dr_1 d\theta_1 \\
    &\times \int_0^{2\pi} \int_0^1 c_{n'k'} n(r_2, \theta_2) J_{n'}(\lambda_{n'k'}r_2) e^{i n' \theta_2}r_2 dr_2 d\theta_2] \\
    &=c_{nk}c_{n'k'}\sigma^2 \int_0^{2\pi}  e^{-i (n - n') \theta)} d\theta \\
    &\times  \int_0^1J_{n}(\lambda_{nk}r) J_{n'}(\lambda_{n'k'}r) \space r dr\\
    &=  c_{nk'}\sigma^2  \delta_{n,n'} (2 \pi)\int_0^1 c_{nk}J_{n}(\lambda_{n,k}r) J_{n}(\lambda_{n,k'}r) \space r dr\\
    &= c_{nk'}\sigma^2  \delta_{n,n'} \frac{1}{c_{nk}} \delta_{k, k'} \\
    &= \sigma^2  \delta_{n,n'}  \delta_{k, k'}.
    \end{align*}

    Finally, we see that $\mathbb{E}[N_{nk} N_{n'k'}] = \mathbb{E}[N^*_{nk} N^*_{n'k'}]$ and $\mathbb{E}[N_{nk} N^*_{n'k'}] = \mathbb{E}[N^*_{nk} N_{n'k'}]$
    \begin{align}
        \mathbb{E}[N^*_{nk} N^*_{n'k'}] & = \mathbb{E}[N_{nk} N_{n'k'}]^* \\ 
        &= \{\sigma^2 \delta_{n,-n'} \delta_{k', k} \gamma_{n, k'} \}^*  \\
        & = \sigma^2 \delta_{n,-n'} \delta_{k', k} \gamma_{n, k'} \\
        & =  \mathbb{E}[N_{nk} N_{n'k'}] 
    \end{align}
    and
    \begin{align}
        \mathbb{E}[N^*_{nk} N_{n'k'}] & = \mathbb{E}[N_{nk} N^*_{n'k'}]^* \\ 
        &= \{\sigma^2   \delta_{k, k'}\delta_{n,n'} \}^*  \\
        & = \sigma^2   \delta_{k, k'}\delta_{n,n'} \\
        & =  \mathbb{E}[N_{nk} N^*_{n'k'}].
    \end{align}
\end{proof}

\begin{proposition}[Fourth Order Moments in Bispectrum Expectation and Variance]
\label{th:fourth-moment}
Selected fourth order moments, seen in the variance proofs, are given by
\begin{align}
    & \mathbb{E}[N_1 N^*_1 N_2 N^*_2] = \mathbb{E}[|N_1|^2 |N_2|^2] =   \mathbb{E}[|N_1|^2] \mathbb{E}[|N_2|^2] + \mathbb{E}[N^*_1 N_2] ^2 + \mathbb{E}[N_1 N_2] ^2 \\
        & \mathbb{E}[N_1 N^*_1 N_2 N_3] =\mathbb{E}[|N_1|^2 N_2 N_3] \\
    & \quad= \mathbb{E}[|N_1|^2] \mathbb{E}[N_2 N_3] + \mathbb{E}[N_1 N_3] \mathbb{E}[N^*_1 N_2]  + \mathbb{E}[N_1 N_2] \mathbb{E}[N^*_1 N_3] \\
    & \\
\end{align}
where in the second example, $N_2$ or $N_3$ can be conjugated, and the statement will still hold.
\end{proposition}
\begin{proof}

    Prop. \ref{th:isserlis-dhc} proves that Isserlis' Theorem \ref{th:isserlis} applies to $\mathbb{E}[N_1, ..., N_n]$. Therefore,
    \begin{align}
        \mathbb{E}[N_1 N_2 N_3 N_4] & = \mathbb{E}[N_1 N_2]\mathbb{E}[N_3 N_4] \\
        & + \mathbb{E}[N_1 N_3]\mathbb{E}[N_2 N_4] \\
        & + \mathbb{E}[N_1 N_4]\mathbb{E}[N_2 N_3].
    \end{align}
Via direct substitution,
\begin{align}
         & \mathbb{E}[|N_1|^2 N_2 N_3] = \mathbb{E}[|N_1|^2] \mathbb{E}[N_2 N_3] + \mathbb{E}[N_1 N_3]\mathbb{E}[N^*_1 N_2]  + \mathbb{E}[N_1 N_2] \mathbb{E}[N^*_1 N_3] \\
    \end{align}
    and
    \begin{align}
            \mathbb{E}[|N_1|^2 |N_2|^2] &  =   \mathbb{E}[|N_1|^2] \mathbb{E}[|N_2|^2] + \mathbb{E}[N_1 N^*_2] \mathbb{E}[N^*_1 N_2] + \mathbb{E}[N_1 N_2] \mathbb{E}[N^*_1 N^*_2]\\
            & =   \mathbb{E}[|N_1|^2] \mathbb{E}[|N_2|^2] + \mathbb{E}[N^*_1 N_2] ^2 + \mathbb{E}[N_1 N_2] ^2,
    \end{align}
    where the last equivalence relies on the complex conjugate relations in Th. \ref{th:second-moment}.
\end{proof}

\begin{proposition}[Sixth Order Moments in Bispectrum expectation]
\label{th:sixth-moment}
    The sixth order moments found in the variance proof is given by 
    \begin{align}
        \mathbb{E}[|N_1|^2 &|N_2|^2 |N_3|^2] = \mathbb{E}[|N_1|^2 ] \mathbb{E}[|N_2|^2 ] \mathbb{E}[|N_3|^2 ] \\
        \\
        & + \mathbb{E}[N_1 N_2]  \\
        & \times \{ \mathbb{E}[N_1 N_2] \mathbb{E}[|N_3|^2]  + 2 \mathbb{E}[N^*_1 N_3] \mathbb{E}[N_2 N_3] + 2\mathbb{E}[N_1 N_3] \mathbb{E}[N^*_2 N_3] \} \\
        & \\
        & + \mathbb{E}[N_1 N_3]  \\
        & \times \{  \mathbb{E}[N_1 N_3] \mathbb{E}[|N_2|^2]  + 2\mathbb{E}[N^*_1 N_2] \mathbb{E}[N_2 N_3]  \} \\
        & \\
        & + \mathbb{E}[N_2 N_3]  \\
        & \times \{ \mathbb{E}[|N_1|^2 ] \mathbb{E}[N_2 N_3] \} \\
        & \\
        & + \mathbb{E}[N^*_2 N_3]  \\
        & \times \{   \mathbb{E}[|N_1|^2 ] \mathbb{E}[N^*_2 N_3]  + 2\mathbb{E}[N^*_1 N_2] \mathbb{E}[N^*_1 N_3]\} \\
        & \\
        & + \mathbb{E}[N^*_1 N_3]  \\
        & \times \{  \mathbb{E}[N^*_1 N_3] \mathbb{E}[|N_2|^2] \} \\
        & \\
        & +  \mathbb{E}[N^*_1 N_2]  \\
        & \times \{ \mathbb{E}[N^*_1 N_2] \mathbb{E}[|N_3|^2] \} \\
    \end{align}
\end{proposition}

\begin{proof}

    Prop. \ref{th:isserlis-dhc} proves that Isserlis' Theorem \ref{th:isserlis} applies to $\mathbb{E}[N_1, ..., N_n]$. Therefore,
    \begin{align}
        & \mathbb{E}[N_1 N_2 N_3 N_4 N_5 N_6]  \\
        & = \mathbb{E}[N_1 N_2]\mathbb{E}[N_3 N_4] \mathbb{E}[N_5 N_6] + \mathbb{E}[N_1 N_2]\mathbb{E}[N_3 N_5] \mathbb{E}[N_4 N_6]  + \mathbb{E}[N_1 N_2]\mathbb{E}[N_3 N_6] \mathbb{E}[N_4 N_5] \\
        & + \mathbb{E}[N_1 N_3]\mathbb{E}[N_2 N_4] \mathbb{E}[N_5 N_6]  + \mathbb{E}[N_1 N_3]\mathbb{E}[N_2 N_5] \mathbb{E}[N_4 N_6]  + \mathbb{E}[N_1 N_3]\mathbb{E}[N_2 N_6] \mathbb{E}[N_4 N_5] \\
        & + \mathbb{E}[N_1 N_4]\mathbb{E}[N_2 N_3] \mathbb{E}[N_5 N_6]  + \mathbb{E}[N_1 N_4]\mathbb{E}[N_2 N_5] \mathbb{E}[N_3 N_6] + \mathbb{E}[N_1 N_4]\mathbb{E}[N_2 N_6] \mathbb{E}[N_3 N_5]  \\
        & + \mathbb{E}[N_1 N_5]\mathbb{E}[N_2 N_3] \mathbb{E}[N_4 N_6]  + \mathbb{E}[N_1 N_5]\mathbb{E}[N_2 N_4] \mathbb{E}[N_3 N_6] + \mathbb{E}[N_1 N_5]\mathbb{E}[N_2 N_6] \mathbb{E}[N_3 N_4]  \\
        & + \mathbb{E}[N_1 N_6]\mathbb{E}[N_2 N_3] \mathbb{E}[N_4 N_5]  + \mathbb{E}[N_1 N_6]\mathbb{E}[N_2 N_4] \mathbb{E}[N_3 N_5]  + \mathbb{E}[N_1 N_6]\mathbb{E}[N_2 N_5] \mathbb{E}[N_3 N_4] \\
    \end{align}

In the variance proof we encounter a sixth order moment terms of the form $\mathbb{E}[N_1 N^*_1 N_2 N^*_2 N_3 N^*_3] = \mathbb{E}[|N_1|^2 |N_2|^2 |N_3|^2]$. We see from substitution that

\begin{align}
        \mathbb{E}[|N_1|^2 &|N_2|^2 |N_3|^2] \\
        & = \mathbb{E}[N_1 N^*_1]\mathbb{E}[N_2 N^*_2] \mathbb{E}[N_3 N^*_3] \\
        & + \mathbb{E}[N_1 N^*_1]\mathbb{E}[N_2 N_3] \mathbb{E}[N^*_2 N^*_3] \\
        & + \mathbb{E}[N_1 N^*_1]\mathbb{E}[N_2 N^*_3] \mathbb{E}[N^*_2 N_3] \\
        & \\
        & + \mathbb{E}[N_1 N_2]\mathbb{E}[N^*_1 N^*_2] \mathbb{E}[N_3 N^*_3] \\
        & + \mathbb{E}[N_1 N_2]\mathbb{E}[N^*_1 N_3] \mathbb{E}[N^*_2 N^*_3] \\
        & + \mathbb{E}[N_1 N_2]\mathbb{E}[N^*_1 N^*_3] \mathbb{E}[N^*_2 N_3] \\
        & \\
        & + \mathbb{E}[N_1 N^*_2]\mathbb{E}[N^*_1 N_2] \mathbb{E}[N_3 N^*_3] \\
        & + \mathbb{E}[N_1 N^*_2]\mathbb{E}[N^*_1 N_3] \mathbb{E}[N_2 N^*_3] \\
        & + \mathbb{E}[N_1 N^*_2]\mathbb{E}[N^*_1 N^*_3] \mathbb{E}[N_2 N_3] \\
        & \\
        & + \mathbb{E}[N_1 N_3]\mathbb{E}[N^*_1 N_2] \mathbb{E}[N^*_2 N^*_3] \\
        & + \mathbb{E}[N_1 N_3]\mathbb{E}[N^*_1 N^*_2] \mathbb{E}[N_2 N^*_3] \\
        & + \mathbb{E}[N_1 N_3]\mathbb{E}[N^*_1 N^*_3] \mathbb{E}[N_2 N^*_2] \\
        & \\
        & + \mathbb{E}[N_1 N^*_3]\mathbb{E}[N^*_1 N_2] \mathbb{E}[N^*_2 N_3] \\
        & + \mathbb{E}[N_1 N^*_3]\mathbb{E}[N^*_1 N^*_2] \mathbb{E}[N_2 N_3] \\
        & + \mathbb{E}[N_1 N^*_3]\mathbb{E}[N^*_1 N_3] \mathbb{E}[N_2 N^*_2]. \\
    \end{align}

First, we sort in an order that will be advantageous for future use, as

\begin{align}
        \mathbb{E}[|N_1|^2 &|N_2|^2 |N_3|^2]  = \mathbb{E}[|N_1|^2 ] \mathbb{E}[|N_2|^2 ] \mathbb{E}[|N_3|^2 ] \\
        \\
        & + \mathbb{E}[N_1 N_2]  \\
        & \times \{ \mathbb{E}[N^*_1 N^*_2] \mathbb{E}[|N_3|^2]  + \mathbb{E}[N^*_1 N_3] \mathbb{E}[N^*_2 N^*_3] + \mathbb{E}[N^*_1 N^*_3] \mathbb{E}[N^*_2 N_3] \} \\
        & \\
        & + \mathbb{E}[N^*_1 N^*_2]  \\
        & \times \{  \mathbb{E}[N_1 N_3] \mathbb{E}[N_2 N^*_3] + \mathbb{E}[N_1 N^*_3] \mathbb{E}[N_2 N_3]  \} \\
        & \\
        & + \mathbb{E}[N_1 N_3]  \\
        & \times \{ \mathbb{E}[N^*_1 N_2] \mathbb{E}[N^*_2 N^*_3] + \mathbb{E}[N^*_1 N^*_3] \mathbb{E}[|N_2|^2]   \} \\
        & \\
        & + \mathbb{E}[N^*_1 N^*_3]  \\
        & \times \{  \mathbb{E}[N_1 N^*_2] \mathbb{E}[N_2 N_3]  \} \\
        & \\
        & + \mathbb{E}[N_2 N_3]  \\
        & \times \{ \mathbb{E}[|N_1|^2 ] \mathbb{E}[N^*_2 N^*_3] \} \\
        & \\
        & + \mathbb{E}[N^*_2 N_3]  \\
        & \times \{   \mathbb{E}[|N_1|^2 ] \mathbb{E}[N_2 N^*_3] + \mathbb{E}[N_1 N^*_3]\mathbb{E}[N^*_1 N_2]\} \\
        & \\
        & + \mathbb{E}[N_2 N^*_3]  \\
        & \times \{  \mathbb{E}[N_1 N^*_2] \mathbb{E}[N^*_1 N_3]\} \\
        & \\
        & +  \mathbb{E}[N_1 N^*_2]  \\
        & \times \{ \mathbb{E}[N^*_1 N_2] \mathbb{E}[|N_3|^2] \} \\
        & \\
        & + \mathbb{E}[N_1 N^*_3]  \\
        & \times \{  \mathbb{E}[N^*_1 N_3] \mathbb{E}[|N_2|^2] \} \\
    \end{align}

Next, recall the complex conjugate relations in Th. \ref{th:second-moment}, which say that 
\begin{align*}
        &\mathbb{E}[N_{nk} N_{n'k'}] = \mathbb{E}[N^*_{nk} N^*_{n'k'}] = \sigma^2 \delta_{n,-n'} \delta_{k', k} \gamma_{n, k'}  \\
        &\mathbb{E}[N_{nk} N^*_{n'k'}] = \mathbb{E}[N^*_{nk} N_{n'k'}] = \sigma^2   \delta_{k, k'}\delta_{n,n'}.
    \end{align*}

Finally, we substitute these relations, standardizing to $\mathbb{E}[N_{nk} N_{n'k'}]$ and $\mathbb{E}[N^*_{nk} N_{n'k'}]$ format, and condensing as
\begin{align}
        \mathbb{E}[|N_1|^2 &|N_2|^2 |N_3|^2] = \mathbb{E}[|N_1|^2 ] \mathbb{E}[|N_2|^2 ] \mathbb{E}[|N_3|^2 ] \\
        \\
        & + \mathbb{E}[N_1 N_2]  \\
        & \times \{ \mathbb{E}[N_1 N_2] \mathbb{E}[|N_3|^2]  + 2 \mathbb{E}[N^*_1 N_3] \mathbb{E}[N_2 N_3] + 2\mathbb{E}[N_1 N_3] \mathbb{E}[N^*_2 N_3] \} \\
        & \\
        & + \mathbb{E}[N_1 N_3]  \\
        & \times \{  \mathbb{E}[N_1 N_3] \mathbb{E}[|N_2|^2]  + 2\mathbb{E}[N^*_1 N_2] \mathbb{E}[N_2 N_3]  \} \\
        & \\
        & + \mathbb{E}[N_2 N_3]  \\
        & \times \{ \mathbb{E}[|N_1|^2 ] \mathbb{E}[N_2 N_3] \} \\
        & \\
        & + \mathbb{E}[N^*_2 N_3]  \\
        & \times \{   \mathbb{E}[|N_1|^2 ] \mathbb{E}[N^*_2 N_3]  + 2\mathbb{E}[N^*_1 N_2] \mathbb{E}[N^*_1 N_3]\} \\
        & \\
        & + \mathbb{E}[N^*_1 N_3]  \\
        & \times \{  \mathbb{E}[N^*_1 N_3] \mathbb{E}[|N_2|^2] \} \\
        & \\
        & +  \mathbb{E}[N^*_1 N_2]  \\
        & \times \{ \mathbb{E}[N^*_1 N_2] \mathbb{E}[|N_3|^2] \} \\
    \end{align}

\end{proof}

\begin{theorem}[Variance]\label{th:variance-app}
Let $a_{n,k}$ be the DH coefficients of $f$ and $\tilde b$ the SDB of $\tilde f = f + \eta$, $\eta \sim \mathcal{N}(0,\sigma^{2})$. The coefficients $a_{0,k}$ are real. For a generic coefficient $b_{2,n,k} = a_{1,1}\,a_{n,1}\,a^*_{n+1,k}$, meaning $n \ge 2$, or $n = 0$ and $k \ge 2$,
\begin{equation}
  \begin{aligned}
  \mathrm{Var}(\tilde b_{2,n,k})
  &= \sigma^{2}\big(\lvert a_{1,1}\rvert^{2}\lvert a_{n,1}\rvert^{2} + \lvert a_{1,1}\rvert^{2}\lvert a_{n+1,k}\rvert^{2} + \lvert a_{n,1}\rvert^{2}\lvert a_{n+1,k}\rvert^{2}\big) \\
  &\quad + \sigma^{4}\big(\lvert a_{1,1}\rvert^{2} + \lvert a_{n,1}\rvert^{2} + \lvert a_{n+1,k}\rvert^{2}\big) + \sigma^{6}.
  \end{aligned}
  \label{eq:variance-generic}
\end{equation}
The remaining coefficients, in which a DH coefficient appears twice, have
\begin{align}
  \mathrm{Var}(\tilde b_{001}) &= 9\sigma^{2}a_{0,1}^{4} + 36\sigma^{4}a_{0,1}^{2} + 15\sigma^{6}, \label{eq:variance-001}\\
  \mathrm{Var}(\tilde b_{00k}) &= \sigma^{2}\big(a_{0,1}^{4} + 4a_{0,1}^{2}a_{0,k}^{2}\big) + \sigma^{4}\big(6a_{0,1}^{2} + 2a_{0,k}^{2}\big) + 3\sigma^{6}, \qquad k \ge 2, \label{eq:variance-00k}\\
  \mathrm{Var}(\tilde b_{201}) &= \sigma^{2}\big(\lvert a_{1,1}\rvert^{4} + 2a_{0,1}^{2}\lvert a_{1,1}\rvert^{2}\big) + \sigma^{4}\big(4\lvert a_{1,1}\rvert^{2} + a_{0,1}^{2}\big) + 2\sigma^{6}, \label{eq:variance-201}\\
  \mathrm{Var}(\tilde b_{21k}) &= \sigma^{2}\big(\lvert a_{1,1}\rvert^{4} + 4\lvert a_{1,1}\rvert^{2}\lvert a_{2,k}\rvert^{2}\big) + \sigma^{4}\big(4\lvert a_{1,1}\rvert^{2} + 2\lvert a_{2,k}\rvert^{2}\big) + 2\sigma^{6}. \label{eq:variance-21k}
\end{align}
\end{theorem}

\begin{proof}
As in Th. \ref{th:expectation-value}, we use $\tilde{b}_{123} = \tilde{a}_1 \tilde{a}_2 \tilde{a}^*_3$ to denote $\tilde{b}_{j_1,j_2, k_3} = \tilde{a}_{n_{j_1} k_{j_1}} \tilde{a}_{n_{j_2} k_{j_2}} \tilde{a}^*_{n_{j_1} + n_{j_2} k_3}$. Here, we compute the variance of the bispectrum of a signal $f$, corrupted with Gaussian noise $\eta \sim \mathcal{N}(0, \sigma^2)$. We denote the bispectrum of the true signal as $b_{123} = a_1 a_2 a^*_3$. 

The variance of the complex random variable $\tilde{b}_{123}$ is given by

\begin{equation}
    \mathrm{Var}(\tilde{b}_{123}) = \mathbb{E}[\tilde{b}_{123}\tilde{b}^*_{123}] - \mathbb{E}[\tilde{b}_{123}]\mathbb{E}[\tilde{b}^*_{123}].
\end{equation}
We start with the second term. As shown in Th. \ref{th:expectation-value}, 
\begin{align}
    & \mathbb{E}[\tilde{b}_{123}]  = a_1 a_2 a^*_3 + a^*_3 \mathbb{E} [N_1N_2] \\
    & + a_2 \mathbb{E} [N_1 N^*_3] + a_1 \mathbb{E} [N_2 N^*_3].
\end{align}
Similarly, 
\begin{align}
    & \mathbb{E}[\tilde{b}^*_{123}]  = a^*_1 a^*_2 a_3 + a_3 \mathbb{E} [N^*_1 N^*_2] \\
    & + a^*_2 \mathbb{E} [N^*_1 N_3] + a^*_1 \mathbb{E} [N^*_2 N_3].
\end{align}
Thus, 
\begin{align}
    \mathbb{E}[\tilde{b}_{123}]&\mathbb{E}[\tilde{b}^*_{123}] \\
    &= |a_1|^2 |a_2|^2 |a_3|^2 + a_1 a_2 |a_3|^2 \mathbb{E} [N^*_1 N^*_2] \\
    &+ a_1 |a_2|^2 a^*_3 \mathbb{E} [N^*_1 N_3] + |a_1|^2 a_2 a^*_3 \mathbb{E} [N^*_2 N_3] \\
    & + a^*_1 a^*_2 |a_3|^2 \mathbb{E} [N_1N_2] + |a_3|^2 \mathbb{E}[N^*_1 N^*_2] \mathbb{E} [N_1N_2] \\
    &+ a^*_3 a^*_2 \mathbb{E} [N_1N_2] \mathbb{E}[N^*_1 N_3] + a^*_3a^*_1\mathbb{E} [N_1N_2] \mathbb{E} [N^*_2 N_3] \\
    & + a^*_1 |a_2|^2 a_3 \mathbb{E} [N_1 N^*_3]+ a_2 a_3 \mathbb{E} [N_1 N^*_3] \mathbb{E} [N^*_1 N^*_2] \\
    &+ |a_2|^2 \mathbb{E} [N_1 N^*_3] \mathbb{E} [N^*_1 N_3] + a_2 a^*_1 \mathbb{E} [N_1 N^*_3] \mathbb{E} [N^*_2 N_3] \\
    & + |a_1|^2 a^*_2 a_3 \mathbb{E} [N_2 N^*_3] + a_1 a_3 \mathbb{E} [N_2 N^*_3] \mathbb{E} [N^*_1 N^*_2]  \\
    &+ a_1 a^*_2 \mathbb{E} [N_2 N^*_3]  \mathbb{E} [N^*_1 N_3] + |a_1|^2 \mathbb{E} [N_2 N^*_3] \mathbb{E} [N^*_2 N_3].
\end{align}
Recall the complex conjugate relations in Th. \ref{th:second-moment}, which say that 
\begin{align*}
        &\mathbb{E}[N_{nk} N_{n'k'}] = \mathbb{E}[N^*_{nk} N^*_{n'k'}] = \sigma^2 \delta_{n,-n'} \delta_{k', k} \gamma_{n, k'}  \\
        &\mathbb{E}[N_{nk} N^*_{n'k'}] = \mathbb{E}[N^*_{nk} N_{n'k'}] = \sigma^2   \delta_{k, k'}\delta_{n,n'}.
    \end{align*}
We substitute these and standardize to $\mathbb{E}[N_{nk} N_{n'k'}]$ and $\mathbb{E}[N^*_{nk} N_{n'k'}]$, resulting in 
\begin{align}
    &\mathbb{E}[\tilde{b}_{123}]\mathbb{E}[\tilde{b}^*_{123}] \\
    &= |a_1|^2 |a_2|^2 |a_3|^2 + a_1 a_2 |a_3|^2 \mathbb{E} [N_1 N_2] \\
    &+ a_1 |a_2|^2 a^*_3 \mathbb{E} [N^*_1 N_3] + |a_1|^2 a_2 a^*_3 \mathbb{E} [N^*_2 N_3] \\
    & + a^*_1 a^*_2 |a_3|^2 \mathbb{E} [N_1N_2] + |a_3|^2 \mathbb{E}[N_1 N_2]^2  \\
    &+ a^*_3 a^*_2 \mathbb{E} [N_1N_2] \mathbb{E}[N^*_1 N_3] + a^*_3a^*_1\mathbb{E} [N_1N_2] \mathbb{E} [N^*_2 N_3] \\
    & + a^*_1 |a_2|^2 a_3 \mathbb{E} [N^*_1 N_3]+ a_2 a_3 \mathbb{E} [N^*_1 N_3] \mathbb{E} [N_1 N_2] \\
    &+ |a_2|^2 \mathbb{E} [N^*_1 N_3]^2 + a_2 a^*_1 \mathbb{E} [N^*_1 N_3] \mathbb{E} [N^*_2 N_3] \\
    & + |a_1|^2 a^*_2 a_3 \mathbb{E} [N^*_2 N_3] + a_1 a_3 \mathbb{E} [N^*_2 N_3] \mathbb{E} [N_1 N_2]  \\
    &+ a_1 a^*_2 \mathbb{E} [N^*_2 N_3]  \mathbb{E} [N^*_1 N_3] + |a_1|^2 \mathbb{E} [N^*_2 N_3]^2.
\end{align}

 Next, we focus on the $\mathbb{E}[\tilde{b}_{123}\tilde{b}^*_{123}]$ term, starting with
 \begin{align}
      & \tilde{b}_{123}\tilde{b}^*_{123} = \tilde{a}_1 \tilde{a}_2 \tilde{a}^*_3 \tilde{a}^*_1 \tilde{a}^*_2 \tilde{a}_3 \\
     & = (a_1 + N_1)(a^*_1 + N^*_1)(a_2 + N_2)(a^*_2 + N^*_2)(a_3 + N_3)(a^*_3 + N^*_3) \\
     & = (|a_1|^2 + a_1 N^*_1 + a^*_1 N_1 +|N_1|^2) \\
     &\quad \times (|a_2|^2 + a_2 N^*_2 + a^*_2 N_2 +|N_2|^2) \\
     &\quad \times (|a_3|^2 + a_3 N^*_3 + a^*_3 N_3 +|N_3|^2). \\
 \end{align}
 Defining $\xi_i = a_i N^*_i + a^*_i N_i$, we can rewrite as
 \begin{align}
      & \tilde{b}_{123}\tilde{b}^*_{123} = \tilde{a}_1 \tilde{a}_2 \tilde{a}^*_3 \tilde{a}^*_1 \tilde{a}^*_2 \tilde{a}_3 \\
     & = (|a_1|^2 + \xi_1 +|N_1|^2) \\
     &\quad \times (|a_2|^2 + \xi_2 +|N_2|^2) \\
     &\quad \times (|a_3|^2 + \xi_3 +|N_3|^2). \\
 \end{align}
 and expand as,
\begin{align}
     \tilde{b}_{123}&\tilde{b}^*_{123} \\
    &= |a_1|^2 |a_2|^2 |a_3|^2 + |a_1|^2 |a_2|^2 \xi_3 + |a_1|^2 |a_2|^2 |N_3|^2 \\
    & + |a_1|^2 \xi_2 |a_3|^2 + |a_1|^2 \xi_2 \xi_3 + |a_1|^2 \xi_2 |N_3|^2 \\
    &+ |a_1|^2 |N_2|^2 |a_3|^2 + |a_1|^2 |N_2|^2 \xi_3 + |a_1|^2 |N_2|^2 |N_3|^2 \\
    & \\
    &+ \xi_1 |a_2|^2 |a_3|^2 + \xi_1 |a_2|^2 \xi_3 + \xi_1 |a_2|^2 |N_3|^2 \\
    & + \xi_1 \xi_2 |a_3|^2 + \xi_1 \xi_2 \xi_3 + \xi_1 \xi_2 |N_3|^2 \\
    &+ \xi_1 |N_2|^2 |a_3|^2 + \xi_1 |N_2|^2 \xi_3 + \xi_1 |N_2|^2 |N_3|^2 \\
    & \\
    &+ |N_1|^2 |a_2|^2 |a_3|^2 + |N_1|^2 |a_2|^2 \xi_3 + |N_1|^2 |a_2|^2 |N_3|^2 \\
    & + |N_1|^2 \xi_2 |a_3|^2 + |N_1|^2 \xi_2 \xi_3 + |N_1|^2 \xi_2 |N_3|^2 \\
    &+ |N_1|^2 |N_2|^2 |a_3|^2 + |N_1|^2 |N_2|^2 \xi_3 + |N_1|^2 |N_2|^2 |N_3|^2 \\
\end{align}
There are sixty four terms above, recalling that $\xi_i$ represents two terms. Thirty two of the terms will have odd-order moments in $N$, and will therefore equal zero according to Th. \ref{th:isserlis-dhc}, which states we can use Isserlis' Theorem Th. \ref{th:isserlis} for DH coefficient noise moments. This leaves thirty two even order terms.

\begin{align}
     \tilde{b}_{123}&\tilde{b}^*_{123} \\
    &= |a_1|^2 |a_2|^2 |a_3|^2  + |a_1|^2 |a_2|^2 |N_3|^2 \\
    &  + |a_1|^2 \xi_2 \xi_3  \\
    &+ |a_1|^2 |N_2|^2 |a_3|^2 +  |a_1|^2 |N_2|^2 |N_3|^2 \\
    & \\
    &+  \xi_1 |a_2|^2 \xi_3  \\
    & + \xi_1 \xi_2 |a_3|^2 +  \xi_1 \xi_2 |N_3|^2 \\
    &+  \xi_1 |N_2|^2 \xi_3  \\
    & \\
    &+ |N_1|^2 |a_2|^2 |a_3|^2  + |N_1|^2 |a_2|^2 |N_3|^2 \\
    & + |N_1|^2 \xi_2 \xi_3  \\
    &+ |N_1|^2 |N_2|^2 |a_3|^2  + |N_1|^2 |N_2|^2 |N_3|^2 \\
\end{align}

Replacing $\xi_i$ and expanding, we find

\begin{align}
    \tilde{b}_{123}&\tilde{b}^*_{123} \\
    &= |a_1|^2 |a_2|^2 |a_3|^2  + |a_1|^2 |a_2|^2 |N_3|^2 \\
    &  + |a_1|^2 (a_2 a_3 N^*_2  N^*_3 + a_2 a^*_3 N^*_2  N_3 + a^*_2 a_3 N_2 N^*_3+ a^*_2 a^*_3 N_2  N_3)  \\
    &+ |a_1|^2 |N_2|^2 |a_3|^2 +  |a_1|^2 |N_2|^2 |N_3|^2 \\
    & \\
    &+  |a_2|^2(a_1 a_3 N^*_1  N^*_3 + a_1 a^*_3 N^*_1  N_3 + a^*_1 a_3 N_1 N^*_3+ a^*_1 a^*_3 N_1  N_3)  \\
    & + |a_3|^2 (a_1 a_2 N^*_1  N^*_2 + a_1 a^*_2 N^*_1  N_2 + a^*_1 a_2 N_1 N^*_2+ a^*_1 a^*_2 N_1  N_2)  \\
    &+ |N_3|^2 (a_1 a_2 N^*_1  N^*_2 + a_1 a^*_2 N^*_1  N_2 + a^*_1 a_2 N_1 N^*_2+ a^*_1 a^*_2 N_1  N_2)  \\
    &+  |N_2|^2(a_1 a_3 N^*_1  N^*_3 + a_1 a^*_3 N^*_1  N_3 + a^*_1 a_3 N_1 N^*_3+ a^*_1 a^*_3 N_1  N_3)  \\
    & \\
    &+ |N_1|^2 |a_2|^2 |a_3|^2  + |N_1|^2 |a_2|^2 |N_3|^2 \\
    & + |N_1|^2 (a_2 a_3 N^*_2  N^*_3 + a_2 a^*_3 N^*_2  N_3 + a^*_2 a_3 N_2 N^*_3+ a^*_2 a^*_3 N_2  N_3)  \\
    &+ |N_1|^2 |N_2|^2 |a_3|^2  + |N_1|^2 |N_2|^2 |N_3|^2 \\
    & \\
\end{align}
 Taking the expectation of $\tilde{b}_{123}\tilde{b}^*_{123}$,
 \begin{align}
     &\mathbb{E}[\tilde{b}_{123}\tilde{b}^*_{123}] \\
    &= |a_1|^2 |a_2|^2 |a_3|^2  + |a_1|^2 |a_2|^2 \mathbb{E}[|N_3|^2] \\
    &  +  |a_1|^2a_2 a_3 \mathbb{E}[N^*_2  N^*_3] + |a_1|^2 a_2 a^*_3 \mathbb{E}[N^*_2  N_3] \\
    &+ |a_1|^2 a^*_2 a_3 \mathbb{E}[N_2 N^*_3]+ |a_1|^2 a^*_2 a^*_3 \mathbb{E}[N_2  N_3]  \\
    &+ |a_1|^2 |a_3|^2 \mathbb{E}[|N_2|^2] +  |a_1|^2 \mathbb{E}[|N_2|^2 |N_3|^2] \\
    & \\
    &+  |a_2|^2 a_1 a_3 \mathbb{E}[N^*_1  N^*_3] + |a_2|^2 a_1 a^*_3 \mathbb{E}[N^*_1  N_3] \\
    &+ |a_2|^2 a^*_1 a_3 \mathbb{E}[N_1 N^*_3]+ |a_2|^2 a^*_1 a^*_3 \mathbb{E}[N_1  N_3]  \\
    & + |a_3|^2 a_1 a_2 \mathbb{E}[N^*_1  N^*_2] + |a_3|^2 a_1 a^*_2 \mathbb{E}[N^*_1  N_2] \\
    &+ |a_3|^2 a^*_1 a_2 \mathbb{E}[N_1 N^*_2]+ |a_3|^2 a^*_1 a^*_2 \mathbb{E}[N_1  N_2]  \\
    &+  a_1 a_2 \mathbb{E}[|N_3|^2 N^*_1  N^*_2] + a_1 a^*_2 \mathbb{E}[|N_3|^2 N^*_1  N_2] \\
    &+ a^*_1 a_2 \mathbb{E}[|N_3|^2 N_1 N^*_2]+ a^*_1 a^*_2 \mathbb{E}[|N_3|^2 N_1  N_2]  \\
    &+  a_1 a_3 \mathbb{E}[|N_2|^2 N^*_1  N^*_3] + a_1 a^*_3 \mathbb{E}[|N_2|^2 N^*_1  N_3] \\
    &+ a^*_1 a_3 \mathbb{E}[|N_2|^2 N_1 N^*_3] + a^*_1 a^*_3 \mathbb{E}[|N_2|^2 N_1  N_3] \\
    & \\
    &+  |a_2|^2 |a_3|^2 \mathbb{E}[|N_1|^2]  +  |a_2|^2 \mathbb{E}[|N_1|^2 |N_3|^2] \\
    & +  a_2 a_3 \mathbb{E}[|N_1|^2 N^*_2  N^*_3] + a_2 a^*_3 \mathbb{E}[|N_1|^2 N^*_2  N_3] \\
    &+ a^*_2 a_3 \mathbb{E}[|N_1|^2 N_2 N^*_3] + a^*_2 a^*_3 \mathbb{E}[|N_1|^2 N_2  N_3]  \\
    &+ |a_3|^2 \mathbb{E}[|N_1|^2 |N_2|^2]   + \mathbb{E}[|N_1|^2 |N_2|^2 |N_3|^2]. \\
 \end{align}
 Next, we recall the fourth order moment expansion in Th. \ref{th:fourth-moment}
\begin{align}
    &  \mathbb{E}[|N_1|^2 |N_2|^2] =   \mathbb{E}[|N_1|^2] \mathbb{E}[|N_2|^2] + \mathbb{E}[N^*_1 N_2] ^2 + \mathbb{E}[N_1 N_2] ^2 \\
        & \mathbb{E}[|N_1|^2 N_2 N_3] = \mathbb{E}[|N_1|^2] \mathbb{E}[N_2 N_3] + \mathbb{E}[N_1 N_3] \mathbb{E}[N^*_1 N_2]  + \mathbb{E}[N_1 N_2] \mathbb{E}[N^*_1 N_3]. \\
\end{align}
\begin{align}
     &\mathbb{E}[\tilde{b}_{123}\tilde{b}^*_{123}] \\
    &= |a_1|^2 |a_2|^2 |a_3|^2  + |a_1|^2 |a_2|^2 \mathbb{E}[|N_3|^2] \\
    &  +  |a_1|^2a_2 a_3 \mathbb{E}[N^*_2  N^*_3] + |a_1|^2 a_2 a^*_3 \mathbb{E}[N^*_2  N_3] \\
    &+ |a_1|^2 a^*_2 a_3 \mathbb{E}[N_2 N^*_3]+ |a_1|^2 a^*_2 a^*_3 \mathbb{E}[N_2  N_3]  \\
    &+ |a_1|^2 |a_3|^2 \mathbb{E}[|N_2|^2] +  |a_1|^2 (\mathbb{E}[|N_2|^2] \mathbb{E}[|N_3|^2] + \mathbb{E}[N^*_2 N_3] ^2 + \mathbb{E}[N_2 N_3] ^2) \\
    & \\
    &+  |a_2|^2 a_1 a_3 \mathbb{E}[N^*_1  N^*_3] + |a_2|^2 a_1 a^*_3 \mathbb{E}[N^*_1 N_3] \\
    &+ |a_2|^2 a^*_1 a_3 \mathbb{E}[N_1 N^*_3]+ |a_2|^2 a^*_1 a^*_3 \mathbb{E}[N_1 N_3]  \\
    & + |a_3|^2 a_1 a_2 \mathbb{E}[N^*_1 N^*_2] + |a_3|^2 a_1 a^*_2 \mathbb{E}[N^*_1 N_2] \\
    &+ |a_3|^2 a^*_1 a_2 \mathbb{E}[N_1 N^*_2]+ |a_3|^2 a^*_1 a^*_2 \mathbb{E}[N_1 N_2]  \\
    &+  a_1 a_2 (\mathbb{E}[|N_3|^2] \mathbb{E}[N^*_1 N^*_2] + \mathbb{E}[N^*_2 N_3] \mathbb{E}[N^*_1 N^*_3]  + \mathbb{E}[N^*_1 N_3] \mathbb{E}[N^*_2 N^*_3]) \\
    &+ a_1 a^*_2 (\mathbb{E}[|N_3|^2] \mathbb{E}[N^*_1 N_2] + \mathbb{E}[N_2 N_3] \mathbb{E}[N^*_1 N^*_3]  + \mathbb{E}[N^*_1 N_3] \mathbb{E}[N_2 N^*_3]) \\
    &+ a^*_1 a_2 (\mathbb{E}[|N_3|^2] \mathbb{E}[N_1 N^*_2] + \mathbb{E}[N^*_2 N_3] \mathbb{E}[N_1 N^*_3]  + \mathbb{E}[N_1 N_3] \mathbb{E}[N^*_2 N^*_3])\\
    &+ a^*_1 a^*_2 (\mathbb{E}[|N_3|^2] \mathbb{E}[N_1 N_2] + \mathbb{E}[N_2 N_3] \mathbb{E}[N_1 N^*_3]  + \mathbb{E}[N_1 N_3] \mathbb{E}[N_2 N^*_3])  \\
    &+  a_1 a_3 (\mathbb{E}[|N_2|^2] \mathbb{E}[N^*_1 N^*_3] + \mathbb{E}[N_2 N^*_3] \mathbb{E}[N^*_2 N^*_1]  + \mathbb{E}[N_2 N^*_1] \mathbb{E}[N^*_2 N^*_3]) \\
    &+ a_1 a^*_3 (\mathbb{E}[|N_2|^2] \mathbb{E}[N^*_1 N_3] + \mathbb{E}[N_2 N_3] \mathbb{E}[N^*_2 N^*_1]  + \mathbb{E}[N_2 N^*_1] \mathbb{E}[N^*_2 N_3]) \\
    &+ a^*_1 a_3 (\mathbb{E}[|N_2|^2] \mathbb{E}[N_1 N^*_3] + \mathbb{E}[N_2 N^*_3] \mathbb{E}[N^*_2 N_1]  + \mathbb{E}[N_2 N_1] \mathbb{E}[N^*_2 N^*_3]) \\
    &+ a^*_1 a^*_3 (\mathbb{E}[|N_2|^2] \mathbb{E}[N_1 N_3] + \mathbb{E}[N_2 N_3] \mathbb{E}[N^*_2 N_1]  + \mathbb{E}[N_2 N_1] \mathbb{E}[N^*_2 N_3]) \\
    & \\
    &+  |a_2|^2 |a_3|^2 \mathbb{E}[|N_1|^2]  +  |a_2|^2 (\mathbb{E}[|N_1|^2] \mathbb{E}[|N_3|^2] + \mathbb{E}[N^*_1 N_3] ^2 + \mathbb{E}[N_1 N_3] ^2) \\
    & +  a_2 a_3 (\mathbb{E}[|N_1|^2] \mathbb{E}[N^*_2 N^*_3] + \mathbb{E}[N_1 N^*_3] \mathbb{E}[N^*_1 N^*_2]  + \mathbb{E}[N_1 N^*_2] \mathbb{E}[N^*_1 N^*_3]) \\
    &+ a_2 a^*_3 (\mathbb{E}[|N_1|^2] \mathbb{E}[N^*_2 N_3] + \mathbb{E}[N_1 N_3] \mathbb{E}[N^*_1 N^*_2]  + \mathbb{E}[N_1 N^*_2] \mathbb{E}[N^*_1 N_3]) \\
    &+ a^*_2 a_3 (\mathbb{E}[|N_1|^2] \mathbb{E}[N_2 N^*_3] + \mathbb{E}[N_1 N^*_3] \mathbb{E}[N^*_1 N_2]  + \mathbb{E}[N_1 N_2] \mathbb{E}[N^*_1 N^*_3]) \\
    &+ a^*_2 a^*_3 (\mathbb{E}[|N_1|^2] \mathbb{E}[N_2 N_3] + \mathbb{E}[N_1 N_3] \mathbb{E}[N^*_1 N_2]  + \mathbb{E}[N_1 N_2] \mathbb{E}[N^*_1 N_3])  \\
    &+ |a_3|^2 (\mathbb{E}[|N_1|^2] \mathbb{E}[|N_2|^2] + \mathbb{E}[N^*_1 N_2] ^2 + \mathbb{E}[N_1 N_2] ^2)   + \mathbb{E}[|N_1|^2 |N_2|^2 |N_3|^2]. \\
 \end{align}
 
 Plugging complex conjugate relations above,
 \begin{align}
     &\mathbb{E}[\tilde{b}_{123}\tilde{b}^*_{123}] \\
    &= |a_1|^2 |a_2|^2 |a_3|^2  + |a_1|^2 |a_2|^2 \mathbb{E}[|N_3|^2] \\
    &  +  |a_1|^2a_2 a_3 \mathbb{E}[N_2 N_3] + |a_1|^2 a_2 a^*_3 \mathbb{E}[N^*_2  N_3] \\
    &+ |a_1|^2 a^*_2 a_3 \mathbb{E}[N^*_2 N_3]+ |a_1|^2 a^*_2 a^*_3 \mathbb{E}[N_2  N_3]  \\
    &+ |a_1|^2 |a_3|^2 \mathbb{E}[|N_2|^2] +  |a_1|^2 (\mathbb{E}[|N_2|^2] \mathbb{E}[|N_3|^2] + \mathbb{E}[N^*_2 N_3] ^2 + \mathbb{E}[N_2 N_3] ^2) \\
    & \\
    &+  |a_2|^2 a_1 a_3 \mathbb{E}[N_1 N_3] + |a_2|^2 a_1 a^*_3 \mathbb{E}[N^*_1 N_3] \\
    &+ |a_2|^2 a^*_1 a_3 \mathbb{E}[N^*_1 N_3]+ |a_2|^2 a^*_1 a^*_3 \mathbb{E}[N_1 N_3]  \\
    & + |a_3|^2 a_1 a_2 \mathbb{E}[N_1 N_2] + |a_3|^2 a_1 a^*_2 \mathbb{E}[N^*_1 N_2] \\
    &+ |a_3|^2 a^*_1 a_2 \mathbb{E}[N^*_1 N_2]+ |a_3|^2 a^*_1 a^*_2 \mathbb{E}[N_1 N_2]  \\
    &+  a_1 a_2 (\mathbb{E}[|N_3|^2] \mathbb{E}[N_1 N_2] + \mathbb{E}[N^*_2 N_3] \mathbb{E}[N_1 N_3]  + \mathbb{E}[N^*_1 N_3] \mathbb{E}[N_2 N_3]) \\
    &+ a_1 a^*_2 (\mathbb{E}[|N_3|^2] \mathbb{E}[N^*_1 N_2] + \mathbb{E}[N_2 N_3] \mathbb{E}[N_1 N_3]  + \mathbb{E}[N^*_1 N_3] \mathbb{E}[N^*_2 N_3]) \\
    &+ a^*_1 a_2 (\mathbb{E}[|N_3|^2] \mathbb{E}[N^*_1 N_2] + \mathbb{E}[N^*_2 N_3] \mathbb{E}[N^*_1 N_3]  + \mathbb{E}[N_1 N_3] \mathbb{E}[N_2 N_3])\\
    &+ a^*_1 a^*_2 (\mathbb{E}[|N_3|^2] \mathbb{E}[N_1 N_2] + \mathbb{E}[N_2 N_3] \mathbb{E}[N^*_1 N_3]  + \mathbb{E}[N_1 N_3] \mathbb{E}[N^*_2 N_3])  \\
    &+  a_1 a_3 (\mathbb{E}[|N_2|^2] \mathbb{E}[N_1 N_3] + \mathbb{E}[N^*_2 N_3] \mathbb{E}[N_2 N_1]  + \mathbb{E}[N_2 N^*_1] \mathbb{E}[N_2 N_3]) \\
    &+ a_1 a^*_3 (\mathbb{E}[|N_2|^2] \mathbb{E}[N^*_1 N_3] + \mathbb{E}[N_2 N_3] \mathbb{E}[N_2 N_1]  + \mathbb{E}[N_2 N^*_1] \mathbb{E}[N^*_2 N_3]) \\
    &+ a^*_1 a_3 (\mathbb{E}[|N_2|^2] \mathbb{E}[N^*_1 N_3] + \mathbb{E}[N^*_2 N_3] \mathbb{E}[N^*_2 N_1]  + \mathbb{E}[N_2 N_1] \mathbb{E}[N_2 N_3]) \\
    &+ a^*_1 a^*_3 (\mathbb{E}[|N_2|^2] \mathbb{E}[N_1 N_3] + \mathbb{E}[N_2 N_3] \mathbb{E}[N^*_2 N_1]  + \mathbb{E}[N_2 N_1] \mathbb{E}[N^*_2 N_3]) \\
    & \\
    &+  |a_2|^2 |a_3|^2 \mathbb{E}[|N_1|^2]  +  |a_2|^2 (\mathbb{E}[|N_1|^2] \mathbb{E}[|N_3|^2] + \mathbb{E}[N^*_1 N_3] ^2 + \mathbb{E}[N_1 N_3] ^2) \\
    & +  a_2 a_3 (\mathbb{E}[|N_1|^2] \mathbb{E}[N_2 N_3] + \mathbb{E}[N^*_1 N_3] \mathbb{E}[N_1 N_2]  + \mathbb{E}[N^*_1 N_2] \mathbb{E}[N_1 N_3]) \\
    &+ a_2 a^*_3 (\mathbb{E}[|N_1|^2] \mathbb{E}[N^*_2 N_3] + \mathbb{E}[N_1 N_3] \mathbb{E}[N_1 N_2]  + \mathbb{E}[N^*_1 N_2] \mathbb{E}[N^*_1 N_3]) \\
    &+ a^*_2 a_3 (\mathbb{E}[|N_1|^2] \mathbb{E}[N^*_2 N_3] + \mathbb{E}[N^*_1 N_3] \mathbb{E}[N^*_1 N_2]  + \mathbb{E}[N_1 N_2] \mathbb{E}[N_1 N_3]) \\
    &+ a^*_2 a^*_3 (\mathbb{E}[|N_1|^2] \mathbb{E}[N_2 N_3] + \mathbb{E}[N_1 N_3] \mathbb{E}[N^*_1 N_2]  + \mathbb{E}[N_1 N_2] \mathbb{E}[N^*_1 N_3])  \\
    &+ |a_3|^2 (\mathbb{E}[|N_1|^2] \mathbb{E}[|N_2|^2] + \mathbb{E}[N^*_1 N_2] ^2 + \mathbb{E}[N_1 N_2] ^2)   + \mathbb{E}[|N_1|^2 |N_2|^2 |N_3|^2]. \\
 \end{align}
 Recalling $\mathbb{E}[\tilde{b}_{123}]\mathbb{E}[\tilde{b}^*_{123}]$, we compute $\mathrm{Var}(\tilde{b}_{123}) = \mathbb{E}[\tilde{b}_{123}\tilde{b}^*_{123}] - \mathbb{E}[\tilde{b}_{123}] \mathbb{E}[\tilde{b}^*_{123}]$ as
 \begin{align}
     &\mathrm{Var}(\tilde{b}_{123}) = \text{-----}  + |a_1|^2 |a_2|^2 \mathbb{E}[|N_3|^2] \\
    &  +  |a_1|^2 a_2 a_3 \mathbb{E}[N_2 N_3] + \text{-----} + \text{-----}+ |a_1|^2 a^*_2 a^*_3 \mathbb{E}[N_2  N_3]  \\
    &+ |a_1|^2 |a_3|^2 \mathbb{E}[|N_2|^2] +  |a_1|^2\mathbb{E}[|N_2|^2] \mathbb{E}[|N_3|^2] + \text{-----} + |a_1|^2 \mathbb{E}[N_2 N_3] ^2 \\
    & \\
    &+  |a_2|^2 a_1 a_3 \mathbb{E}[N_1 N_3] + \text{-----} + \text{-----}+ |a_2|^2 a^*_1 a^*_3 \mathbb{E}[N_1 N_3]  \\
    & + \text{-----} + |a_3|^2 a_1 a^*_2 \mathbb{E}[N^*_1 N_2] + |a_3|^2 a^*_1 a_2 \mathbb{E}[N^*_1 N_2]+ \text{-----}  \\
    &+  a_1 a_2 \mathbb{E}[|N_3|^2] \mathbb{E}[N_1 N_2] + a_1 a_2 \mathbb{E}[N^*_2 N_3] \mathbb{E}[N_1 N_3]  + a_1 a_2 \mathbb{E}[N^*_1 N_3] \mathbb{E}[N_2 N_3] \\
    &+ a_1 a^*_2 \mathbb{E}[|N_3|^2] \mathbb{E}[N^*_1 N_2] + a_1 a^*_2 \mathbb{E}[N_2 N_3] \mathbb{E}[N_1 N_3]  + \text{-----} \\
    &+ a^*_1 a_2 \mathbb{E}[|N_3|^2] \mathbb{E}[N^*_1 N_2] + \text{-----}  + a^*_1 a_2 \mathbb{E}[N_1 N_3] \mathbb{E}[N_2 N_3]\\
    &+ a^*_1 a^*_2 \mathbb{E}[|N_3|^2] \mathbb{E}[N_1 N_2] + a^*_1 a^*_2 \mathbb{E}[N_2 N_3] \mathbb{E}[N^*_1 N_3]  + a^*_1 a^*_2 \mathbb{E}[N_1 N_3] \mathbb{E}[N^*_2 N_3]  \\
    &+ a_1 a_3 \mathbb{E}[|N_2|^2] \mathbb{E}[N_1 N_3] + \text{-----}  + a_1 a_3 \mathbb{E}[N_2 N^*_1] \mathbb{E}[N_2 N_3] \\
    &+ a_1 a^*_3 \mathbb{E}[|N_2|^2] \mathbb{E}[N^*_1 N_3] + a_1 a^*_3 \mathbb{E}[N_2 N_3] \mathbb{E}[N_2 N_1]  + a_1 a^*_3 \mathbb{E}[N_2 N^*_1] \mathbb{E}[N^*_2 N_3] \\
    &+ a^*_1 a_3 \mathbb{E}[|N_2|^2] \mathbb{E}[N^*_1 N_3] + a^*_1 a_3 \mathbb{E}[N^*_2 N_3] \mathbb{E}[N^*_2 N_1]  + a^*_1 a_3 \mathbb{E}[N_2 N_1] \mathbb{E}[N_2 N_3] \\
    &+ a^*_1 a^*_3 \mathbb{E}[|N_2|^2] \mathbb{E}[N_1 N_3] + a^*_1 a^*_3 \mathbb{E}[N_2 N_3] \mathbb{E}[N^*_2 N_1]  + \text{-----} \\
    & \\
    &+  |a_2|^2 |a_3|^2 \mathbb{E}[|N_1|^2]  +  |a_2|^2 \mathbb{E}[|N_1|^2] \mathbb{E}[|N_3|^2] + \text{-----} + |a_2|^2 \mathbb{E}[N_1 N_3]^2) \\
    & + a_2 a_3 \mathbb{E}[|N_1|^2] \mathbb{E}[N_2 N_3] + \text{-----}  + a_2 a_3 \mathbb{E}[N^*_1 N_2] \mathbb{E}[N_1 N_3] \\
    &+ a_2 a^*_3 \mathbb{E}[|N_1|^2] \mathbb{E}[N^*_2 N_3] + a_2 a^*_3 \mathbb{E}[N_1 N_3] \mathbb{E}[N_1 N_2]  + a_2 a^*_3 \mathbb{E}[N^*_1 N_2] \mathbb{E}[N^*_1 N_3] \\
    &+ a^*_2 a_3 \mathbb{E}[|N_1|^2] \mathbb{E}[N^*_2 N_3] + a^*_2 a_3 \mathbb{E}[N^*_1 N_3] \mathbb{E}[N^*_1 N_2]  + a^*_2 a_3 \mathbb{E}[N_1 N_2] \mathbb{E}[N_1 N_3] \\
    &+ a^*_2 a^*_3 \mathbb{E}[|N_1|^2] \mathbb{E}[N_2 N_3] + a^*_2 a^*_3 \mathbb{E}[N_1 N_3] \mathbb{E}[N^*_1 N_2]  + \text{-----}  \\
    &+ |a_3|^2 \mathbb{E}[|N_1|^2] \mathbb{E}[|N_2|^2] + |a_3|^2 \mathbb{E}[N^*_1 N_2]^2 + \text{-----}   + \mathbb{E}[|N_1|^2 |N_2|^2 |N_3|^2], \\
 \end{align}
 where $\text{-----}$ indicates a term in $\mathbb{E}[\tilde{b}_{123}\tilde{b}^*_{123}]$ that cancels with $\mathbb{E}[\tilde{b}_{123}] \mathbb{E}[\tilde{b}^*_{123}]$. Note that after this cancellation, there are no surviving terms from $\mathbb{E}[\tilde{b}_{123}] \mathbb{E}[\tilde{b}^*_{123}]$. Next, we remove cancellation placeholders and substitute $\mathbb{E}[N_iN^*_i] = \mathbb{E}[|N_i|^2] = \sigma^2$, finding
\begin{align}
     \mathrm{Var}(\tilde{b}_{123}) &=  + |a_1|^2 |a_2|^2 \sigma^2 \\
    &  +  |a_1|^2 a_2 a_3 \mathbb{E}[N_2 N_3] + |a_1|^2 a^*_2 a^*_3 \mathbb{E}[N_2  N_3]  \\
    &+ |a_1|^2 |a_3|^2 \sigma^2 +  |a_1|^2\sigma^2 \sigma^2 + |a_1|^2 \mathbb{E}[N_2 N_3] ^2 \\
    & \\
    &+  |a_2|^2 a_1 a_3 \mathbb{E}[N_1 N_3] + |a_2|^2 a^*_1 a^*_3 \mathbb{E}[N_1 N_3]  \\
    & + |a_3|^2 a_1 a^*_2 \mathbb{E}[N^*_1 N_2] + |a_3|^2 a^*_1 a_2 \mathbb{E}[N^*_1 N_2]\\
    &+  a_1 a_2 \sigma^2 \mathbb{E}[N_1 N_2] + a_1 a_2 \mathbb{E}[N^*_2 N_3] \mathbb{E}[N_1 N_3]  + a_1 a_2 \mathbb{E}[N^*_1 N_3] \mathbb{E}[N_2 N_3] \\
    &+ a_1 a^*_2 \sigma^2 \mathbb{E}[N^*_1 N_2] + a_1 a^*_2 \mathbb{E}[N_2 N_3] \mathbb{E}[N_1 N_3]  \\
    &+ a^*_1 a_2 \sigma^2 \mathbb{E}[N^*_1 N_2] + a^*_1 a_2 \mathbb{E}[N_1 N_3] \mathbb{E}[N_2 N_3]\\
    &+ a^*_1 a^*_2 \sigma^2 \mathbb{E}[N_1 N_2] + a^*_1 a^*_2 \mathbb{E}[N_2 N_3] \mathbb{E}[N^*_1 N_3]  + a^*_1 a^*_2 \mathbb{E}[N_1 N_3] \mathbb{E}[N^*_2 N_3]  \\
    &+ a_1 a_3 \sigma^2 \mathbb{E}[N_1 N_3] + a_1 a_3 \mathbb{E}[N_2 N^*_1] \mathbb{E}[N_2 N_3] \\
    &+ a_1 a^*_3 \sigma^2 \mathbb{E}[N^*_1 N_3] + a_1 a^*_3 \mathbb{E}[N_2 N_3] \mathbb{E}[N_2 N_1]  + a_1 a^*_3 \mathbb{E}[N_2 N^*_1] \mathbb{E}[N^*_2 N_3] \\
    &+ a^*_1 a_3 \sigma^2 \mathbb{E}[N^*_1 N_3] + a^*_1 a_3 \mathbb{E}[N^*_2 N_3] \mathbb{E}[N^*_2 N_1]  + a^*_1 a_3 \mathbb{E}[N_2 N_1] \mathbb{E}[N_2 N_3] \\
    &+ a^*_1 a^*_3 \sigma^2 \mathbb{E}[N_1 N_3] + a^*_1 a^*_3 \mathbb{E}[N_2 N_3] \mathbb{E}[N^*_2 N_1]  \\
    & \\
    &+  |a_2|^2 |a_3|^2 \sigma^2  +  |a_2|^2 \sigma^2 \sigma^2 + |a_2|^2 \mathbb{E}[N_1 N_3]^2) \\
    & + a_2 a_3 \sigma^2 \mathbb{E}[N_2 N_3] + a_2 a_3 \mathbb{E}[N^*_1 N_2] \mathbb{E}[N_1 N_3] \\
    &+ a_2 a^*_3 \sigma^2 \mathbb{E}[N^*_2 N_3] + a_2 a^*_3 \mathbb{E}[N_1 N_3] \mathbb{E}[N_1 N_2]  + a_2 a^*_3 \mathbb{E}[N^*_1 N_2] \mathbb{E}[N^*_1 N_3] \\
    &+ a^*_2 a_3 \sigma^2 \mathbb{E}[N^*_2 N_3] + a^*_2 a_3 \mathbb{E}[N^*_1 N_3] \mathbb{E}[N^*_1 N_2]  + a^*_2 a_3 \mathbb{E}[N_1 N_2] \mathbb{E}[N_1 N_3] \\
    &+ a^*_2 a^*_3 \sigma^2 \mathbb{E}[N_2 N_3] + a^*_2 a^*_3 \mathbb{E}[N_1 N_3] \mathbb{E}[N^*_1 N_2]  \\
    &+ |a_3|^2 \sigma^2 \sigma^2 + |a_3|^2 \mathbb{E}[N^*_1 N_2]^2 + \mathbb{E}[|N_1|^2 |N_2|^2 |N_3|^2]. \\
 \end{align}
 Reorganizing,
 \begin{align}
     &\mathrm{Var}(\tilde{b}_{123}) =  \sigma^2 \{|a_1|^2 |a_2|^2 + |a_1|^2 |a_3|^2 + |a_2|^2 |a_3|^2 \}  \\
     & +  \sigma^4 \{|a_1|^2 +  |a_2|^2  + |a_3|^2 \} \\
     &\\
     & + \mathbb{E}[N_1 N_2] \{  \sigma^2 (a_1 a_2  + a^*_1 a^*_2 ) + \mathbb{E}[N_1 N_3] (a^*_2 a_3  + a_2 a^*_3 ) + \mathbb{E}[N_2 N_3] (a^*_1 a_3  + a_1 a^*_3 ) \}  \\
     &\\
     & +  \mathbb{E}[N_1 N_3] \{|a_2|^2 ( a_1 a_3 + a^*_1 a^*_3 + \mathbb{E}[N_1 N_3])+ \sigma^2 (a_1 a_3  + a^*_1 a^*_3)  \\
     & \quad \quad +  \mathbb{E}[N_2 N_3] (a_1 a^*_2  + a^*_1 a_2) + \mathbb{E}[N^*_2 N_3] (a^*_1 a^*_2 + a_1 a_2) + \mathbb{E}[N^*_1 N_2](a_2 a_3  + a^*_2 a^*_3) \} \\
     &\\
     & +  \mathbb{E}[N_2 N_3] \{|a_1|^2 ( a_2 a_3 + a^*_2 a^*_3  + \mathbb{E}[N_2 N_3] )  + \sigma^2 (a_2 a_3 + a^*_2 a^*_3 ) \\
     & \quad \quad + \mathbb{E}[N^*_1 N_3] (a_1 a_2 + a^*_1 a^*_2) + \mathbb{E}[N^*_1 N_2](a_1 a_3 + a^*_1 a^*_3)  \} \\
     &\\
     & + \mathbb{E}[N^*_2 N_3] \{ \sigma^2 (a_2 a^*_3 + a^*_2 a_3) + \mathbb{E}[N^*_1 N_2]( a_1 a^*_3   + a^*_1 a_3)  \} \\
     &\\
     & + \mathbb{E}[N^*_1 N_3] \{ \sigma^2 (a_1 a^*_3   + a^*_1 a_3) + \mathbb{E}[N^*_1 N_2](a_2 a^*_3 + a^*_2 a_3) \} \\
     &\\
     & +  \mathbb{E}[N^*_1 N_2] \{|a_3|^2 (a_1 a^*_2  + a^*_1 a_2 + \mathbb{E}[N^*_1 N_2]) + \sigma^2(a_1 a^*_2 + a^*_1 a_2)  \}  \\
     &\\
     & + \mathbb{E}[|N_1|^2 |N_2|^2 |N_3|^2]. \\
 \end{align}
 Recall the sixth order moment from Th. \ref{th:sixth-moment}, 
 \begin{align}
        \mathbb{E}[|N_1|^2 &|N_2|^2 |N_3|^2] = \mathbb{E}[|N_1|^2 ] \mathbb{E}[|N_2|^2 ] \mathbb{E}[|N_3|^2 ] \\
        & + \mathbb{E}[N_1 N_2]   \{ \mathbb{E}[N_1 N_2] \mathbb{E}[|N_3|^2]  + 2 \mathbb{E}[N^*_1 N_3] \mathbb{E}[N_2 N_3] + 2\mathbb{E}[N_1 N_3] \mathbb{E}[N^*_2 N_3] \} \\
        & + \mathbb{E}[N_1 N_3]   \{  \mathbb{E}[N_1 N_3] \mathbb{E}[|N_2|^2]  + 2\mathbb{E}[N^*_1 N_2] \mathbb{E}[N_2 N_3]  \} \\
        & + \mathbb{E}[N_2 N_3]   \{ \mathbb{E}[|N_1|^2 ] \mathbb{E}[N_2 N_3] \} \\
        & + \mathbb{E}[N^*_2 N_3] \{   \mathbb{E}[|N_1|^2 ] \mathbb{E}[N^*_2 N_3]  + 2\mathbb{E}[N^*_1 N_2] \mathbb{E}[N^*_1 N_3]\} \\
        & + \mathbb{E}[N^*_1 N_3] \{  \mathbb{E}[N^*_1 N_3] \mathbb{E}[|N_2|^2] \} \\
        & +  \mathbb{E}[N^*_1 N_2] \{ \mathbb{E}[N^*_1 N_2] \mathbb{E}[|N_3|^2] \} \\
    \end{align}
    and plug in to find 
 
 \begin{align}
     &\mathrm{Var}(\tilde{b}_{123}) =  \sigma^2 \{|a_1|^2 |a_2|^2 + |a_1|^2 |a_3|^2 + |a_2|^2 |a_3|^2 \}  \\
     & +  \sigma^4 \{|a_1|^2 +  |a_2|^2  + |a_3|^2 \} + \sigma^6\\
     &\\
     & + \mathbb{E}[N_1 N_2] \{  \sigma^2 (a_1 a_2  + a^*_1 a^*_2 + \mathbb{E}[N_1 N_2]) + \mathbb{E}[N_1 N_3] (a^*_2 a_3  + a_2 a^*_3 + 2\mathbb{E}[N^*_2 N_3]) \\
     & \quad \quad + \mathbb{E}[N_2 N_3] (a^*_1 a_3  + a_1 a^*_3 + 2 \mathbb{E}[N^*_1 N_3] ) \}\\
     &\\
     & + \mathbb{E}[N_1 N_3] \{(|a_2|^2 + \sigma^2) ( a_1 a_3 + a^*_1 a^*_3 + \mathbb{E}[N_1 N_3])\\
     & \quad \quad  +  \mathbb{E}[N_2 N_3] (a_1 a^*_2  + a^*_1 a_2 + 2\mathbb{E}[N^*_1 N_2])  \\
     & \quad \quad + \mathbb{E}[N^*_2 N_3] (a^*_1 a^*_2 + a_1 a_2) + \mathbb{E}[N^*_1 N_2](a_2 a_3  + a^*_2 a^*_3) \} \\
     &\\
     & +  \mathbb{E}[N_2 N_3] \{(|a_1|^2 + \sigma^2) ( a_2 a_3 + a^*_2 a^*_3  + \mathbb{E}[N_2 N_3] ) \\
     & \quad \quad + \mathbb{E}[N^*_1 N_3] (a_1 a_2 + a^*_1 a^*_2) + \mathbb{E}[N^*_1 N_2](a_1 a_3 + a^*_1 a^*_3) \} \\
     &\\
     & + \mathbb{E}[N^*_2 N_3] \{ \sigma^2 (a_2 a^*_3 + a^*_2 a_3 + \mathbb{E}[N^*_2 N_3]) + \mathbb{E}[N^*_1 N_2]( a_1 a^*_3   + a^*_1 a_3 + 2\mathbb{E}[N^*_1 N_3])  \} \\
     &\\
     & + \mathbb{E}[N^*_1 N_3] \{ \sigma^2 (a_1 a^*_3  + a^*_1 a_3 + \mathbb{E}[N^*_1 N_3]) + \mathbb{E}[N^*_1 N_2](a_2 a^*_3 + a^*_2 a_3 ) \} \\
     &\\
     & +  \mathbb{E}[N^*_1 N_2] \{(|a_3|^2 + \sigma^2) (a_1 a^*_2  + a^*_1 a_2 + \mathbb{E}[N^*_1 N_2]) \}. \\
 \end{align}

We recall second moment identities from Th. \ref{th:second-moment} and remember that the SDB only considers terms for $n \geq 0$, so $\delta_{n,-n'} \to \delta_{n , 0}\delta_{n' , 0}$ and $\gamma_{0k'} = 1$. With these conditions, 
\begin{align*}
        &\mathbb{E}[N_{nk} N_{n'k'}] = \sigma^2 \delta_{n , 0}\delta_{n' , 0} \delta_{k', k}  \\
        &\mathbb{E}[N_{nk} N^*_{n'k'}] = \sigma^2   \delta_{n , n'} \delta_{k , k'}.
    \end{align*}
 Plugging in second order moments for the SDB,
 \begin{align}
     & \mathrm{Var}(\tilde{b}_{123}) \\
     & = \sigma ^6  + \sigma^4 (|a_1|^2   + |a_2|^2  + |a_3|^2 )  \\
      &  +\sigma^2 (|a_1|^2 |a_2|^2  + |a_1|^2 |a_3|^2  + |a_2|^2 |a_3|^2) \\
      & \\
      & + \sigma^2 \delta_{n _1, 0}\delta_{n_2 , 0} \delta_{k_1, k_2}  \times \{ \sigma^2 (\sigma^2 \delta_{n _1, 0}\delta_{n_2 , 0} \delta_{k_1, k_2}  +a_1 a_2 + a^*_1 a^*_2 )  \\
      & \quad \quad + \sigma^2 \delta_{n _1, 0}\delta_{n_3 , 0} \delta_{k_1, k_3}  (a_2 a^*_3 + a^*_2 a_3 + 2 \sigma^2 \delta_{n_2 , n_3} \delta_{k_2 , k_3} )\\
      & \quad \quad +  \sigma^2 \delta_{n _2, 0}\delta_{n_3 , 0} \delta_{k_2, k_3}  (a_1 a^*_3 +a^*_1 a_3 + 2\sigma^2 \delta_{n_1 , n_3} \delta_{k_1 , k_3}) \}  \\
      & \\
      &  +  \sigma^2 \delta_{n _1, 0}\delta_{n_3 , 0} \delta_{k_1, k_3}  \times \{ (|a_2|^2 + \sigma^2) (\sigma^2 \delta_{n _1, 0}\delta_{n_3 , 0} \delta_{k_1, k_3}  + a_1 a_3 + a^*_1 a^*_3 )   \\
      & \quad \quad  + \sigma^2 \delta_{n _2, 0}\delta_{n_3 , 0} \delta_{k_2, k_3} (a_1 a^*_2 + a^*_1 a_2 + 2\sigma^2 \delta_{n_1 , n_2} \delta_{k_1 , k_2} ) \\
      & \quad \quad + \sigma^2 \delta_{n_1 , n_2} \delta_{k_1 , k_2} (a_2 a_3 + a^*_2 a^*_3) + \sigma^2 \delta_{n_2 , n_3} \delta_{k_2 , k_3} (a_1 a_2 + a^*_1 a^*_2) \} \\
      &  \\
      & + \sigma^2 \delta_{n _2, 0}\delta_{n_3 , 0} \delta_{k_2, k_3}  \times \{(|a_1|^2 + \sigma^2)(\sigma^2 \delta_{n _2, 0}\delta_{n_3 , 0} \delta_{k_2, k_3}  + a_2 a_3 + a^*_2 a^*_3 ) \\
      & \quad \quad + \sigma^2 \delta_{n_1 , n_2} \delta_{k_1 , k_2} (a_1 a_3 + a^*_1 a^*_3) + \sigma^2 \delta_{n_1 , n_3} \delta_{k_1 , k_3} (a_1 a_2 + a^*_1 a^*_2)  \} \\
      & \\
      & + \sigma^2 \delta_{n_2 , n_3} \delta_{k_2 , k_3} \times \{ \sigma^2 (\sigma^2 \delta_{n_2 , n_3} \delta_{k_2 , k_3} + a_2 a^*_3 + a^*_2 a_3) \\
      & \quad \quad +\sigma^2 \delta_{n_1 , n_2} \delta_{k_1 , k_2}  (a_1 a^*_3 +a^*_1 a_3 + 2 \sigma^2 \delta_{n_1 , n_3} \delta_{k_1 , k_3}) \} \\
      & \\
      & + \sigma^2 \delta_{n_1 , n_3} \delta_{k_1 , k_3} \times \{\sigma^2  (\sigma^2 \delta_{n_1 , n_3} \delta_{k_1 , k_3} + a_1 a^*_3 + a^*_1 a_3) \\
      & \quad \quad +\sigma^2 \delta_{n_1 , n_2} \delta_{k_1 , k_2} (a_2 a^*_3 + a^*_2 a_3) \} \\
      &  \\
      & + \sigma^2 \delta_{n_1 , n_2} \delta_{k_1 , k_2} \times \{(|a_3|^2 + \sigma^2)  (\sigma^2 \delta_{n_1 , n_2} \delta_{k_1 , k_2} + a_1 a^*_2 + a^*_1 a_2) \}. \\
 \end{align}

 Next, we remember that our notation $\tilde{b}_{123} = \tilde{a}_1 \tilde{a}_2 \tilde{a}^*_3$ is shorthand for $\tilde{b}_{j_1,j_2, k_3} = \tilde{a}_{n_{j_1} k_{j_1}} \tilde{a}_{n_{j_2} k_{j_2}} \tilde{a}^*_{n_{j_1} + n_{j_2} k_3}$. Consequently, $\delta_{n _1, 0}\delta_{n_3 , 0}$ and $\delta_{n _2, 0}\delta_{n_3 , 0}$, are equivalent to $\delta_{n _1, 0}\delta_{n_2 , 0}$. The SDB only considers coefficients where $k_{j_1} = k_{j_2} = 1$. Therefore, $\delta_{n _1, 0}\delta_{n_2 , 0} \delta_{k_1, k_2} \to \delta_{n _1, 0}\delta_{n_2 , 0}$, and $\delta_{n _2, 0}\delta_{n_3 , 0} \delta_{k_2, k_3} \to \delta_{n _1, 0}\delta_{n_2 , 0} \delta_{k_3,1}$ for example. Additionally, $\delta_{n_2 , n_3} = \delta_{n_2 , n_1 + n_2} = \delta_{n_1 , 0}$ and $\delta_{n_1 , n_3} = \delta_{n_1 , n_1 + n_2} = \delta_{n_2 , 0}$. With this,  and simplifying redundant deltas 

 \begin{align}
     & \mathrm{Var}(\tilde{b}_{123}) \\
     & = \sigma ^6  + \sigma^4 (|a_1|^2   + |a_2|^2  + |a_3|^2 )  \\
      &  +\sigma^2 (|a_1|^2 |a_2|^2  + |a_1|^2 |a_3|^2  + |a_2|^2 |a_3|^2) \\
      & \\
      & + \sigma^2 \delta_{n _1, 0}\delta_{n_2 , 0}  \times \{ \sigma^2 (\sigma^2  +a_1 a_2 + a^*_1 a^*_2 ) \\
      & \quad \quad + \sigma^2 \delta_{k_3, 1}  (a_2 a^*_3 + a^*_2 a_3 + 2 \sigma^2)\\
      & \quad \quad +  \sigma^2 \delta_{k_3, 1}  (a_1 a^*_3 +a^*_1 a_3 + 2\sigma^2) \}  \\
      & \\
      &  +  \sigma^2 \delta_{n _1, 0}\delta_{n_2 , 0} \delta_{k_3, 1}  \times \{ (|a_2|^2 + \sigma^2) (\sigma^2  + a_1 a_3 + a^*_1 a^*_3 )   \\
      & \quad \quad  + \sigma^2  (a_1 a^*_2 + a^*_1 a_2 + 2\sigma^2 ) \\
      & \quad \quad + \sigma^2  (a_2 a_3 + a^*_2 a^*_3) + \sigma^2 (a_1 a_2 + a^*_1 a^*_2) \} \\
      &  \\
      & + \sigma^2 \delta_{n _1, 0}\delta_{n_2 , 0} \delta_{k_3, 1}  \times \{(|a_1|^2 + \sigma^2)(\sigma^2  + a_2 a_3 + a^*_2 a^*_3 ) \\
      & \quad \quad + \sigma^2 (a_1 a_3 + a^*_1 a^*_3) + \sigma^2 (a_1 a_2 + a^*_1 a^*_2)  \} \\
      & \\
      & + \sigma^2 \delta_{n_1 , 0} \delta_{k_3, 1} \times \{ \sigma^2 (\sigma^2 + a_2 a^*_3 + a^*_2 a_3) \\
      & \quad \quad +\sigma^2 \delta_{n_2, 0}  (a_1 a^*_3 +a^*_1 a_3 + 2 \sigma^2  ) \} \\
      & \\
      & + \sigma^2 \delta_{n_2, 0} \delta_{k_3, 1} \times \{\sigma^2  (\sigma^2 + a_1 a^*_3 + a^*_1 a_3) \\
      & \quad \quad +\sigma^2 \delta_{n_1 , 0} (a_2 a^*_3 + a^*_2 a_3) \} \\
      &  \\
      & + \sigma^2 \delta_{n_1 , n_2} \times \{(|a_3|^2 + \sigma^2)  (\sigma^2 + a_1 a^*_2 + a^*_1 a_2) \}. \\
 \end{align}

Next, we combine terms. For the SDB, $n_1 = 0$ only for the first set of terms, where $n_2=0$ also. Therefore, $\delta_{n_1 , 0} \delta_{k_3, 1}$ in the third to last term is equivalent to $\delta_{n _1, 0}\delta_{n_2 , 0} \delta_{k_3, 1}$, and we will write it as the latter in order to combine it with the terms above. With that, only the last two terms have deltas not equivalent to $\delta_{n _1, 0}\delta_{n_2 , 0} \delta_{k_3, 1}$. In the second to last term, $\delta_{n_2 , 0}$ could happen when $n_1 = 0$ in the first set of SDB coefficients or $n_1=1$ in the second set of SDB coefficients. In the last term above, $\delta_{n_1 , n_2}$
 could happen in the first set of coefficients when $n_1 = n_2 = 0$ or in the second set of coefficients when $n_1 = n_2 = 1$. With this in mind, we split the variance cases, temporarily keeping $a_1, a_2, a_3$ notation for ease of readability,
\begin{align}
     & \mathrm{Var}(\tilde{b}_{j_1 j_2 k_3}) \text{ for } \tilde{b}_{j_1 j_2 k_3}\notin \{\tilde{b}_{001}, \tilde{b}_{00k}, \tilde{b}_{201}, \tilde{b}_{21k}  \}  \\
     & = \sigma ^6  + \sigma^4 (|a_1|^2   + |a_2|^2  + |a_3|^2 )  \\
      &  +\sigma^2 (|a_1|^2 |a_2|^2  + |a_1|^2 |a_3|^2  + |a_2|^2 |a_3|^2) \\
 \end{align}
 \begin{align}
     & \mathrm{Var}(\tilde{b}_{001})  \\
     & = \sigma ^6  + \sigma^4 (|a_1|^2   + |a_2|^2  + |a_3|^2 )  \\
      &  +\sigma^2 (|a_1|^2 |a_2|^2  + |a_1|^2 |a_3|^2  + |a_2|^2 |a_3|^2) \\
      & + \sigma^2   \times \{ \sigma^2 (\sigma^2  +a_1 a_2 + a^*_1 a^*_2 ) \\
      & \quad \quad + \sigma^2   (a_2 a^*_3 + a^*_2 a_3 + 2 \sigma^2)\\
      & \quad \quad +  \sigma^2   (a_1 a^*_3 +a^*_1 a_3 + 2\sigma^2) \}  \\
      &  +  \sigma^2  \times \{ (|a_2|^2 + \sigma^2) (\sigma^2  + a_1 a_3 + a^*_1 a^*_3 )   \\
      & \quad \quad  + \sigma^2  (a_1 a^*_2 + a^*_1 a_2 + 2\sigma^2 ) \\
      & \quad \quad + \sigma^2  (a_2 a_3 + a^*_2 a^*_3) + \sigma^2 (a_1 a_2 + a^*_1 a^*_2) \} \\
      & + \sigma^2  \times \{(|a_1|^2 + \sigma^2)(\sigma^2  + a_2 a_3 + a^*_2 a^*_3 ) \\
      & \quad \quad + \sigma^2 (a_1 a_3 + a^*_1 a^*_3) + \sigma^2 (a_1 a_2 + a^*_1 a^*_2)  \} \\
      & + \sigma^4 (\sigma^2 + a_2 a^*_3 + a^*_2 a_3) \\
      & \quad \quad +\sigma^4  (a_1 a^*_3 +a^*_1 a_3 + 2 \sigma^2  )  \\
      & +\sigma^4  (\sigma^2 + a_1 a^*_3 + a^*_1 a_3) \\
      & \quad \quad +\sigma^4  (a_2 a^*_3 + a^*_2 a_3) \\
        & + \sigma^2 (|a_3|^2 + \sigma^2)  (\sigma^2 + a_1 a^*_2 + a^*_1 a_2)
 \end{align}
 
 \begin{align}
     & \mathrm{Var}(\tilde{b}_{00k})  \\
     & = \sigma ^6  + \sigma^4 (|a_1|^2   + |a_2|^2  + |a_3|^2 )  \\
      &  +\sigma^2 (|a_1|^2 |a_2|^2  + |a_1|^2 |a_3|^2  + |a_2|^2 |a_3|^2) \\
      & + \sigma^4 (\sigma^2  +a_1 a_2 + a^*_1 a^*_2 ) \\
      & + \sigma^2 (|a_3|^2 + \sigma^2)  (\sigma^2 + a_1 a^*_2 + a^*_1 a_2)
 \end{align}
 \begin{align}
     & \mathrm{Var}(\tilde{b}_{201})  \\
     & = \sigma ^6  + \sigma^4 (|a_1|^2   + |a_2|^2  + |a_3|^2 )  \\
      &  +\sigma^2 (|a_1|^2 |a_2|^2  + |a_1|^2 |a_3|^2  + |a_2|^2 |a_3|^2) \\
      & + \sigma^4  (\sigma^2 + a_1 a^*_3 + a^*_1 a_3) \\
 \end{align}
 
\begin{align}
     & \mathrm{Var}(\tilde{b}_{21k})  \\
     & = \sigma ^6  + \sigma^4 (|a_1|^2   + |a_2|^2  + |a_3|^2 )  \\
      &  +\sigma^2 (|a_1|^2 |a_2|^2  + |a_1|^2 |a_3|^2  + |a_2|^2 |a_3|^2) \\
      & + \sigma^2 (|a_3|^2 + \sigma^2)  (\sigma^2 + a_1 a^*_2 + a^*_1 a_2)
 \end{align}

 Next, sorting each variance expression into second, fourth, and sixth order noise terms, 
 \begin{align}
     & \mathrm{Var}(\tilde{b}_{j_1 j_2 k_3}) \text{ for } \tilde{b}_{j_1 j_2 k_3}\notin \{\tilde{b}_{001}, \tilde{b}_{00k}, \tilde{b}_{201}, \tilde{b}_{21k}  \}  \\
     & = \sigma ^6  + \sigma^4 (|a_1|^2   + |a_2|^2  + |a_3|^2 )  \\
      &  +\sigma^2 (|a_1|^2 |a_2|^2  + |a_1|^2 |a_3|^2  + |a_2|^2 |a_3|^2) \\
 \end{align}
 \begin{align}
\mathrm{Var}(\tilde{b}_{001}) &= 15\sigma^6 \\
&+ \sigma^4\{2|a_1|^2 + 2|a_2|^2 + 2|a_3|^2 \\
&\quad + 3(a_1a_2 + a_1^*a_2^*) + 2(a_1a_2^* + a_1^*a_2) \\
&\quad + 3(a_2a_3^* + a_2^*a_3) + 2(a_2a_3 + a_2^*a_3^*) \\
&\quad + 3(a_1a_3^* + a_1^*a_3) + 2(a_1a_3 + a_1^*a_3^*)\} \\
&+ \sigma^2\{|a_1|^2|a_2|^2 + |a_1|^2|a_3|^2 + |a_2|^2|a_3|^2 \\
&\quad + |a_2|^2(a_1 a_3 + a_1^*a_3^*) \\
&\quad + |a_1|^2(a_2 a_3 + a_2^*a_3^*) \\
&\quad + |a_3|^2(a_1 a_2^* + a_1^*a_2)\}
\end{align}
 \begin{align}
\mathrm{Var}(\tilde{b}_{00k}) &= 3\sigma^6 \\
&+ \sigma^4\{|a_1|^2 + |a_2|^2 + 2|a_3|^2 \\
&\quad + (a_1a_2 + a_1^*a_2^*) + (a_1a_2^* + a_1^*a_2)\} \\
&+ \sigma^2\{|a_1|^2|a_2|^2 + |a_1|^2|a_3|^2 
+ |a_2|^2|a_3|^2 \\
&\quad + |a_3|^2(a_1a_2^* + a_1^*a_2)\}
\end{align}

 \begin{align}
     & \mathrm{Var}(\tilde{b}_{201})  \\
     & = 2 \sigma ^6  \\
     & + \sigma^4 \{|a_1|^2   + |a_2|^2  + |a_3|^2 + a_1 a^*_3 + a^*_1 a_3\}  \\
      &  +\sigma^2 \{|a_1|^2 |a_2|^2  + |a_1|^2 |a_3|^2  + |a_2|^2 |a_3|^2\} \\
 \end{align}
 \begin{align}
\mathrm{Var}(\tilde{b}_{21k}) &= 2\sigma^6 \\
&+ \sigma^4\{|a_1|^2 + |a_2|^2 + 2|a_3|^2 
+ (a_1a_2^* + a_1^*a_2)\} \\
&+ \sigma^2\{|a_1|^2|a_2|^2 + |a_1|^2|a_3|^2 
+ |a_2|^2|a_3|^2 \\
&\quad + |a_3|^2(a_1a_2^* + a_1^*a_2)\}
\end{align}

Finally, we reintroduce full notation ($a_1 \to a_{n_{j_1}, k_{j_1}}$, $a_2 \to a_{n_{j_2}, k_{j_2}}$, and $a^*_3 \to a^*_{n_{j_1}+ n_{j_2}, k_{j_3}}$). We recognize that $x+x^* = 2Re(x)$. For the SDB, we use,
\begin{align}
    & j= 0 \to n_0 = 0, k_0 = 1 \\
    & j = 2 \to n_2 = 1, k_2 = 1, 
\end{align}
so that, for several of the terms, we recognize that $a_{n,k} = |a_{n,k}|e^{-in\phi}$, so $a_{01} = |a_{0,1}|e^{-i0 \cdot \phi} = |a_{0,1}| = Re(a_{01})$, and $a^2_{01} = |a_{0,1}|^2e^{-i2\cdot 0 \cdot \phi} = |a_{0,1}|^2 = Re(a^2_{01}) = Re(a_{0,1})^2$. Rewriting in true SDB notation, 
 \begin{align}
     & \mathrm{Var}(\tilde{b}_{j_1 n_2 k_3}) \text{ for } \tilde{b}_{j_1 n_2 k_3}\notin \{\tilde{b}_{001}, \tilde{b}_{00k}, \tilde{b}_{201}, \tilde{b}_{21k}  \}  \\
     & = \sigma ^6  + \sigma^4 (|a_{n_{j_1}, k_{j_1}}|^2   + |a_{n_2, 1}|^2  + |a^*_{n_{j_1}+ n_2, k_{j_3}}|^2 )  \\
      &  +\sigma^2 (|a_{n_{j_1}, k_{j_1}}|^2 |a_{n_2, 1}|^2  + |a_{n_{j_1}, k_{j_1}}|^2 |a^*_{n_{j_1}+ n_2, k_{j_3}}|^2  + |a_{n_2, 1}|^2 |a^*_{n_{j_1}+ n_2, k_{j_3}}|^2) \\
      & \mathrm{Var}(\tilde{b}_{001}) = 15\sigma^6 
+ 36\sigma^4|a_{0,1}|^2
+ 9\sigma^2|a_{0,1}|^4\\
& \mathrm{Var}(\tilde{b}_{00k}) = 3\sigma^6 
+ \sigma^4\{6|a_{0,1}|^2 + 2|a_{0,k}|^2\}
+ \sigma^2\{|a_{0,1}|^4 + 4|a_{0,1}|^2|a_{0,k}|^2\}\\
& \mathrm{Var}(\tilde{b}_{201})  = 2 \sigma ^6   + \sigma^4 \{4|a_{1,1}|^2   + |a_{0,1}|^2  \}    +\sigma^2 \{|a_{1,1}|^4  + 2|a_{0,1}|^2 |a_{1,1}|^2\} \\
& \mathrm{Var}(\tilde{b}_{21k}) = 2\sigma^6 + \sigma^4\{4|a_{1,1}|^2 + 2|a_{2,k}|^2\} + \sigma^2\{|a_{1,1}|^4 + 4|a_{1,1}|^2|a_{2,k}|^2\}
 \end{align}

\end{proof}


\newpage
\section*{NeurIPS Paper Checklist}

The checklist is designed to encourage best practices for responsible machine learning research, addressing issues of reproducibility, transparency, research ethics, and societal impact. Do not remove the checklist: {\bf The papers not including the checklist will be desk rejected.} The checklist should follow the references and follow the (optional) supplemental material.  The checklist does NOT count towards the page
limit. 

Please read the checklist guidelines carefully for information on how to answer these questions. For each question in the checklist:
\begin{itemize}
    \item You should answer \answerYes{}, \answerNo{}, or \answerNA{}.
    \item \answerNA{} means either that the question is Not Applicable for that particular paper or the relevant information is Not Available.
    \item Please provide a short (1--2 sentence) justification right after your answer (even for \answerNA). 
\end{itemize}

{\bf The checklist answers are an integral part of your paper submission.} They are visible to the reviewers, area chairs, senior area chairs, and ethics reviewers. You will also be asked to include it (after eventual revisions) with the final version of your paper, and its final version will be published with the paper.

The reviewers of your paper will be asked to use the checklist as one of the factors in their evaluation. While \answerYes{} is generally preferable to \answerNo{}, it is perfectly acceptable to answer \answerNo{} provided a proper justification is given (e.g., error bars are not reported because it would be too computationally expensive'' or ``we were unable to find the license for the dataset we used''). In general, answering \answerNo{} or \answerNA{} is not grounds for rejection. While the questions are phrased in a binary way, we acknowledge that the true answer is often more nuanced, so please just use your best judgment and write a justification to elaborate. All supporting evidence can appear either in the main paper or the supplemental material, provided in appendix. If you answer \answerYes{} to a question, in the justification please point to the section(s) where related material for the question can be found.

IMPORTANT, please:
\begin{itemize}
    \item {\bf Delete this instruction block, but keep the section heading ``NeurIPS Paper Checklist"},
    \item  {\bf Keep the checklist subsection headings, questions/answers and guidelines below.}
    \item {\bf Do not modify the questions and only use the provided macros for your answers}.
\end{itemize}


\begin{enumerate}

\item {\bf Claims}
    \item[] Question: Do the main claims made in the abstract and introduction accurately reflect the paper's contributions and scope?
    \item[] Answer: \answerYes{} 
    \item[] Justification: We clearly state that our technical contribution is the disk bispectrum inversion, enabling the selection of a minimal subset of coefficients neccessary for complete image description.
    \item[] Guidelines:
    \begin{itemize}
        \item The answer \answerNA{} means that the abstract and introduction do not include the claims made in the paper.
        \item The abstract and/or introduction should clearly state the claims made, including the contributions made in the paper and important assumptions and limitations. A \answerNo{} or \answerNA{} answer to this question will not be perceived well by the reviewers. 
        \item The claims made should match theoretical and experimental results, and reflect how much the results can be expected to generalize to other settings. 
        \item It is fine to include aspirational goals as motivation as long as it is clear that these goals are not attained by the paper. 
    \end{itemize}

\item {\bf Limitations}
    \item[] Question: Does the paper discuss the limitations of the work performed by the authors?
    \item[] Answer: \answerYes{} 
    \item[] Justification: The limitations of the SDB are the same as the limitations of other bispectra. Bispectra are only invariant to one group action, and the SDB is only invariant to rotation. Additionally, its invariance is global, so it is not well suited for tasks requiring local invariance.
    \item[] Guidelines:
    \begin{itemize}
        \item The answer \answerNA{} means that the paper has no limitation while the answer \answerNo{} means that the paper has limitations, but those are not discussed in the paper. 
        \item The authors are encouraged to create a separate ``Limitations'' section in their paper.
        \item The paper should point out any strong assumptions and how robust the results are to violations of these assumptions (e.g., independence assumptions, noiseless settings, model well-specification, asymptotic approximations only holding locally). The authors should reflect on how these assumptions might be violated in practice and what the implications would be.
        \item The authors should reflect on the scope of the claims made, e.g., if the approach was only tested on a few datasets or with a few runs. In general, empirical results often depend on implicit assumptions, which should be articulated.
        \item The authors should reflect on the factors that influence the performance of the approach. For example, a facial recognition algorithm may perform poorly when image resolution is low or images are taken in low lighting. Or a speech-to-text system might not be used reliably to provide closed captions for online lectures because it fails to handle technical jargon.
        \item The authors should discuss the computational efficiency of the proposed algorithms and how they scale with dataset size.
        \item If applicable, the authors should discuss possible limitations of their approach to address problems of privacy and fairness.
        \item While the authors might fear that complete honesty about limitations might be used by reviewers as grounds for rejection, a worse outcome might be that reviewers discover limitations that aren't acknowledged in the paper. The authors should use their best judgment and recognize that individual actions in favor of transparency play an important role in developing norms that preserve the integrity of the community. Reviewers will be specifically instructed to not penalize honesty concerning limitations.
    \end{itemize}

\item {\bf Theory assumptions and proofs}
    \item[] Question: For each theoretical result, does the paper provide the full set of assumptions and a complete (and correct) proof?
    \item[] Answer: \answerYes{} 
    \item[] Justification: All noise proofs assume i.i.d. Gaussian noise is added at each point. The bispectrum inversion relies on non-zero DH coefficients, which is a canonical assumption in bispectrum inversion literature and can be easily satisfied if some amount of noise is present on the image.
    \item[] Guidelines:
    \begin{itemize}
        \item The answer \answerNA{} means that the paper does not include theoretical results. 
        \item All the theorems, formulas, and proofs in the paper should be numbered and cross-referenced.
        \item All assumptions should be clearly stated or referenced in the statement of any theorems.
        \item The proofs can either appear in the main paper or the supplemental material, but if they appear in the supplemental material, the authors are encouraged to provide a short proof sketch to provide intuition. 
        \item Inversely, any informal proof provided in the core of the paper should be complemented by formal proofs provided in appendix or supplemental material.
        \item Theorems and Lemmas that the proof relies upon should be properly referenced. 
    \end{itemize}

    \item {\bf Experimental result reproducibility}
    \item[] Question: Does the paper fully disclose all the information needed to reproduce the main experimental results of the paper to the extent that it affects the main claims and/or conclusions of the paper (regardless of whether the code and data are provided or not)?
    \item[] Answer: \answerYes{} 
    \item[] Justification: We provide information on noise, seed, MLP parameter counts and number of hidden layers. The code will be publicly available on Github upon acceptance.
    \item[] Guidelines:
    \begin{itemize}
        \item The answer \answerNA{} means that the paper does not include experiments.
        \item If the paper includes experiments, a \answerNo{} answer to this question will not be perceived well by the reviewers: Making the paper reproducible is important, regardless of whether the code and data are provided or not.
        \item If the contribution is a dataset and\slash or model, the authors should describe the steps taken to make their results reproducible or verifiable. 
        \item Depending on the contribution, reproducibility can be accomplished in various ways. For example, if the contribution is a novel architecture, describing the architecture fully might suffice, or if the contribution is a specific model and empirical evaluation, it may be necessary to either make it possible for others to replicate the model with the same dataset, or provide access to the model. In general. releasing code and data is often one good way to accomplish this, but reproducibility can also be provided via detailed instructions for how to replicate the results, access to a hosted model (e.g., in the case of a large language model), releasing of a model checkpoint, or other means that are appropriate to the research performed.
        \item While NeurIPS does not require releasing code, the conference does require all submissions to provide some reasonable avenue for reproducibility, which may depend on the nature of the contribution. For example
        \begin{enumerate}
            \item If the contribution is primarily a new algorithm, the paper should make it clear how to reproduce that algorithm.
            \item If the contribution is primarily a new model architecture, the paper should describe the architecture clearly and fully.
            \item If the contribution is a new model (e.g., a large language model), then there should either be a way to access this model for reproducing the results or a way to reproduce the model (e.g., with an open-source dataset or instructions for how to construct the dataset).
            \item We recognize that reproducibility may be tricky in some cases, in which case authors are welcome to describe the particular way they provide for reproducibility. In the case of closed-source models, it may be that access to the model is limited in some way (e.g., to registered users), but it should be possible for other researchers to have some path to reproducing or verifying the results.
        \end{enumerate}
    \end{itemize}

\item {\bf Open access to data and code}
    \item[] Question: Does the paper provide open access to the data and code, with sufficient instructions to faithfully reproduce the main experimental results, as described in supplemental material?
    \item[] Answer: \answerNo{} 
    \item[] Justification: Code will be publicly available on Github upon paper acceptance.
    \item[] Guidelines:
    \begin{itemize}
        \item The answer \answerNA{} means that paper does not include experiments requiring code.
        \item Please see the NeurIPS code and data submission guidelines (\url{https://neurips.cc/public/guides/CodeSubmissionPolicy}) for more details.
        \item While we encourage the release of code and data, we understand that this might not be possible, so \answerNo{} is an acceptable answer. Papers cannot be rejected simply for not including code, unless this is central to the contribution (e.g., for a new open-source benchmark).
        \item The instructions should contain the exact command and environment needed to run to reproduce the results. See the NeurIPS code and data submission guidelines (\url{https://neurips.cc/public/guides/CodeSubmissionPolicy}) for more details.
        \item The authors should provide instructions on data access and preparation, including how to access the raw data, preprocessed data, intermediate data, and generated data, etc.
        \item The authors should provide scripts to reproduce all experimental results for the new proposed method and baselines. If only a subset of experiments are reproducible, they should state which ones are omitted from the script and why.
        \item At submission time, to preserve anonymity, the authors should release anonymized versions (if applicable).
        \item Providing as much information as possible in supplemental material (appended to the paper) is recommended, but including URLs to data and code is permitted.
    \end{itemize}

\item {\bf Experimental setting/details}
    \item[] Question: Does the paper specify all the training and test details (e.g., data splits, hyperparameters, how they were chosen, type of optimizer) necessary to understand the results?
    \item[] Answer: \answerYes{} 
    \item[] Justification: In addition to providing adequate details in the core paper, we provide details on MLP architecture and hyperparameters in the appendix. Additionally, code will be made publicly available upon acceptance.
    \item[] Guidelines:
    \begin{itemize}
        \item The answer \answerNA{} means that the paper does not include experiments.
        \item The experimental setting should be presented in the core of the paper to a level of detail that is necessary to appreciate the results and make sense of them.
        \item The full details can be provided either with the code, in appendix, or as supplemental material.
    \end{itemize}

\item {\bf Experiment statistical significance}
    \item[] Question: Does the paper report error bars suitably and correctly defined or other appropriate information about the statistical significance of the experiments?
    \item[] Answer: \answerYes{} 
    \item[] Justification: The experiments are run across 5 seeds. Graph markers are the mean, and error bars show standard deviation.
    \item[] Guidelines:
    \begin{itemize}
        \item The answer \answerNA{} means that the paper does not include experiments.
        \item The authors should answer \answerYes{} if the results are accompanied by error bars, confidence intervals, or statistical significance tests, at least for the experiments that support the main claims of the paper.
        \item The factors of variability that the error bars are capturing should be clearly stated (for example, train/test split, initialization, random drawing of some parameter, or overall run with given experimental conditions).
        \item The method for calculating the error bars should be explained (closed form formula, call to a library function, bootstrap, etc.)
        \item The assumptions made should be given (e.g., Normally distributed errors).
        \item It should be clear whether the error bar is the standard deviation or the standard error of the mean.
        \item It is OK to report 1-sigma error bars, but one should state it. The authors should preferably report a 2-sigma error bar than state that they have a 96\% CI, if the hypothesis of Normality of errors is not verified.
        \item For asymmetric distributions, the authors should be careful not to show in tables or figures symmetric error bars that would yield results that are out of range (e.g., negative error rates).
        \item If error bars are reported in tables or plots, the authors should explain in the text how they were calculated and reference the corresponding figures or tables in the text.
    \end{itemize}

\item {\bf Experiments compute resources}
    \item[] Question: For each experiment, does the paper provide sufficient information on the computer resources (type of compute workers, memory, time of execution) needed to reproduce the experiments?
    \item[] Answer: \answerYes{} 
    \item[] Justification: We report that our experiments are run on one NVIDIA A100 GPU, and average train times are reported for each feature representation.
    \item[] Guidelines:
    \begin{itemize}
        \item The answer \answerNA{} means that the paper does not include experiments.
        \item The paper should indicate the type of compute workers CPU or GPU, internal cluster, or cloud provider, including relevant memory and storage.
        \item The paper should provide the amount of compute required for each of the individual experimental runs as well as estimate the total compute. 
        \item The paper should disclose whether the full research project required more compute than the experiments reported in the paper (e.g., preliminary or failed experiments that didn't make it into the paper). 
    \end{itemize}
    
\item {\bf Code of ethics}
    \item[] Question: Does the research conducted in the paper conform, in every respect, with the NeurIPS Code of Ethics \url{https://neurips.cc/public/EthicsGuidelines}?
    \item[] Answer: \answerYes{} 
    \item[] Justification: We follow the code of ethics.
    \item[] Guidelines:
    \begin{itemize}
        \item The answer \answerNA{} means that the authors have not reviewed the NeurIPS Code of Ethics.
        \item If the authors answer \answerNo, they should explain the special circumstances that require a deviation from the Code of Ethics.
        \item The authors should make sure to preserve anonymity (e.g., if there is a special consideration due to laws or regulations in their jurisdiction).
    \end{itemize}

\item {\bf Broader impacts}
    \item[] Question: Does the paper discuss both potential positive societal impacts and negative societal impacts of the work performed?
    \item[] Answer: \answerYes{} 
    \item[] Justification: The proposed method could offer faster training on some tasks, which could reduce energy expenditure. 
    \item[] Guidelines:
    \begin{itemize}
        \item The answer \answerNA{} means that there is no societal impact of the work performed.
        \item If the authors answer \answerNA{} or \answerNo, they should explain why their work has no societal impact or why the paper does not address societal impact.
        \item Examples of negative societal impacts include potential malicious or unintended uses (e.g., disinformation, generating fake profiles, surveillance), fairness considerations (e.g., deployment of technologies that could make decisions that unfairly impact specific groups), privacy considerations, and security considerations.
        \item The conference expects that many papers will be foundational research and not tied to particular applications, let alone deployments. However, if there is a direct path to any negative applications, the authors should point it out. For example, it is legitimate to point out that an improvement in the quality of generative models could be used to generate Deepfakes for disinformation. On the other hand, it is not needed to point out that a generic algorithm for optimizing neural networks could enable people to train models that generate Deepfakes faster.
        \item The authors should consider possible harms that could arise when the technology is being used as intended and functioning correctly, harms that could arise when the technology is being used as intended but gives incorrect results, and harms following from (intentional or unintentional) misuse of the technology.
        \item If there are negative societal impacts, the authors could also discuss possible mitigation strategies (e.g., gated release of models, providing defenses in addition to attacks, mechanisms for monitoring misuse, mechanisms to monitor how a system learns from feedback over time, improving the efficiency and accessibility of ML).
    \end{itemize}
    
\item {\bf Safeguards}
    \item[] Question: Does the paper describe safeguards that have been put in place for responsible release of data or models that have a high risk for misuse (e.g., pre-trained language models, image generators, or scraped datasets)?
    \item[] Answer: \answerNA{}{} 
    \item[] Justification: This does not apply to our paper. we do not release data, pre-trained language models, or image generators.
    \item[] Guidelines:
    \begin{itemize}
        \item The answer \answerNA{} means that the paper poses no such risks.
        \item Released models that have a high risk for misuse or dual-use should be released with necessary safeguards to allow for controlled use of the model, for example by requiring that users adhere to usage guidelines or restrictions to access the model or implementing safety filters. 
        \item Datasets that have been scraped from the Internet could pose safety risks. The authors should describe how they avoided releasing unsafe images.
        \item We recognize that providing effective safeguards is challenging, and many papers do not require this, but we encourage authors to take this into account and make a best faith effort.
    \end{itemize}

\item {\bf Licenses for existing assets}
    \item[] Question: Are the creators or original owners of assets (e.g., code, data, models), used in the paper, properly credited and are the license and terms of use explicitly mentioned and properly respected?
    \item[] Answer: \answerYes{} 
    \item[] Justification: To the best of our knowledge, we have cited all appropriate works relevant to our paper.
    \item[] Guidelines:
    \begin{itemize}
        \item The answer \answerNA{} means that the paper does not use existing assets.
        \item The authors should cite the original paper that produced the code package or dataset.
        \item The authors should state which version of the asset is used and, if possible, include a URL.
        \item The name of the license (e.g., CC-BY 4.0) should be included for each asset.
        \item For scraped data from a particular source (e.g., website), the copyright and terms of service of that source should be provided.
        \item If assets are released, the license, copyright information, and terms of use in the package should be provided. For popular datasets, \url{paperswithcode.com/datasets} has curated licenses for some datasets. Their licensing guide can help determine the license of a dataset.
        \item For existing datasets that are re-packaged, both the original license and the license of the derived asset (if it has changed) should be provided.
        \item If this information is not available online, the authors are encouraged to reach out to the asset's creators.
    \end{itemize}

\item {\bf New assets}
    \item[] Question: Are new assets introduced in the paper well documented and is the documentation provided alongside the assets?
    \item[] Answer: \answerNA{}{} 
    \item[] Justification: We will release the code upon acceptance.
    \item[] Guidelines:
    \begin{itemize}
        \item The answer \answerNA{} means that the paper does not release new assets.
        \item Researchers should communicate the details of the dataset\slash code\slash model as part of their submissions via structured templates. This includes details about training, license, limitations, etc. 
        \item The paper should discuss whether and how consent was obtained from people whose asset is used.
        \item At submission time, remember to anonymize your assets (if applicable). You can either create an anonymized URL or include an anonymized zip file.
    \end{itemize}

\item {\bf Crowdsourcing and research with human subjects}
    \item[] Question: For crowdsourcing experiments and research with human subjects, does the paper include the full text of instructions given to participants and screenshots, if applicable, as well as details about compensation (if any)? 
    \item[] Answer: \answerNA{} 
    \item[] Justification: Our work does not include human subjects.
    \item[] Guidelines:
    \begin{itemize}
        \item The answer \answerNA{} means that the paper does not involve crowdsourcing nor research with human subjects.
        \item Including this information in the supplemental material is fine, but if the main contribution of the paper involves human subjects, then as much detail as possible should be included in the main paper. 
        \item According to the NeurIPS Code of Ethics, workers involved in data collection, curation, or other labor should be paid at least the minimum wage in the country of the data collector. 
    \end{itemize}

\item {\bf Institutional review board (IRB) approvals or equivalent for research with human subjects}
    \item[] Question: Does the paper describe potential risks incurred by study participants, whether such risks were disclosed to the subjects, and whether Institutional Review Board (IRB) approvals (or an equivalent approval/review based on the requirements of your country or institution) were obtained?
    \item[] Answer: \answerNA{}{} 
    \item[] Justification: Our work does not include human subjects.
    \item[] Guidelines:
    \begin{itemize}
        \item The answer \answerNA{} means that the paper does not involve crowdsourcing nor research with human subjects.
        \item Depending on the country in which research is conducted, IRB approval (or equivalent) may be required for any human subjects research. If you obtained IRB approval, you should clearly state this in the paper. 
        \item We recognize that the procedures for this may vary significantly between institutions and locations, and we expect authors to adhere to the NeurIPS Code of Ethics and the guidelines for their institution. 
        \item For initial submissions, do not include any information that would break anonymity (if applicable), such as the institution conducting the review.
    \end{itemize}

\item {\bf Declaration of LLM usage}
    \item[] Question: Does the paper describe the usage of LLMs if it is an important, original, or non-standard component of the core methods in this research? Note that if the LLM is used only for writing, editing, or formatting purposes and does \emph{not} impact the core methodology, scientific rigor, or originality of the research, declaration is not required.
    \item[] Answer: \answerYes{} 
    \item[] Justification: Our paper does not use LLMs for important, original, or non-standard components of the core methods in the research.
    \item[] Guidelines:
    \begin{itemize}
        \item The answer \answerNA{} means that the core method development in this research does not involve LLMs as any important, original, or non-standard components.
        \item Please refer to our LLM policy in the NeurIPS handbook for what should or should not be described.
    \end{itemize}

\end{enumerate}

\end{document}